\documentclass[aip,jcp,reprint,superscriptaddress]{revtex4-2}

\usepackage{graphicx}

\usepackage{amsmath,amssymb}

\usepackage{bm}

\usepackage{booktabs}
\usepackage{tabularx}

\usepackage{xcolor}

\usepackage[
 colorlinks=true,
 linkcolor=blue,
 citecolor=blue,
 urlcolor=blue,
 hypertexnames=false
]{hyperref}

\usepackage{here}

\graphicspath{{./}}

\begin{document}
\title{Machine learning evaluation of structural descriptors for
supercooled water}

\author{Kohei Yoshikawa}
\affiliation{Division of Chemical Engineering, Department of Materials Engineering Science, Graduate School of Engineering Science, The University of Osaka, Toyonaka, Osaka 560-8531, Japan}

\author{Kokoro Shikata}
\affiliation{Division of Chemical Engineering, Department of Materials Engineering Science, Graduate School of Engineering Science, The University of Osaka, Toyonaka, Osaka 560-8531, Japan}

\author{Kang Kim}
\email{kk@cheng.es.osaka-u.ac.jp}
\affiliation{Division of Chemical Engineering, Department of Materials Engineering Science, Graduate School of Engineering Science, The University of Osaka, Toyonaka, Osaka 560-8531, Japan}

\author{Nobuyuki Matubayasi}
\email{nobuyuki@cheng.es.osaka-u.ac.jp}
\affiliation{Division of Chemical Engineering, Department of Materials Engineering Science, Graduate School of Engineering Science, The University of Osaka, Toyonaka, Osaka 560-8531, Japan}

\date{\today}

\begin{abstract}
The anomalous behavior of liquid water is widely associated with a liquid-liquid
 phase transition between high- and low-density states in the
 supercooled regime. 
At the microscopic level, tetrahedral hydrogen-bond networks
 govern these properties, motivating
 structural descriptors that characterize local molecular
 environments.
These structural descriptors quantify features such as tetrahedral
 order, local density, and the separation between the first and second
 coordination shells; however, they have largely been proposed 
 independently, with limited
 systematic comparison. 
Here we evaluate 16 previously proposed 
 descriptors using a
 neural-network-based temperature classification framework, enabling 
 an objective assessment of their ability to distinguish
 temperature-dependent structural changes in supercooled water.
We further apply an explainable artificial intelligence method that
 identifies the structural features responsible for the model
 predictions.
This approach reveals how different descriptors encode local structural
 information and establishes a data-driven framework for benchmarking
 structural descriptors in liquid water.
\end{abstract}
\maketitle

\section*{Introduction}

Liquid water exhibits significant anomalies, including a
density maximum near 4 $^\circ$C, 
in response to variations in pressure and temperature; 
these anomalies 
arise from rearrangements in the
local molecular structure associated with the H-bond
network~\cite{gallo2016Water, gallo2021Advances}.
When liquid water enters the supercooled regime, its anomalous properties 
become extensively noticeable~\cite{angell1983Supercooled,
debenedetti2003Supercooleda}.
In particular, in the supercooled regime, both isothermal
compressibility and specific heat sharply increase with cooling,
indicative of divergent behavior~\cite{tanaka2000Simple, holten2012Thermodynamics}.

This phenomenon is commonly interpreted within the framework of the
liquid-liquid phase transition (LLPT) hypothesis, which proposes that
water exists in two liquid states: low-density liquid (LDL)
and a high-density liquid (HDL)~\cite{poole1992Phase, 
handle2017Supercooled, palmer2018Advances, debenedetti2020Second}.
Transformations between low-
and high-density amorphous ices are further
considered to represent the
arrested form of the liquid-liquid transition occurring under deeper
supercooled conditions, terminating at a
liquid–liquid critical point
(LLCP)~\cite{mishima1985Apparently, mishima1998Relationship,
kim2017Maxima, kim2020Experimental}.
The LLPT hypothesis has become an important 
framework that provides a consistent and comprehensive explanation for a wide range of
anomalous behaviors observed in supercooled water.

The two-state model describing the competition between HDL and
LDL structures is extensively used as an effective framework for
elucidating the structural transformations in supercooled
water based on molecular dynamics (MD)
simulations~\cite{sciortino1997Line, starr1999Dynamics, xu2006Relationship, cuthbertson2011Mixturelike,
holten2012Entropydriven, overduin2013Analysis, yagasaki2014Spontaneous,
palmer2014Metastable, 
singh2016Twostate, biddle2017Twostructure, kawasaki2017Identifying,
guillaud2017Decoupling, galamba2017Hydrogenbond, saito2018Crucial,
saito2019Thermodynamic, martelli2019Unravelling, 
neophytou2022Topological, sciortino2024Freeenergy, malosso2025Dynamical}.
To quantitatively examine this framework, scalar structural descriptors
have been diversely introduced to characterize the local environment of individual
water molecules~\cite{duboue-dijon2015Characterization,
shi2018Microscopic, tanaka2019Revealing, verde2019Comparing,
verde2022Journey}.

Representative examples include $q_\mathrm{tet}$, which quantifies the
tetrahedral ordering among the four nearest neighbors of a water
molecule~\cite{matubayasi1994MatchingMismatching, chau1998New, errington2001Relationship}, and $\zeta$,
which measures the degree of order within H-bond 
networks~\cite{russo2014Understanding}.
These descriptors enable clear differentiation between the highly
ordered open structure associated with LDL and the compact densely
packed structure characteristic of HDL.
Furthermore, Local Structure Index (LSI) serves as another structural
descriptor that distinguishes HDL-like and LDL-like environments based
on local structural fluctuations~\cite{shiratani1996Growth, shiratani1998Molecular}.
Recently, Node Total Communicability (NTC) has been proposed as a
graph-theoretical structural descriptor for supercooled water that
characterizes not only local arrangements within the first 
coordination shell, but also medium- and long-range structural orders~\cite{faccio2022Low}.
Furthermore, $\Psi$ is a descriptor to exploits information on the connectivity of
H-bond networks~\cite{foffi2023Correlated, foffi2024Identification}.

\begin{figure*}[htbp]
\centering
\includegraphics[width=\textwidth]{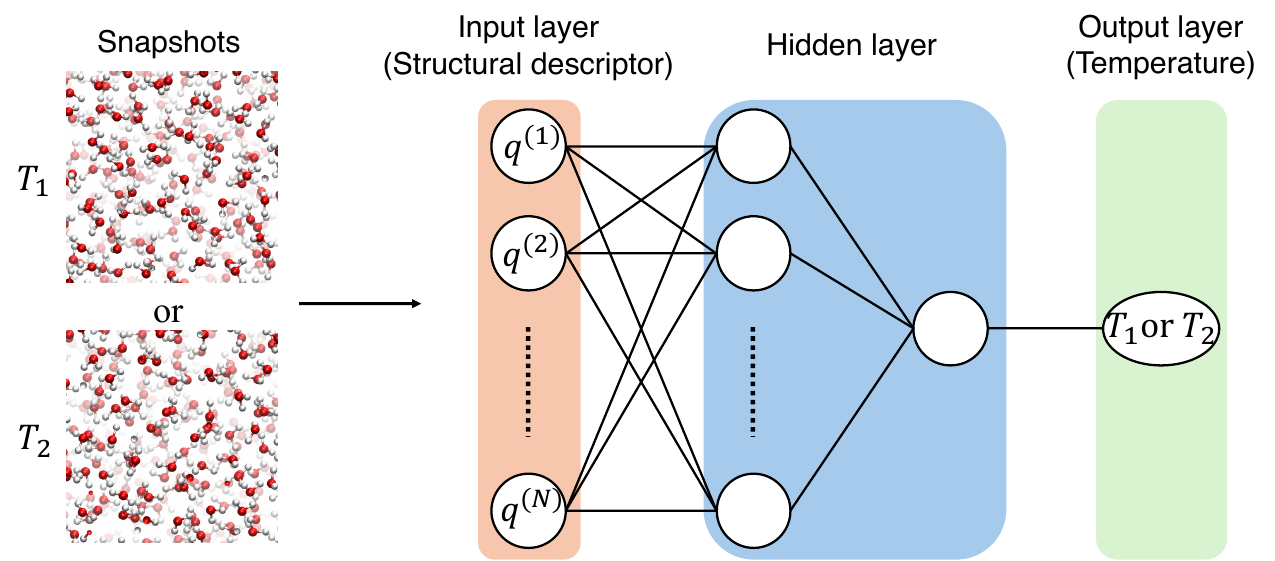}
\caption{\textbf{Fully connecteed Neural network model for temperature classification.} 
Schematic illustration of the neural network architecture used in this study.
Structural descriptors $(q^{(1)}, q^{(2)}, \cdots, q^{(N)})$, assigned to
individual molecules and computed from MD
 simulations of the two temperatures, $T_1$ and $T_2$, are provided as the input variables.
Number of input nodes corresponds to the number of molecules in the
 MD snapshots ($N=1000$).
Therefore, the input is a 1000 $\times$ 1 matrix.
Network consists of a single hidden layer with the 
same number of nodes as that of the input layer, followed by a hidden
 layer with a single node, and employs a 
sigmoid function. 
Output layer produced a
binary value (0 or 1, corresponding to $T_1$ or $T_2$, respectively), 
 enabling the model to learn temperature-dependent structural differences.
Note that the node representing the bias term is omitted from the diagram.}
\label{fig:ML}
\end{figure*}

Notably, Foffi \textit{et al.} investigated the correlation between
critical density fluctuations near
the LLCP in supercooled water and various structural descriptors using MD
simulations~\cite{foffi2023Correlated}.
Their results reveal that most structural descriptors exhibit strong
correlations with density, indicating a close relationship between
critical fluctuations and the distinct structural characteristics of 
H-bond networks in HDL and LDL.

In contrast, 
studies have demonstrated the occurrence of a ``structural crossover,'' 
away from the LLCP under ambient pressure ($\sim$1 bar), 
in which the ratio of HDL-like to LDL-like structures continuously varies 
across the Widom line, that is, the locus of maxima in thermodynamic
response functions~\cite{xu2005Relation, stanley2008Liquid}.
Although this crossover does not correspond to a distinct phase
transition, it represents the conditions under which anomalies in density,
specific heat, and related properties become most evident, providing
key insight into the structural origins of supercooled water. 
Accordingly, quantitative evaluation of the structural crossover under 
ambient pressure is essential for elucidating the universal
structural anomalies of liquid water.
Therefore, examining the performances of existing
structural descriptors under ambient pressure is crucial. 

However, because these descriptors qualitatively differ in their
definitions and quantitatively vary in their dimensions and scales, a direct
comparison is not straightforward.
Thus, a consistent framework for estimating and comparing their
performances of these descriptors is
required, along with assessment methods capable of reliably
describing structural variations under supercooled conditions.

This study evaluates the performances of 16 available structural
descriptors in distinguishing HDL-like and LDL-like
structures within the structural crossover region defined by the Widom
line using machine learning.
At first, MD simulations of TIP4P/2005 water were 
performed under isochoric (1 g/cm$^3$) and isobaric (1 bar)
conditions, with temperatures ranging from 200 to 300 K,  adequately
away from the LLCP.
Note that the temperature, pressure, and mass density at the LLCP of TIP4P/2005 water
were estimated as $(T_\mathrm{C}, P_\mathrm{C}, \rho_\mathrm{C})\approx
(182~\mathrm{K}, 170~\mathrm{MPa},
1.017~\mathrm{g/cm^3})$, respectively.\cite{singh2016Twostate, biddle2017Twostructure}.
Additionally, 
the crossover between HDL-like and LDL-like states occurs near the Widom line
at approximately 230 K under 1 bar~\cite{singh2016Twostate,
biddle2017Twostructure}.
Details of the MD simulations are provided in the Methods section, and static
and dynamic
properties are shown in Supplementary Figs.~1 and 2.

Machine learning using structural descriptors
has proven to be a powerful
approach for classifying local structures in liquid water, 
glasses, and colloids~\cite{geiger2013Neural, 
boattini2019Unsupervised, boattini2020Autonomously,
martelli2020Connection, doi2021Searching}.
In this study, we propose an alternative machine-learning framework, in which
structural data obtained from MD simulations of supercooled water were
used to construct a neural network model that takes per-molecule structural
descriptor values 
as inputs and temperature as the output.
The objective of this study is not merely to identify HDL-like and
LDL-like structural states, but rather to quantitatively evaluate,
within a unified framework, the extent to which different structural
descriptors are sensitive to temperature-dependent structural changes. 
From this perspective, temperature classification is used as a
controlled proxy task for measuring the sensitivity of structural
descriptors to such changes. 
In other words, we assess the extent to which each descriptor captures
intrinsic structural information by examining whether it can distinguish
structures sampled at different temperatures.
Note that we recently employed graph neural networks to perform a
similar machine learning-based temperature classification using
structural snapshots of glass-forming liquids~\cite{yoshikawa2025Graph}.

Figure~\ref{fig:ML} schematically depicts the fully connected neural
network (FCNN) model
employed in this study.
This FCNN model quantitatively assesses the ability of each
structural descriptor to distinguish between two different temperatures
using the area under 
the Receiver Operating Characteristic (ROC) curve (ROC-AUC, hereinafter
referred to as 
AUC) as the evaluation metric.
A ROC curve is a graphical tool used to evaluate classification models
in machine learning. 
The AUC quantifies the overall classification performance. 
An AUC value close to 1 indicates that the model can accurately
distinguish between the two classes, whereas an AUC value close to 0.5
corresponds to random classification.
Moreover, to enhance the interpretability of neural networks,
which are inherently black-box predictive models, we employed
explainable artificial intelligence (XAI)~\cite{adadi2018Peeking,
molnar2020Interpretable, holzinger2022Explainable}.
Specifically, we applied
Local Interpretable Model-Agnostic Explanations (LIME)~\cite{ribeiro2016Why}, a post-hoc
interpretability method, to examine whether the decision-making of the
model 
is consistent with physically meaningful structural changes.
Therefore, the advance of this study lies not only in
reaffirming the importance of each structural descriptor in
distinguishing between HDL-like and LDL-like structures, but also in
quantitatively comparing their respective contributions using machine
learning and in uniformly evaluating the performance of structural
descriptors with different dimensions and scales.

\begin{figure}[t]
\centering
\includegraphics[width=0.85\linewidth]{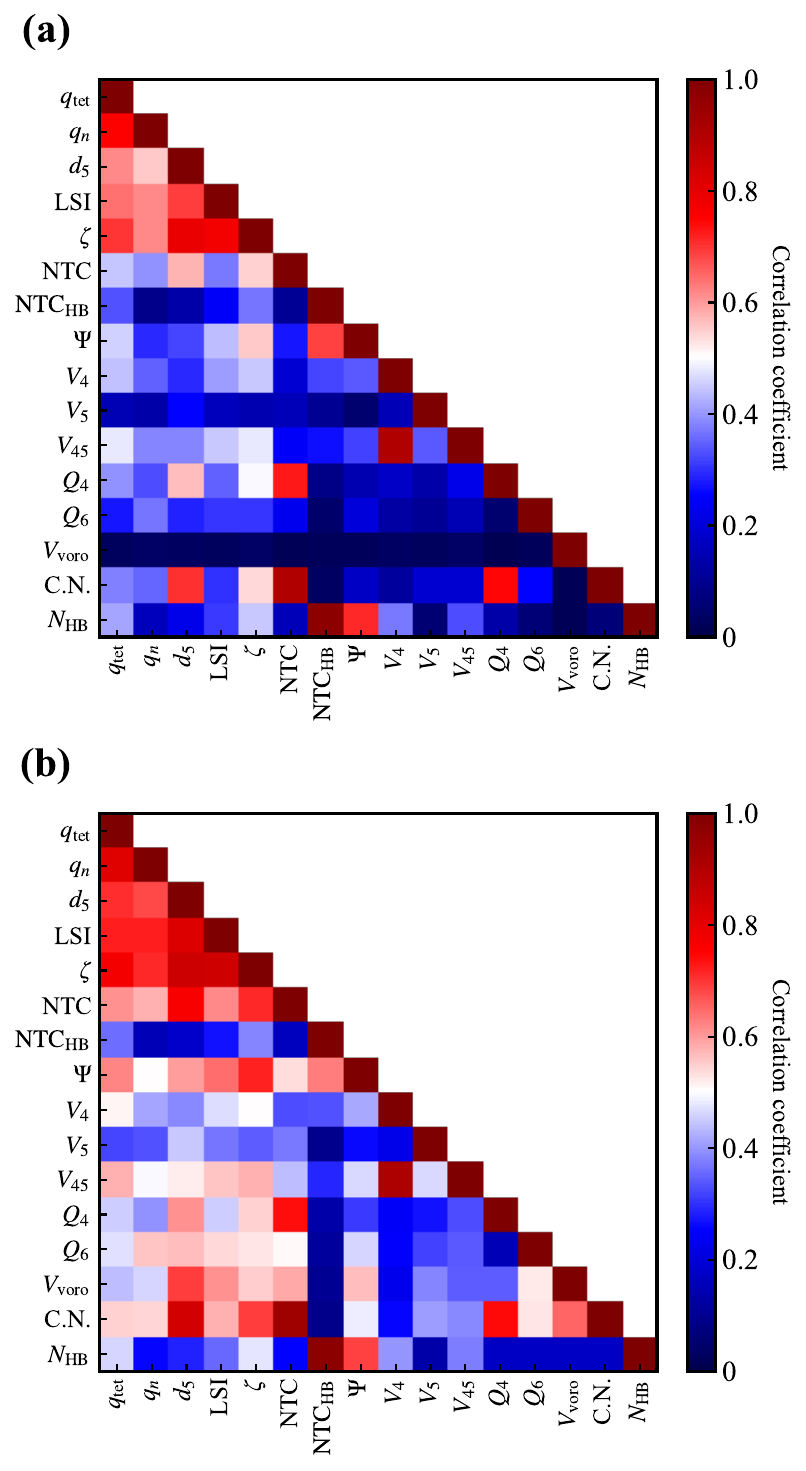}
\caption{\textbf{Correlation coefficients among structural descriptors.}
(a) isochoric and (b) isobaric conditions.
\label{fig:correlation_coefficient}
}
\end{figure}

\begin{figure*}[htbp]
\centering
\includegraphics[width=\textwidth]{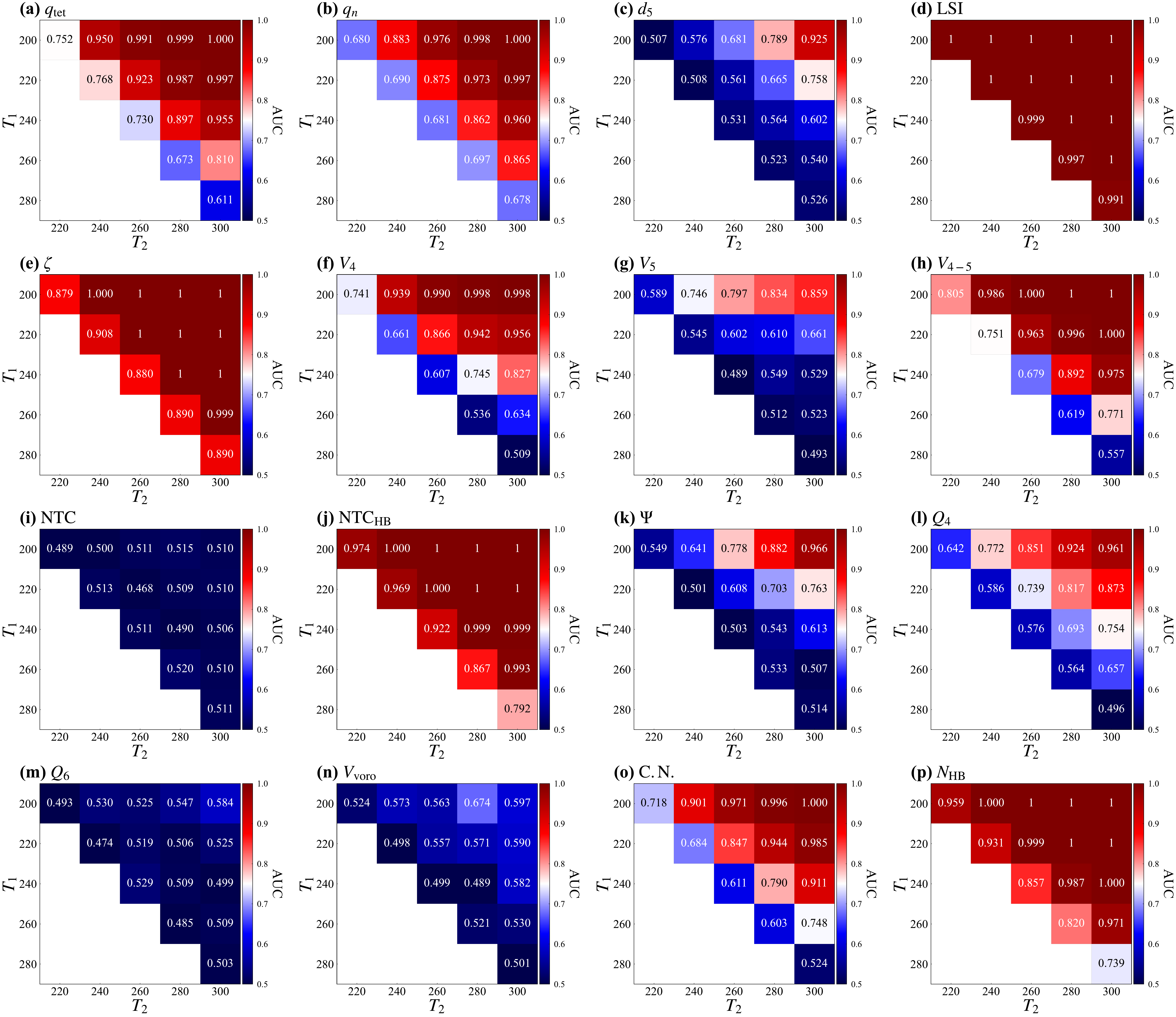}
\caption{
\textbf{Classification performance of structural descriptors under
 isochoric conditions.}
Combinations of AUC values for each structural descriptor under
 isochoric conditions.
$T_1$ and $T_2$ denote temperature pairs with $T_1 < T_2$.
Color bar represents AUC values in the 
$0.5\le \mathrm{AUC}\le 1.0$ range.
An entry of 
 AUC$=1.000$ represents a value of $0.999\dots$.
Panels show results for 
(a) $q_\mathrm{tet}$, (b) $q_n$, (c) $d_5$, (d) LSI, (e) $\zeta$, (f)
 $V_4$, (g) $V_5$, (h) $V_{4-5}$, (i) NTC, (j) NTC$_\mathrm{HB}$, (k)
 $\Psi$, (l) $Q_4$, (m) $Q_6$, (n) $V_\mathrm{voro}$, (o) C.N., and (p) $N_\mathrm{HB}$.}
\label{fig:NVTAUC}
\end{figure*}

\begin{figure*}[htbp]
\centering
\includegraphics[width=\textwidth]{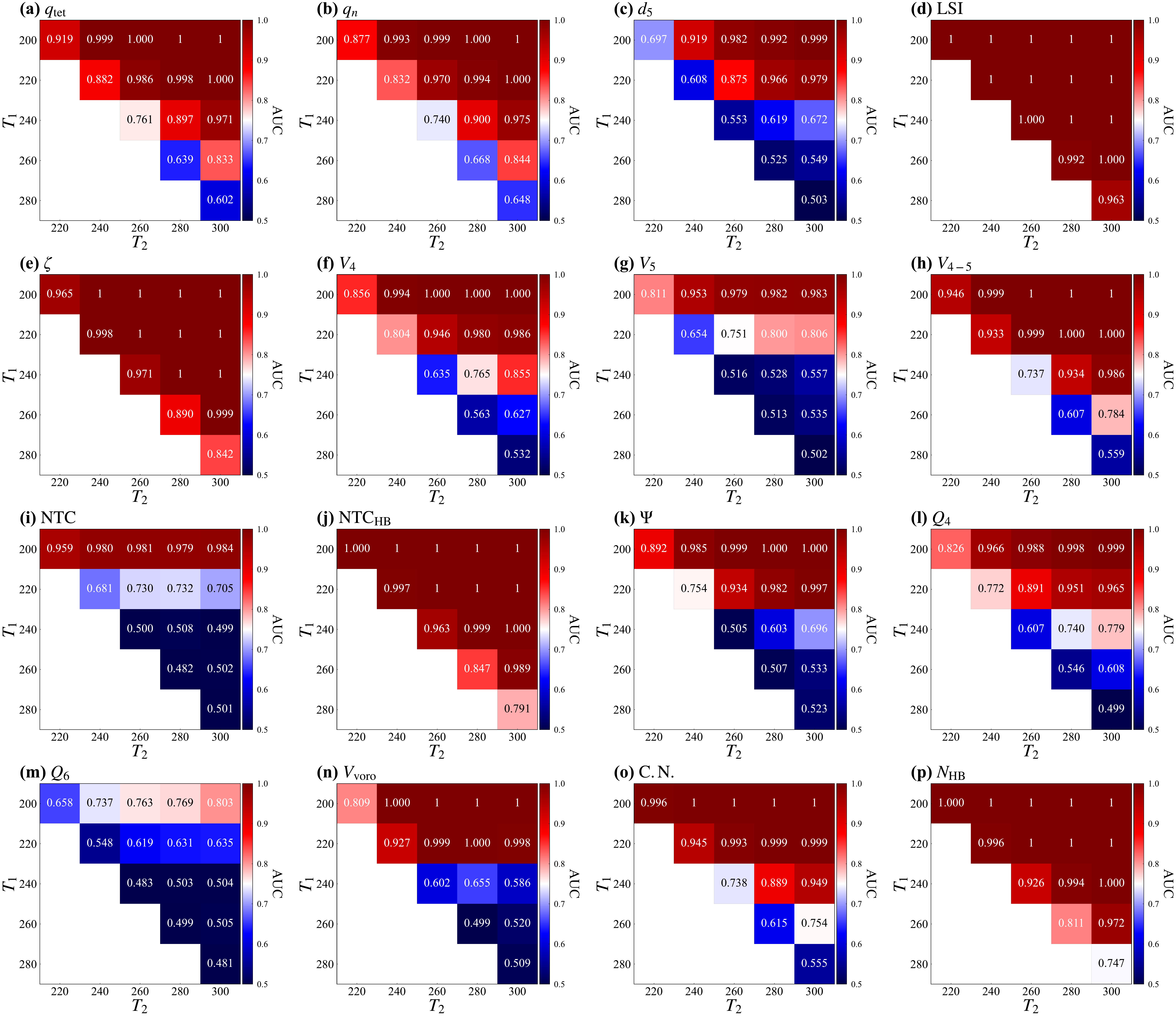}
\caption{\textbf{Classification performances of structural descriptors under
 isobaric conditions.}
Combinations of AUC values for each structural descriptor under
 isobaric conditions.
$T_1$ and $T_2$ denote temperature pairs with $T_1 < T_2$.
Color bar represents AUC values in the 
$0.5\le \mathrm{AUC}\le 1.0$ range.
An entry of 
 AUC$=1.000$ represents a value of $0.999\dots$.
Panels show results for 
(a) $q_\mathrm{tet}$, (b) $q_n$, (c) $d_5$, (d) LSI, (e) $\zeta$, (f)
 $V_4$, (g) $V_5$, (h) $V_{4-5}$, (i) NTC, (j) NTC$_\mathrm{HB}$, (k)
 $\Psi$, (l) $Q_4$, (m) $Q_6$, (n) $V_\mathrm{voro}$, (o) C.N., and (p) $N_\mathrm{HB}$.}
\label{fig:NPTAUC}
\end{figure*}

\section*{Results and Discussion}

\subsection*{Correlation coefficients among structural descriptors}
\label{correlation_coefficient}

Structural descriptors quantify the degree of local order around each
molecule; for liquid water, they are often defined in terms of
tetrahedrality, 
local density, interaction energy, and topological properties of H-bond networks. 
Herein, 16 available structural descriptors were calculated from
MD simulations. 
The definition of each descriptor is provided in the Methods section,
and corresponding 
temperature-dependent distributions are shown in
Supplementary Figs.~3 and 4.

Initially, correlation coefficients were computed to examine the
similarities among the structural descriptors.
In this study, the correlation coefficient was evaluated in the absolute form of
the Pearson product-moment correlation coefficient, denoted as 
\begin{align}
|r|=\left| \frac{\sum_{i=1}^{N}\left(X_i-\overline{X_i}\right)\left(Y_i-\overline{Y_i}\right)}{\sqrt{\sum_{i=1}^{N}\left(X_i-\overline{X}\right)^2\sum_{j=1}^{N}\left(Y_j-\overline{Y}\right)^2}}\right|,
\end{align}
where $X_i$ and $Y_i$ represents the values assigned to the molecule $i$
by the two structural
descriptors $X$ and $Y$, respectively.
Note that $\overline{X}$ and $\overline{Y}$ represent the corresponding mean values.

For each simulated temperature in the 200-300 K range, 
$|r|$ was calculated and subsequently averaged over the 1000 or 500
configurations sampled, and the resulting values obtained were then
averaged over all temperatures.
Figure~\ref{fig:correlation_coefficient} depicts $|r|$ among the
structural descriptors for isochoric
(a) and isobaric (b) conditions.
Structural descriptors with large $|r|$ values ($|r| \ge 0.7$)
under isochoric conditions were classified into four main groups.

First group comprises $q_\mathrm{tet}$, $q_n$, $d_5$, LSI, and $\zeta$.
Among them, $q_\mathrm{tet}$ and $q_n$ 
characterize the degree of tetrahedral order according to angular
information, whereas
$d_5$, LSI, and $\zeta$
evaluate 
the degree of separation between the first and second
coordination shells 
using distance-based information.

Second group consists of NTC, $Q_4$, and coordination number (C.N.).
Among them, NTC, and C.N. quantify local structural fluctuations using
distance-based information, whereas $Q_4$ describes 
orientational symmetry without
explicitly considering local density. 
The observation that these evidently distinct descriptors are mutually
correlated yet uncorrelated with the first group suggests that the
second group assesses the ordering associated with local structural
fluctuations and lacks
geometric information related to the tetrahedral structures formed
through H-bonds.

Third group comprises $\Psi$, $N_\mathrm{HB}$ and
NTC$_\mathrm{HB}$, which
characterize the topological features of H-bond networks.
In particular, $N_\mathrm{HB}$ and
NTC$_\mathrm{HB}$ quantify the 
connectivity of the H-bond network via graph-theoretical approaches.
Their relatively weak correlations with the descriptors in the
first group imply that the topological properties of H-bond networks are not
necessarily strongly related to the tetrahedral structures.

Fourth group consists of $V_4$ and $V_{4-5}$, both of which represent
energetic quantities associated with negative interactions arising from
H-bonds.

Contrary to these four groups, 
$V_\mathrm{voro}$ and $Q_6$ exhibit almost no correlation with other
descriptors under isochoric conditions, whereas
weak correlations emerges under isobaric conditions.
$V_\mathrm{voro}$ estimates local volume fluctuations, whreas 
$Q_6$ 
describes orientational symmetry in a manner similar to that of $Q_4$.
These results suggest that density fluctuations enhance the ordering of
local structures, specifically under isobaric conditions.
Moreover, $\Psi$ in the third group demonstrates stronger 
correlations with the structural descriptors 
in the first group under isobaric conditions, implying 
that
variations in H-bond networks are influenced by density changes.
Finally, $V_5$ exhibits no correlations with other descriptors under either
isochoric or isobaric conditions.

\subsection*{Machine learning evaluation of structural descriptors}
\label{results:nvt_vs_npt}

Figures~\ref{fig:NVTAUC} and \ref{fig:NPTAUC} show the performances of
the structural descriptors that
classify two different temperatures, namely, $T_1$ and $T_2$, through the FCNN
under isochoric and isobaric conditions, respectively.
For each combination, temperatures were selected such that $T_1 < T_2$,
and 
determinations of $T_1$ and $T_2$ were set as binary 
classification tasks when 
output values were $< 0.5$ and $\ge 0.5$, respectively.
The classification performances of the models were quantitatively
assessed using AUC values under both isochoric and isobaric conditions.
Table~\ref{table:average_AUC} summarizes the average AUC values for each
structural descriptor depicted in Figs.~\ref{fig:NVTAUC} and
\ref{fig:NPTAUC}. 
Based on these average values, the descriptors were categorized into four
groups:
(1) AUC $\ge$ 0.9: 
LSI, $\zeta$, NTC$_\mathrm{HB}$, and $N_\mathrm{HB}$;
(2) $0.75 \le \mathrm{AUC} < 0.9$: $q_\mathrm{tet}$, 
$V_{4-5}$, 
$q_n$, C.N., $V_4$, 
and $Q_4$;
(3) $0.65 \le \mathrm{AUC} < 0.75$: 
$\Psi$, $d_5$, $V_\mathrm{voro}$, and $V_5$; 
and (4) AUC $<$ 0.65: NTC and $Q_6$.

\begin{table}[t]
\centering
\caption{\textbf{Ranking of structural descriptors based on AUC values.}
Descriptors are sorted by the average AUC over isochoric and isobaric conditions.}
\small
\setlength{\tabcolsep}{6pt}
\renewcommand{\arraystretch}{1.1}

\begin{tabular}{c l c c c}
\toprule
Rank & Descriptor & Average & Isochoric & Isobaric \\
\midrule
1  & LSI              & 0.998 & 0.999 & 0.997 \\
2  & $\zeta$          & 0.970 & 0.963 & 0.978 \\
3  & NTC$_\mathrm{HB}$& 0.970 & 0.968 & 0.972 \\
4  & $N_\mathrm{HB}$  & 0.957 & 0.951 & 0.963 \\
5  & $q_\mathrm{tet}$ & 0.884 & 0.869 & 0.899 \\
6  & $V_{4-5}$        & 0.883 & 0.866 & 0.899 \\
7  & $q_n$            & 0.875 & 0.854 & 0.896 \\
8  & C.N.             & 0.855 & 0.815 & 0.895 \\
9  & $V_4$            & 0.816 & 0.796 & 0.836 \\
10 & $Q_4$            & 0.768 & 0.727 & 0.809 \\
11 & $\Psi$           & 0.717 & 0.640 & 0.794 \\
12 & $d_5$            & 0.690 & 0.617 & 0.763 \\
13 & $V_\mathrm{voro}$& 0.679 & 0.551 & 0.806 \\
14 & $V_5$            & 0.674 & 0.622 & 0.725 \\
15 & NTC              & 0.610 & 0.505 & 0.715 \\
16 & $Q_6$            & 0.563 & 0.516 & 0.609 \\
\bottomrule
\end{tabular}

\label{table:average_AUC}
\end{table}

\begin{figure*}[t]
\centering
\includegraphics[width=\textwidth]{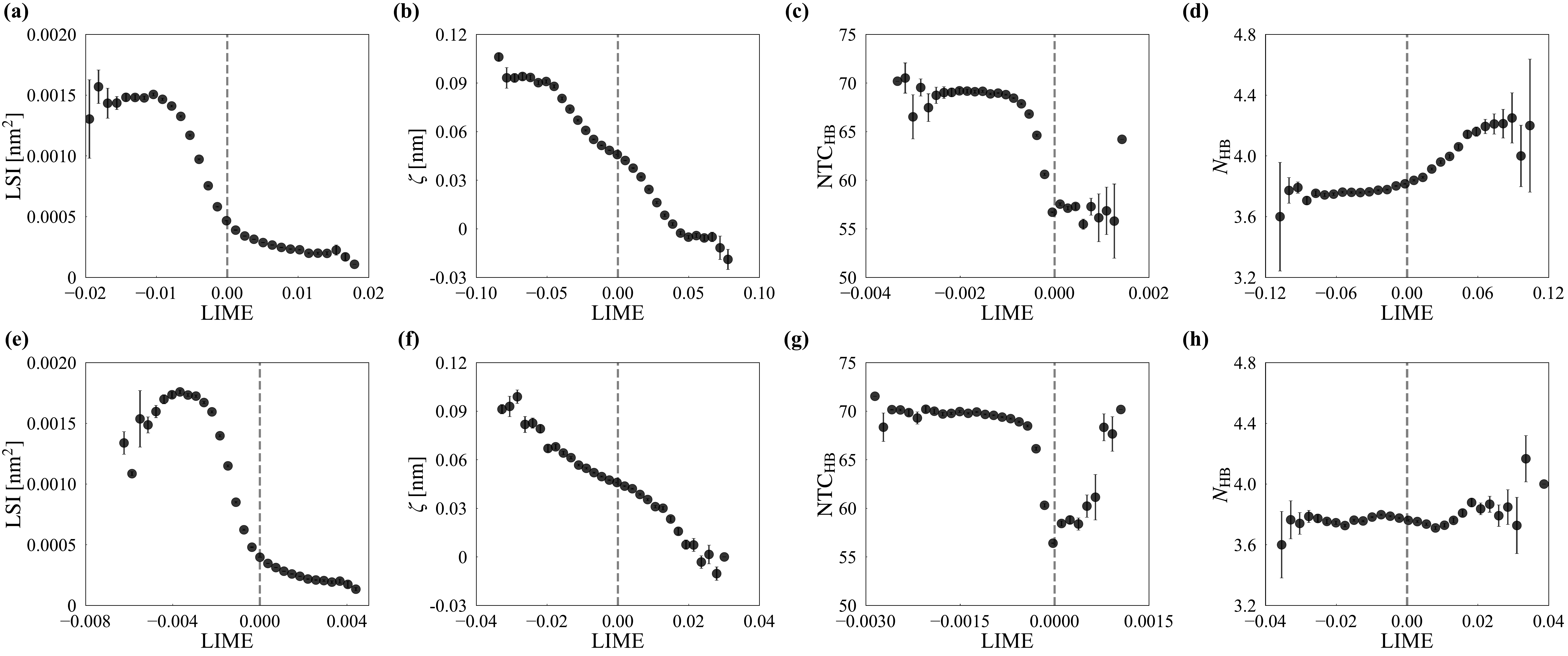}
\caption{\textbf{LIME-based interpretation of the model predictions.}
 Average values of structural descriptors as functions of LIME:
 LSI [(a) and (e)], $\zeta$ [(b) and (f)], NTC
 [(c) and (g)], and $N_\mathrm{HB}$ [(d) and (h)]. 
Panels (a)-(d) correspond to isochoric conditions, whereas panels
 (e)-(h) are related to isobaric conditions.
}
\label{fig:LIME}
\end{figure*}

Herein, we sequentially discuss the characteristics of each group.
(1) LSI, $\zeta$, $\mathrm{NTC}_\mathrm{HB}$, and $N_\mathrm{HB}$ all
demonstrate high classification performances across all temperature
combinations. 
In particular, LSI achieves nearly perfect discrimination, with AUC
$\ge$ 0.991 under isochoric conditions and AUC $\ge$ 0.963 under
isobaric conditions.
For $\zeta$, the minimum AUC value is 0.842, indicating that this 
descriptor also maintains high accuracy.
These results suggest that the degree of separation between the first and second
coordination shells, a feature common to both
descriptors, is a key factor in discriminating temperature-dependent
structural changes.
Furthermore, compared to $\zeta$, LSI incorporates information from a
larger number of surrounding molecules relative to the central molecule,
which accounts for its higher classification performance.
NTC$_\mathrm{HB}$ quantifies the graph-based topological feature, in which O
atoms and H-bonds are treated as nodes and edges, respectively, whereas
$N_\mathrm{HB}$ represents the number of H-bonds.
Both descriptors exhibit high AUC values
(AUC $\ge$ 0.791 and AUC $\ge$ 0.739, respectively). 
As shown in Fig.~\ref{fig:correlation_coefficient},
NTC$_\mathrm{HB}$ and $N_\mathrm{HB}$ exhibit weak correlations
with LSI and $\zeta$, revealing that H-bond network connectivity provides
an additional structural feature for distinguishing
temperature-dependent structural changes, independent of the 
separation between the first and second coordination shells.

(2) Classification performances of $q_\mathrm{tet}$, 
$V_{4-5}$, 
$q_n$, C.N., $V_4$, 
and $Q_4$
degrade at high temperatures.
Specifically, under isochoric conditions, the AUC
values approached those of random classification.
This is because 
these structural descriptors are defined based on angles and
distances, and are thus more sensitive to thermal fluctuations at high
temperatures than 
LSI and $\zeta$, which directly include information on H-bonds.
In fact, as shown in Supplementary Figs.~3 and 4, the overlap of the distribution
functions increases at high temperatures, reducing classification
accuracy.
Although C.N. is relatively less affected by thermal fluctuations, its
distribution differs from that of 
$N_\mathrm{HB}$ at high temperatures as compared to the case at low
temperatures (see 
Supplementary Figs.~3 and 4), 
implying that it does not fully acquire the H-bond information, which reduces
the classification accuracy.
In contrast, for 
these structural descriptors, the AUC values under isobaric
conditions are generally higher than those under isochoric conditions,
indicating that the FCNN model can effectively learn the tendency of local
structures to become more ordered due to density changes.
In particular, 
AUC values higher than 0.75 are 
obtained for classification between low temperatures or between low and
high temperatures, suggesting that the relationship between structure
and temperature is more clearly described in the low-temperature region.

(3) $\Psi$, $d_5$, $V_\mathrm{voro}$, and $V_5$
overall demonstrate low classification performances.
Under isochoric conditions, only structures at low and high
temperatures are distinguishable.
$d_5$ and $V_5$ characterize the fifth nearest
molecule from the central molecule and thereby quantify overcoordination.
$\Psi$ also serves as an indirect measure of overcoordination
characterized by 
the H-bond network connectivity.
Accordingly, these descriptors acquire a common structural
feature associated with overcoordination, and significant classification
is possible only when the temperature differences are sufficiently large.
Notably, although $d_5$ correlates with LSI and $\zeta$, its classification
performance is inferior to those of LSI and $\zeta$ because $d_5$
accounts only for overcoordination and does not characterize broader
features of the degree of separation between the first and second coordination shells.
Classification performance
of $V_\mathrm{voro}$ is close to random under isochoric conditions 
probably because the local Voronoi volume 
demonstrates slight temperature dependence when the system volume is fixed.
Contrarily, under isobaric
conditions, the classification accuracy of $V_\mathrm{voro}$
for $T_1=$200 K and 220 K considerably increases.
This temperature range is close to and below the Widom line.
Specifically, the high accuracy for temperatures below 220 K implies 
that $V_\mathrm{voro}$ is sensitive to the HDL-like and LDL-like
structure crossover when
the density is allowed to fluctuate.

(4) NTC and $Q_6$ exhibit low classification
performances, particularly under isochoric conditions, where their
performances are close to that of random classification.
However, in
classifications comparing the 200 K state and higher
temperatures under isobaric conditions, NTC yields AUC values of at
least 0.95.
Under isobaric conditions, $Q_6$ also demonstrates AUC
values of 0.85 or lower for the 200 and 300 K classifications, and low
accuracy is observed for other temperature combinations.
In NTC, O atoms are represented as nodes and edges are defined
between nearest neighbor O atoms, in contrast to NTC$_\mathrm{HB}$.
While the H-bond-based variant NTC$_\mathrm{HB}$ in group (1)
achieves high classification accuracy, the performance of NTC degrades
when edges are defined solely based on distance.
This contrast highlights the crucial role of H-bond connectivity in
accurately describing the underlying structural features.
$Q_6$ demonstrates inferior classification performance than $Q_4$.
Note that the values of $Q_4$, $Q_6$, and C.N. are sensitive
to 
the cutoff distance $r_\mathrm{cut}$, and the adopted values are
typically based on empirical considerations with respect to the 
first coordination shell.
In this study, $r_\mathrm{cut}=0.35$ nm was adopted to be consistent
with the distance criterion for the H-bond state (details are provides
in the Methods section).
As these descriptors are sensitive to the choice of $r_\mathrm{cut}$,
variations in this parameter are expected to substantially 
affect machine learning accuracy.
Consistent with this expectation, Supplementary Fig.~5
shows that 
the AUC
values of $Q_4$, $Q_6$, and C.N. for the 200 and 300 K temperature
classifications under isochoric conditions peak at around 
$r_\mathrm{cut}=0.3$ nm, which approximately corresponds to the first peak position of the
radial distribution function $g(r)$, as depicted in
Supplementary Figs.~1 and 2.

Overall, the isobaric systems exhibited higher AUC values than those of the
isochoric systems, and specifically evident improvements in
classification performance are noticed at low temperatures. 
This indicates that density changes under isobaric conditions are
closely related to 
the ordering of local structures associated with the HDL-like and
LDL-like structural
crossover, and that the machine learning model successfully describes this
behavior. 
In contrast, in the high-temperature range of 260-300 K, where 
density variations are small (as shown in Supplementary Fig.~2(f)), the
classification performance
decreases for many structural descriptors. 
Even under these conditions, LSI, $\zeta$, NTC$_\mathrm{HB}$,
and $N_\mathrm{HB}$ retain high classification performances,
confirming
their robustness as structural descriptors for quantifying local structural
order in supercooled water.

In addition, to examine the limits of the classification
performance of the four descriptors, LSI, $\zeta$, NTC$_\mathrm{HB}$,
and, $N_\mathrm{HB}$ from group (1), which showed good performance with
the FCNN, we evaluated their classification performance using logistic
regression, a linear classification model.
The results under isochoric and isobaric conditions are shown in 
Supplementary Figs.~6 and 7, respectively.
The AUC values of LSI remained largely unchanged, while a slight decline
was observed for $\zeta$
In contrast, the performance of NTC$_\mathrm{HB}$ and $N_\mathrm{HB}$
decreased substantially, with AUC values close to 0.5 for all
temperature combinations.
These results indicate that the influence of structural descriptors on
classification performance depends on the machine-learning model used.
From this perspective, LSI and $\zeta$ appear to be particular robust
among the structural descriptors examined.

\subsection*{Interpretability of machine learning predictions Using LIME}
\label{results:LIME}

Finally, we applied LIME as XAI to analyze the
contribution of each
structural descriptor to temperature classification in the FCNN model. 
Specifically, the LIME values correspond to the coefficients of each input
feature when the original FCNN is locally approximated by a
linear regression model around the target prediction, representing both
the magnitude and the sign of each input's contribution to the
classification result.
Thus, the absolute value of the LIME value indicates the importance of an
input, while its sign indicates the direction of its contribution to the
classification outcome.
The LIME values 
were calculated separately 
for the FCNN models 
using LSI,
$\zeta$, NTC$_\mathrm{HB}$, and $N_\mathrm{HB}$ as target descriptors
belonging to group (1), using data at 200 and 300 K.
In the FCNN, the model was trained to predict the high-temperature
state $T_2$ using binary cross entropy as the loss function. 
Accordingly, positive LIME values indicate contributions to the
high-temperature prediction, whereas negative values represent 
contributions
to the low-temperature prediction.

Figure~\ref{fig:LIME} depicts the relationship between LIME values and
structural descriptors.
In practice, the LIME values on the horizontal axis are binned using a
fixed bin width, and the average value of the corresponding structural
descriptor is evaluated for each bin.
LSI and $\zeta$ exhibit a clear trend in which the descriptor values
increase when the LIME values become more negative and decreased when the
LIME values become more positive. 
This indicates that LDL-like local structures contribute to the predictions
of low temperatures, whereas HDL-like local structures contribute to
the predictions of high temperatures.
These results demonstrate that the classification rationale learned by
the machine learning model are consistent with the temperature
dependences of the corresponding distribution functions of LSI and
$\zeta$ (see Supplementary Figs.~3 and 4), particularly highlighting 
the degree of separation between the first and second coordination shells.

NTC$_\mathrm{HB}$ exhibits a trend similar to those of LSI and $\zeta$. 
Physically, this behavior is attributable to the development of H-bond
networks at lower temperatures, as shown in the distribution functions
in Supplementary Figs.~3 and 4, which leads
to significantly larger NTC$_\mathrm{HB}$ values with a decrease in temperature. 
Nevertheless, unlike the cases of LSI and $\zeta$, the data are biased toward negative
LIME values, implying that machine learning models using
NTC$_\mathrm{HB}$ places considerable emphasis on predicting low temperatures. 
Additionally, large dispersions are observed in the positive LIME region
corresponding to high-temperature predictions, and some values overlap
with those at low temperatures where the LIME values are negative,
specifically under isobaric conditions. 
These features are consistent with the reduced classification accuracy
between high-temperature states noticed in Figs.~\ref{fig:NVTAUC} and
\ref{fig:NPTAUC}, suggesting that NTC$_\mathrm{HB}$ is predominantly
sensitive to low-temperature structures associated with H-bond networks.

$N_\mathrm{HB}$
values remain close and slightly below 4 across the entire LIME range under both
isochoric and isobaric 
conditions, and 
only a slight increase is detected with an increase in the LIME values.
This behavior implies that LIME emphasizes overcoordination at high
temperatures. 
This interpretation is
consistent with the increasing population of
$N_\mathrm{HB}=5$ observed in the distribution functions depicted 
in Supplementary Figs.~3 and 4; however, 
the population of $N_\mathrm{HB}=3$ remains larger than that of $N_\mathrm{HB}=5$.
Nevertheless, the overall variation with respect to the LIME value is
marginal possibly because 
$N_\mathrm{HB}$ takes discrete integer values in a narrow range from 2 to 5.

These LIME results are consistent with the
temperature-dependent changes observed in the distribution functions of
the structural descriptors, suggesting that the machine learning models
are trained in a manner that does not contradict the known trends in
temperature-induced structural changes.
While statistical measures such as distribution functions represent
trends across the entire system, LIME is characterized by its ability to
quantify which molecules contribute to classification predictions within
a single snapshot.
Evaluating these local contributions provides a complementary
perspective for interpreting machine learning models in addition to
conventional statistical analyses.

\section*{Conclusions}

In this study, we examined 16 previously proposed structural descriptors
that quantify local structures in supercooled water under isochoric
(1 g/cm$^3$) and isobaric (1 bar) conditions
using MD
simulations.
In particular, we assessed the abilities of these descriptors to
distinguish temperature-dependent
structural changes via neural network-based machine learning.

Specifically, we performed a temperature classification task using the 
FCNN, with each structural descriptor used as
an input feature, and quantitatively evaluated the classification
performance using the AUC.
The temperature classification in this study serves as a
prove for measuring the ``structural sensitivity'' of the descriptors.
It is well documented that in supercooled water the contribution of LDL-like
structures increases at lower temperatures, whereas that of HDL-like
structures increases at higher temperatures; thus, structural
differences associated with temperature changes are closely related to
these variations in local structure.
Consequently, the classification performance between different
temperatures can be interpreted as an indicator of how effectively the
descriptors capture the difference between HDL-like and LDL-like environments.
Furthermore, the results obtained in this study are consistent with
distribution functions, as shown in Supplementary Figs.~3 and 4, indicating
that the machine learning model
describes structural changes in a manner consistent with established
understanding.

The results reveal that LSI and $\zeta$ achieve the highest
classification accuracies, demonstrating that these descriptors
effectively describe temperature dependent-structural changes in
supercooled water.
NTC$_\mathrm{HB}$ and $N_\mathrm{HB}$, which exhibit the next highest
classification performances, quantify the connectivity of the H-bond network. 
Their weak correlations with LSI and $\zeta$ suggests that these
descriptors obtain 
complementary structural features for differentiating 
temperature-dependent structural changes, independent of the degree of
separation between the first and second coordination shells.
Overall, these results indicate that structural changes in supercooled
water under 
ambient pressure are
governed by both the geometric order of the coordination shell and 
connectivity of the H-bond networks.

The comparison between isochoric and isobaric conditions aims
to examine whether the physical insights derived from distribution
functions, as shown in Supplementary Figs.~3 and 4, are
consistent with the classification performance obtained through machine learning. 
Specifically, by considering both isochoric conditions, where density
fluctuations are suppressed, and isobaric conditions, where density
varies, we investigate how the distinguishability of structural changes
differs between these ensembles. 
Based on the distribution functions, structural changes are expected to
be enhanced under isobaric conditions, making classification easier. 
In this study, an overall improvement in classification performance was
observed under isobaric conditions, suggesting that density fluctuations
provide additional information useful for discrimination.
It is also interesting to note that even for structural descriptors that
do not explicitly include density information, such as tetrahedral order
parameters, $q_\mathrm{tet}$ and $q_n$, classification performance also
improves under isobaric conditions.
These results suggest that while density fluctuations contribute to improved
performance under isobaric conditions, sensitivity to local tetrahedral structural
changes also plays an important role in the classification.

The machine learning model in this study uses descriptor
values corresponding to
each molecule as input, and the final classification result is
obtained by nonlinearly integrating this information in the hidden
layers. 
In this sense, 
the model 
performs classification based on the entire snapshot and does not
directly identify individual local environments, including HDL/LDL
heterogeneous structures or the bimodal nature of descriptor distributions. 
However, 
the XAI analysis using LIME elucidates
the principle underlying the learned temperature classification of the
neural network model.
In particular, applying LIME enables quantification of the
contribution of each input (\textit{i.e.}, each molecule) to the
classification result for a given snapshot. 
From this perspective, our machine-learning framework can be viewed as a
method for indirectly analyzing the relationship between local
structures and global classification outcomes. 

In fact, the LIME results show that molecules with larger LSI and $\zeta$
values preferentially contribute 
to low-temperature (LDL-like) predictions, whereas molecules with smaller values
contribute to high-temperature (HDL-like) predictions. 
This reveals that these descriptors describe the  structural changes in
supercooled water through the separation of the first and second
coordination shells, yielding a physically plausible interpretation
consistent with the temperature dependences of the corresponding
distribution functions.
Moreover, molecules with large NTC$_\mathrm{HB}$ values contribute to
the 
low-temperature predictions, highlighting the development of H-bond
networks as a
key factor driving structural variations with a decrease in temperature.
In contrast, although the correspondence between $N_\mathrm{HB}$ and the
LIME values is limited, the analysis emphasizes an enhanced contribution
from overcoordination in the high-temperature regime.

In conclusion, this study assessed 16 existing structural descriptors
for liquid water in a unified machine-learning framework, 
explaining the structural features characterized by each descriptor and
their temperature dependence. 
In particular, LSI and $\zeta$ were identified as highly efficient 
descriptors for characterizing local structural changes in supercooled
water. 
Their combination with H-bond network descriptors affords to a more
comprehensive understanding of the HDL-like and LDL-like structural crossover in supercooled
water.

Based on these findings, the development of new structural
descriptors that integrate the length scales and information content of
available structures is expected.
Such descriptors may enable the direct identification of HDL-like and
LDL-like environments, establishing a framework to distinguish them, and
addressing the transferability across different water models.
Additionally, extending the analysis to higher-order structural
representations using nonlinear models, such as graph neural networks,
is estimated to further advance this field~\cite{ishiai2023Graph,
ishiai2024GraphNeuralNetworkBased}.
Another promising direction is to explore the features of H-bond networks by
establishing connections between descriptors used in machine-learning
potential frameworks, including 
atom-centered symmetry functions (ACSF) and smooth overlap of atomic
positions (SOAP), and physically inspired and hand-crafted structural descriptors. 
Studies along these lines have been reported not only at the liquid–gas
interface~\cite{donkor2023MachineLearning} but also in supercooled water~\cite{donkor2024Local}.

\section*{Methods}

\subsection*{MD simulations}
\label{method:MD}

MD simulations were conducted using the TIP4P/2005
water model~\cite{abascal2005General, abascal2010Widom} with the GROMACS
2020.6 package~\cite{abraham2015GROMACS}.
Simulated system consisted of $N=1000$
water molecules in a cubic box with periodic boundary conditions, and a 
linear dimension of 
approximately 3.1 nm, corresponding to a mass density of 1 g/cm$^3$.
Simulations were performed at the target temperatures of $T=300$, 280, 260,
240, 220, and 200 K.

We prepared 1000 independent initial structures and performed
equilibration for each under $NVT$ conditions while gradually lowering
the temperature from 300 K to 200 K. 
At each temperature, the systems were equilibrated using the 
Nos\'{e}--Hoover thermostat for up to 100 ns, followed by a 1 ps
simulation in the microcanonical
ensemble to obtain structural data for machine learning. 
By repeating this procedure at each temperature, we generated a dataset
of 1000 structures under isochoric conditions.

Moreover, equilibration under isobaric conditions under 1 bar was
performed using the Nos\'{e}--Hoover thermostat and the Parrinello--Rahman
barostat following equilibration in the $NVT$ ensemble, with
an equilibration time of up to 1 $\mu$s.
Furthermore, a 10 ps simulation was conducted in the $NPT$ ensemble, and
the structure exhibiting the smallest deviation from the average density
during this run was selected. 
Thereafter, a 10 ps simulation in the canonical ensemble followed by a
1 ps simulation in the microcanonical ensemble was performed to
formulate structural data under isobaric conditions.
A total of 1000 structural datasets were produced at each temperature,
except for 200 K, where the computational cost limited the dataset size
to 500.

Supplementary Figs. 1 and 2 depict the
temperature dependences of the O-O radial distribution
function $g(r)$, mean-square displacement (MSD) of O atoms, H-bond time correlation
function $P_\mathrm{HB}(t)$, HB lifetime $\tau_\mathrm{HB}$, diffusion coefficient
$D$, potential energy $V$, and pressure $p$ or mass density $\rho$ under isochoric
and isobaric conditions, respectively. 
Although gradual temperature dependence is observed for $g(r)$ and $V$,
dynamical properties such as
$D$ and $\tau_\mathrm{HB}$ demonstrate evident changes with a decrease
in temperature.

According to the conventionally accepted definition, a pair of water
molecules is considered to be in a H-bond state when the O-O
interatomic distance is within 0.35 nm and the H-O$\cdots$O angle is
within 30$^\circ$~\cite{luzar1996Effect, luzar1996Hydrogenbond}.
$P_\mathrm{HB}(t)$ is defined using the 
H-bond index $h_{ij}(t)$ for the molecular pair of $(i, j)$ as
follows~\cite{rapaport1983Hydrogen, luzar1996Effect,
luzar1996Hydrogenbond, kumar2007Hydrogen}:
\begin{align}
P_\mathrm{HB} (t) = \frac{\langle h_{ij}(t)h_{ij}(0)\rangle}{\langle h_{ij}(0)\rangle}, 
\end{align}
where $h_{ij}(t)$ is 1 if the molecules $i$ and $j$ form the H-bond
at the time $t$, and 0 otherwise.
Herein, $\langle \cdots \rangle$ represents the ensemble average over all
molecular pairs and initial times.
From the fitting of $P_\mathrm{HB}(t)$ to the
Kohlrausch--Williams--Watts (KWW) function, namely, $P_\mathrm{HB}(t)
=\exp[-(t/\tau_\mathrm{KWW})^{\beta_\mathrm{KWW}}]$, 
the HB lifetime $\tau_\mathrm{HB}$ was evaluated by $\tau_\mathrm{HB}
=(\tau_\mathrm{KWW}/\beta_\mathrm{KWW})\Gamma(1/\beta_\mathrm{KWW})$
with the Gamma function $\Gamma(\cdots)$.
Moreover, the diffusion coefficient $D$ was obtained from dividing the
long-time limit of the MSD by $6t$.
Note that the equilibrium times of 100 ns and 1 $\mu$s under isochoric
and isobaric conditions, respectively, correspond to the durations
required for $P_\mathrm{HB}(t)$ to decay to zero at 200 K.

\subsection*{Structural descriptors}
\label{method:SI}

Structural descriptors are metrics that quantify the degree of order of
a molecular structure by assigning scalar values based on the local
environment of each molecule.
For liquid systems, the structure has generally been characterized using
two-body correlation functions for example, radial distribution
functions and
static structure factors. 
As these quantities are averaged over spherical symmetry, they
cannot adequately describe the local geometric order. 
This limitation is particularly noticeable in supercooled liquids, in
which 
temperature-dependent variations in the local structure play crucial roles. 
To address this issue, various structural descriptors have been proposed
that incorporate local many-body correlations based on geometric
features such as angular arrangements and
intermolecular distance~\cite{duboue-dijon2015Characterization,
shi2018Microscopic, tanaka2019Revealing, verde2019Comparing,
verde2022Journey}.
Furthermore, in liquid water, the production of a
tetrahedral network is regarded as a low-density highly ordered state, 
whereas coordination defects and more compact configurations correspond
to high-density disordered states~\cite{soper2000Structures, yan2007Structure}.
Hereinafter, we summarize the definitions of the 16 structural descriptors
employed in this study.
The temperature dependences of the descriptor distributions under
isochoric and isobaric conditions are shown in
Supplementary Figs.~3 and 4, respectively.
Note that in addition to the structural descriptors considered herein, 
additional structural descriptors, including helicity~\cite{neophytou2022Topological} and local
order metric~\cite{martelli2018Localorder, martelli2019Unravelling},
have been proposed.

(a) $q_{\mathrm{tet}}$ measures the degree of tetrahedral coordination
of liquid water and is defined by~\cite{chau1998New, errington2001Relationship}
\begin{align}
q_{\mathrm{tet}}=1-\frac{3}{8}\sum_{j=1}^{3}\sum_{k=j+1}^{4}\left(\cos\psi_{jk}+\frac{1}{3}\right)^2,
\end{align}
where 
$\psi_{jk}$ is the angle formed by lines connecting the O atom of the
central molecule to the O atoms of its two nearest neighbor molecules,
that is, 
$j$ and $k$.
The summation is taken over all unique pairs among the four nearest
neighbor molecules.
If a molecule is located at the center of a regular tetrahedron created
by its four nearest neighbors, 
$\cos \psi_{jk} = 1/3$, yielding $q_\mathrm{tet}=1$.
In contrast, for a random molecular arrangement such as that in an ideal gas,
the six angles associated with the central molecule are statistically
independent, and the average value of $q_\mathrm{tet}$ approaches 0.

(b) $q_n$ is a generalized form of $q_\mathrm{tet}$ that accounts for
the possibility that the four nearest neighbor molecules do not
necessarily belong to the first coordination shell~\cite{santis2024Descriptors}.
This quantifies the extent to which the local H-bond environment approaches
tetrahedral coordination.
The definition is presented by 
\begin{align}
q_n=1-\frac{9}{2n(n-1)}\sum_{j=1}^{n-1}\sum_{k=j+1}^{n}\left(\cos\psi_{jk}+\frac{1}{3}\right)^2,
\end{align}
where $n$
is the number of H-bonds of the central water molecule, \textit{i.e.}, $n=2$, 3, 4,
and 5.

(c) $d_5$ is expressed as the O-O distance between the central molecule
and its fifth nearest neighbor~\cite{cuthbertson2011Mixturelike}.
$d_5$ classifies molecules as HDL-like when its value 
exceeds 3.5 Å. 
This descriptor contains only distance-based information and
characterizes the local structure up to the first coordination shell.

(d) LSI (Local Structure Index) for the molecule $i$ is defined by ordering the O-O
nearest-neighbor distances between the central molecule $i$ and its
neighboring molecules, denoted as $r_j$, such that
$r_1 < r_2 < \cdots < r_j < \cdots < r_{n(i)} < 0.37\ \mathrm{nm} < r_{n(i)+1}$.
Herein, $n(i)$ is chosen such that $r_{n(i)} < 0.37\ \mathrm{nm} < r_{n(i)+1}$.
Based on this ordered set of distances, the LSI is expressed 
as~\cite{shiratani1996Growth, shiratani1998Molecular}
\begin{align}
\mathrm{LSI}=\frac{1}{n(i)}\sum_{j=1}^{n(i)}\left[\Delta(j)-\bar{\Delta}\right], 
\end{align}
where $\Delta(j)=r_{j+1}-r_j$ and $\bar\Delta$ is the average of
$\Delta(j)$ over all the nearest neighbors $j$ of the molecule $i$.
Herein, $0.37$ nm lies between the first and second coordination shells,
which is determined to optimize the sensitivity of the distribution function.
LSI quantifies the degree of separation between the first and second
coordination shells of a target water molecule, thereby characterizing
the local structure beyond the first coordination shell. 
Large LSI value indicates a well-ordered first coordination shell,
with no water molecules located between the first and second
coordination shells. 
In contrast, small LSI value implies either a disordered first
coordination shell or the presence of water molecules between the first
and second coordination shells.

(e) $\zeta$ is defined as the difference between the distance to the
nearest no H-bond molecule ($r_\mathrm{NHB}$) and the distance to the
farthest H-bond molecule ($r_\mathrm{HB}$), \textit{i.e.},
$\zeta=r_\mathrm{NHB} - r_\mathrm{HB}$~\cite{russo2014Understanding, shi2018Common}.
$\zeta$ is a descriptor designed to measure local order
extending to the second coordination shell by explicitly incorporating
the existance or absence of H-bonds.
$\zeta \approx 0$ nm indicates a locally disordered (HDL-like)
environment in which the second coordination shell is disrupted, whereas
$\zeta \approx 1$ nm corresponds to a locally ordered (LDL-like)
structure with a clear separation between the first and second
coordination shells.

(f) $V_4$ is expressed as the fourth-largest negative interaction energy
among the interaction energies between a central water molecule and
its surrounding molecules~\cite{montesdeoca2020Structural}.
If a local tetrahedral structure is approximately formed, $V_4$ is expected to
correspond to a typical H-bond energy; when the tetrahedral arrangement is
disrupted, $V_4$ approaches 0.
Therefore, $V_4$ serves as an descriptor for identifying coordination defects.

(g) Similar to $V_4$, $V_5$ is defined as the fifth largest negative
interaction energy to primarily
highlights overcoordinated molecules, contrary to the case of
$V_4$~\cite{montesdeoca2020Structural}.

(h) $V_{4-5}$ is the difference between $V_4$ and $V_5$, \textit{i.e.},
$V_{4-5}=V_4-V_5$, which is used to 
discriminate the four-fold tetrahedrally coordinated water molecules
from both the three-fold-coordinated and overcoordinated molecules~\cite{loubet2024Role}.

(i) NTC (Node Total Communicability) 
is a graph-theoretical metric that evaluates the cumulative
connectivity of each water molecules~\cite{faccio2022Low}.
To apply graph-based analysis to molecular structures, atomic coordinates are
converted into a graph representation.
Following 
the literature,
O atoms are treated as nodes, and an edge is assigned between two nodes
when the O-O distance is less than or equal to 0.35 nm~\cite{faccio2022Low}.
This procedure yields an undirected, unweighted graph.
NTC of the node $v_i$ is expressed as 
\begin{align}
\mathrm{NTC}=\left[e^{\beta A}\mathbf{1}\right]_i=1+\beta\left[A\mathbf{1}\right]_i+\frac{\beta^2}{2!}\left[A^2\mathbf{1}\right]_i+\frac{\beta^3}{3!}\left[A^3\mathbf{1}\right]_i+\cdots,
\end{align}
where $A$ is the adjacency matrix of the graph consisting of 
$N$ nodes, where $N$ is the number of molecules in the system.
Thus, $A$ is an $N \times N$ square matrix with the element $a_{ij}$. 
If an edge exists between the nodes $v_i$ and $v_j$, then $a_{ij} = 1$; otherwise, $a_{ij} = 0$.
Additionally, $\mathbf{1}$ is the vector of all ones and $\beta$ is a parameter that
weights medium- and long-range effects.
In this study, $\beta=1$ was employed based on the corresponding
literature~\cite{faccio2022Low}.
This descriptor includes the influence of network connectivity over
medium- and long-ranges, and a large NTC value indicates a highly
connected and dense network.

(j) We define an alternative descriptor, namely, NTC$_\mathrm{HB}$, as a
variation of NTC, in which edges 
are constructed based on the H-bond expression rather than the distance-based
criterion used in NTC.

(k) $\Psi$ is a structural descriptor that characterizes topological changes
in H-bond networks across the liquid-liquid transition in
water~\cite{foffi2021Structural,
foffi2021Structure}.
It is defined as follows~\cite{foffi2023Correlated, foffi2024Identification}:
for a specific molecule $i$, molecules connected to $i$
by the chemical distance $D=4$, that is, molecules separated by 
four H-bond links along the
network, are identified at first. 
These molecules constitute the fourth bond-coordination shell of $i$.
O-O distances 
among the molecules in this shell are evaluated, and
the shortest of these distances is expressed as $\Psi$ for molecule $i$.
In the LDL-like state, the tetrahedral H-bond network is adequately
developed, affording 
$\Psi\approx 0.6$ nm. 
In contrast, in the HDL-like state, the H-bond network adopts a disrupted
structure, resulting in $\Psi \approx 0.35$ nm.

(l) and (m) 
Bond-orientational order (BOO) describes the degree of local
orientational order in liquids~\cite{steinhardt1983Bondorientational}.
To quantify the orientational symmetry of neighboring particles around
particle $i$, the complex order parameter 
$q_{lm}(i)$ is defined as follows:
\begin{align}
q_{lm}(i) = \frac{1}{N_b(i)} \sum_{j=1}^{N_b(i)} Y_{lm}(\hat{\bm{r}}_{ij}), 
\end{align}
where
$N_b(i)$ represents the number of neighboring particles within the first
coordination shell of $i$.
In this study, the first coordination shell was expressed as the region
within the cutoff distance of $r_\mathrm{cut}=0.35$ nm, consistent with
the H-bond definition.
Herein, $Y_{lm}(\cdots)$ denotes the spherical harmonic function of the degree
$l$ and order $m$, with 
$m = -l, \cdots, +l$, and 
$\hat{\bm{r}}_{ij}$ is the unit vector pointing from particle $i$ to
particle $j$.
To ensure rotational invariance, the BOO is described as
\begin{align}
Q_l=\sqrt{\frac{4\pi}{2l+1}\sum_{m=-l}^{l}|q_{lm}(i)|^2}.
\end{align}
Generally, 
the choice of $l$ is arbitrary.
In this study, we employed $Q_4$ (l) and $Q_6$ (m), which are sensitive
to cubic and hexagonal symmetries, respectively.
The combination of these parameters is widely used to distinguish
between liquid-like and crystalline local structures. 
All $Q_4$ and $Q_6$ calculations were performed using the Pyboo code~\cite{Pyboo}.

(n) $V_\mathrm{voro}$ is the local volume associated with each molecule
based on Voronoi tessellation~\cite{montoro1993Voronoi, shih1994Voronoi}.
Herein, $V_\mathrm{voro}$ was evaluated using the Pyvoro code~\cite{Pyvoro}.

(o) C.N. is defined as the number of
molecules within the first coordination shell, determined by the O-O 
distance within the cutoff $r_\mathrm{cut}=0.35$ nm utilized in the H-bond definition.
To guarantee that C.N. is smooth and continuous, it is
defined as follows:~\cite{donkor2023MachineLearning}:
\begin{align}
\mathrm{C.N.} = \sum_{j=1}^N \frac{1-
 \left(\tfrac{r_j}{r_\mathrm{cut}}\right)^{12}}{1-
 \left(\tfrac{r_j}{r_\mathrm{cut}}\right)^{28}}, 
\end{align}
where $r_j$ denotes the distance between a central O atom and the $j$-th
O atom.

(p) $N_\mathrm{HB}$ is described as the number of H-bonds of the central
molecule.

\subsection*{Neural network model}
\label{method:ML}

Fully connected neural network (FCNN) was employed to classify the 
structures between two temperatures. 
As shown in Fig.~\ref{fig:ML}, the network architecture comprises an
input layer with 1000 nodes, two hidden layers (with 1000 and 1 nodes,
respectively) using the LeakyReLU
activation function (leak parameter =0.01), and an output layer with a
sigmoid activation function to predict the temperature, $T_1$ or $T_2$. 
Note that each node contains a bias term.
For each structural descriptor, we constructed an independent
neural network model for a total of 15 temperature combinations.
For all models, the outputs were assigned as $T_1=0$ and $T_2=1$, with
$T_1<T_2$, and the network was trained by minimizing the binary
cross-entropy loss function.
In general, the model's representation capacity can be
improved by adjusting hyperparameters such as the number of hidden
layers and the number of nodes.
This implies that high-precision classification may be achievable for
many of the structural descriptors considered in this study.
However, the primary objective of this study is to evaluate the
classification performance of structural descriptors within a unified
framework while introducing minimal nonlinearity between inputs and
outputs; therefore, we adopted a relatively simple FCNN with two hidden layers.

Training was conducted using TensorFlow~\cite{tensorflow2015-whitepaper} with the optimization
algorithm Adam, a learning rate of $10^{-6}$, and a batch size of 32. 
At each temperature, the dataset was divided into training, validation,
and test datasets at an 8:1:1 ratio.
Epoch dependences of the training and validation loss functions for the
case using $q_\mathrm{tet}$ as the structural descriptor are depicted in
Supplementary Figs.~8 and 9.
To enable
performance comparisons under equal conditions across temperature
combinations for each descriptor, we determined the epoch at which the validation loss
shifted from decreasing to increasing and evaluated the model's AUC at
that point.
In addition, to prevent biases arising from the arbitrary assignment of molecular
indices in MD simulations, the molecular indices used as input were
randomly shuffled. 
To account for the permutation-invariant
nature of the data, training and evaluation involving this shuffling
procedure were performed independently ten times for each structural
descriptor, and the AUC was calculated as the average of these values.

For comparison, we also used logistic regression as a linear
classification model without nonlinearity.
As shown in Supplementary Fig.~10, 
the model has an input layer with 1000 nodes, similar to the FCNN,
followed by a single output layer performing binary classification using
a sigmoid activation function.
As with the FCNN, the model was trained separately for each structural
descriptor.
Training was performed using TensorFlow with the Adam optimization
algorithm, a learning rate of $10^{-6}$, and a batch size of 32,
identical to those used for the FCNN.
The loss function was binary cross-entropy.
Training was repeated 10 times with randomly shuffled molecular indices,
and the average AUC was calculated.

\subsection*{LIME analysis}
\label{method:XAI}

To interpret the temperature classification predictions generated by 
FCNNs, we applied LIME, an XAI method~\cite{ribeiro2016Why}.
Because FCNNs are typically black-box models, identifying input features
that contribute to their predictions is difficult.
LIME addresses this issue by locally approximating the original model 
with an interpretable model, thereby quantifying the
contribution of each input variable. 
Specifically, LIME explains the prediction for a certain input instance 
by creating a local linear regression model
that reproduces the behavior of the original classifier in the
neighborhood of the instance.
Herein, LIME value computations were performed using the available Python
code~\cite{ribeiro2026Marcotcr}.


\begin{acknowledgments}
We acknowledge support from the Maruho Collaborative Project for
 Theoretical Pharmaceutics. 
Numerical calculations were performed at D3 Center, The University of
 Osaka (Project: 2026-Stu-02), and 
the
 Research Center for Computational Science, Okazaki, National Institutes
 of Natural Sciences (Projects: 25-IMS-C052 and 26-IMS-C051).
\end{acknowledgments}

\section*{Funding}

Y.K. discloses support for the research of this work from the Japan
Science and Technology Agency (JST) SPRING (Grant No. JPMJSP2138).

K.S. discloses support for the research of this work from the Japan
Society for the Promotion of Science (JSPS) KAKENHI (Grant
No. JP25KJ1764).

K.K. discloses support for the research of this work from the Japan
Society for the Promotion of Science (JSPS) KAKENHI (Grant
Nos. JP25K00968, JP24H01719, and JP22K03550).

N.M. discloses support for the research of this work from the Ministry
of Education, Culture, Sports, Science and Technology (MEXT) (Grant
Nos. JPMXP1020230325 and JPMXP1122714694).

\section*{AUTHOR DECLARATIONS}

\section*{Competing interests}

The authors have no conflicts to disclose.

\section*{Code availability}

The codes that support the findings of this study are openly
available in Zenodo at
\url{https://doi.org/10.5281/zenodo.19938178}.

\section*{Data availability}

The data that support the findings of this study are openly
available in Zenodo at
\url{https://doi.org/10.5281/zenodo.19938177}.

%


\begin{thebibliography}{86}%
\makeatletter
\providecommand \@ifxundefined [1]{%
 \@ifx{#1\undefined}
}%
\providecommand \@ifnum [1]{%
 \ifnum #1\expandafter \@firstoftwo
 \else \expandafter \@secondoftwo
 \fi
}%
\providecommand \@ifx [1]{%
 \ifx #1\expandafter \@firstoftwo
 \else \expandafter \@secondoftwo
 \fi
}%
\providecommand \natexlab [1]{#1}%
\providecommand \enquote  [1]{``#1''}%
\providecommand \bibnamefont  [1]{#1}%
\providecommand \bibfnamefont [1]{#1}%
\providecommand \citenamefont [1]{#1}%
\providecommand \href@noop [0]{\@secondoftwo}%
\providecommand \href [0]{\begingroup \@sanitize@url \@href}%
\providecommand \@href[1]{\@@startlink{#1}\@@href}%
\providecommand \@@href[1]{\endgroup#1\@@endlink}%
\providecommand \@sanitize@url [0]{\catcode `\\12\catcode `\$12\catcode
  `\&12\catcode `\#12\catcode `\^12\catcode `\_12\catcode `\%12\relax}%
\providecommand \@@startlink[1]{}%
\providecommand \@@endlink[0]{}%
\providecommand \url  [0]{\begingroup\@sanitize@url \@url }%
\providecommand \@url [1]{\endgroup\@href {#1}{\urlprefix }}%
\providecommand \urlprefix  [0]{URL }%
\providecommand \Eprint [0]{\href }%
\providecommand \doibase [0]{https://doi.org/}%
\providecommand \selectlanguage [0]{\@gobble}%
\providecommand \bibinfo  [0]{\@secondoftwo}%
\providecommand \bibfield  [0]{\@secondoftwo}%
\providecommand \translation [1]{[#1]}%
\providecommand \BibitemOpen [0]{}%
\providecommand \bibitemStop [0]{}%
\providecommand \bibitemNoStop [0]{.\EOS\space}%
\providecommand \EOS [0]{\spacefactor3000\relax}%
\providecommand \BibitemShut  [1]{\csname bibitem#1\endcsname}%
\let\auto@bib@innerbib\@empty
\bibitem [{\citenamefont {Gallo}\ \emph {et~al.}(2016)\citenamefont {Gallo},
  \citenamefont {{Amann-Winkel}}, \citenamefont {Angell}, \citenamefont
  {Anisimov}, \citenamefont {Caupin}, \citenamefont {Chakravarty},
  \citenamefont {Lascaris}, \citenamefont {Loerting}, \citenamefont
  {Panagiotopoulos}, \citenamefont {Russo}, \citenamefont {Sellberg},
  \citenamefont {Stanley}, \citenamefont {Tanaka}, \citenamefont {Vega},
  \citenamefont {Xu},\ and\ \citenamefont {Pettersson}}]{gallo2016Water}%
  \BibitemOpen
  \bibfield  {author} {\bibinfo {author} {\bibfnamefont {P.}~\bibnamefont
  {Gallo}}, \bibinfo {author} {\bibfnamefont {K.}~\bibnamefont
  {{Amann-Winkel}}}, \bibinfo {author} {\bibfnamefont {C.~A.}\ \bibnamefont
  {Angell}}, \bibinfo {author} {\bibfnamefont {M.~A.}\ \bibnamefont
  {Anisimov}}, \bibinfo {author} {\bibfnamefont {F.}~\bibnamefont {Caupin}},
  \bibinfo {author} {\bibfnamefont {C.}~\bibnamefont {Chakravarty}}, \bibinfo
  {author} {\bibfnamefont {E.}~\bibnamefont {Lascaris}}, \bibinfo {author}
  {\bibfnamefont {T.}~\bibnamefont {Loerting}}, \bibinfo {author}
  {\bibfnamefont {A.~Z.}\ \bibnamefont {Panagiotopoulos}}, \bibinfo {author}
  {\bibfnamefont {J.}~\bibnamefont {Russo}}, \bibinfo {author} {\bibfnamefont
  {J.~A.}\ \bibnamefont {Sellberg}}, \bibinfo {author} {\bibfnamefont {H.~E.}\
  \bibnamefont {Stanley}}, \bibinfo {author} {\bibfnamefont {H.}~\bibnamefont
  {Tanaka}}, \bibinfo {author} {\bibfnamefont {C.}~\bibnamefont {Vega}},
  \bibinfo {author} {\bibfnamefont {L.}~\bibnamefont {Xu}},\ and\ \bibinfo
  {author} {\bibfnamefont {L.~G.~M.}\ \bibnamefont {Pettersson}},\ }\bibfield
  {title} {\enquote {\bibinfo {title} {Water: {{A Tale}} of {{Two Liquids}}},}\
  }\href {https://doi.org/10.1021/acs.chemrev.5b00750} {\bibfield  {journal}
  {\bibinfo  {journal} {Chem. Rev.}\ }\textbf {\bibinfo {volume} {116}},\
  \bibinfo {pages} {7463--7500} (\bibinfo {year} {2016})}\BibitemShut {NoStop}%
\bibitem [{\citenamefont {Gallo}\ \emph {et~al.}(2021)\citenamefont {Gallo},
  \citenamefont {Bachler}, \citenamefont {Bove}, \citenamefont {B{\"o}hmer},
  \citenamefont {Camisasca}, \citenamefont {Coronas}, \citenamefont {Corti},
  \citenamefont {{de Almeida Ribeiro}}, \citenamefont {{de Koning}},
  \citenamefont {Franzese}, \citenamefont {{Fuentes-Landete}}, \citenamefont
  {Gainaru}, \citenamefont {Loerting}, \citenamefont {{de Oca}}, \citenamefont
  {Poole}, \citenamefont {Rovere}, \citenamefont {Sciortino}, \citenamefont
  {Tonauer},\ and\ \citenamefont {Appignanesi}}]{gallo2021Advances}%
  \BibitemOpen
  \bibfield  {author} {\bibinfo {author} {\bibfnamefont {P.}~\bibnamefont
  {Gallo}}, \bibinfo {author} {\bibfnamefont {J.}~\bibnamefont {Bachler}},
  \bibinfo {author} {\bibfnamefont {L.~E.}\ \bibnamefont {Bove}}, \bibinfo
  {author} {\bibfnamefont {R.}~\bibnamefont {B{\"o}hmer}}, \bibinfo {author}
  {\bibfnamefont {G.}~\bibnamefont {Camisasca}}, \bibinfo {author}
  {\bibfnamefont {L.~E.}\ \bibnamefont {Coronas}}, \bibinfo {author}
  {\bibfnamefont {H.~R.}\ \bibnamefont {Corti}}, \bibinfo {author}
  {\bibfnamefont {I.}~\bibnamefont {{de Almeida Ribeiro}}}, \bibinfo {author}
  {\bibfnamefont {M.}~\bibnamefont {{de Koning}}}, \bibinfo {author}
  {\bibfnamefont {G.}~\bibnamefont {Franzese}}, \bibinfo {author}
  {\bibfnamefont {V.}~\bibnamefont {{Fuentes-Landete}}}, \bibinfo {author}
  {\bibfnamefont {C.}~\bibnamefont {Gainaru}}, \bibinfo {author} {\bibfnamefont
  {T.}~\bibnamefont {Loerting}}, \bibinfo {author} {\bibfnamefont {J.~M.~M.}\
  \bibnamefont {{de Oca}}}, \bibinfo {author} {\bibfnamefont {P.~H.}\
  \bibnamefont {Poole}}, \bibinfo {author} {\bibfnamefont {M.}~\bibnamefont
  {Rovere}}, \bibinfo {author} {\bibfnamefont {F.}~\bibnamefont {Sciortino}},
  \bibinfo {author} {\bibfnamefont {C.~M.}\ \bibnamefont {Tonauer}},\ and\
  \bibinfo {author} {\bibfnamefont {G.~A.}\ \bibnamefont {Appignanesi}},\
  }\bibfield  {title} {\enquote {\bibinfo {title} {Advances in the study of
  supercooled water},}\ }\href
  {https://doi.org/10.1140/epje/s10189-021-00139-1} {\bibfield  {journal}
  {\bibinfo  {journal} {Eur. Phys. J. E}\ }\textbf {\bibinfo {volume} {44}},\
  \bibinfo {pages} {143} (\bibinfo {year} {2021})}\BibitemShut {NoStop}%
\bibitem [{\citenamefont {Angell}(1983)}]{angell1983Supercooled}%
  \BibitemOpen
  \bibfield  {author} {\bibinfo {author} {\bibfnamefont {C.~A.}\ \bibnamefont
  {Angell}},\ }\bibfield  {title} {\enquote {\bibinfo {title} {Supercooled
  {{Water}}},}\ }\href {https://doi.org/10.1146/annurev.pc.34.100183.003113}
  {\bibfield  {journal} {\bibinfo  {journal} {Annu. Rev. Phys. Chem.}\ }\textbf
  {\bibinfo {volume} {34}},\ \bibinfo {pages} {593--630} (\bibinfo {year}
  {1983})}\BibitemShut {NoStop}%
\bibitem [{\citenamefont {Debenedetti}\ and\ \citenamefont
  {Stanley}(2003)}]{debenedetti2003Supercooleda}%
  \BibitemOpen
  \bibfield  {author} {\bibinfo {author} {\bibfnamefont {P.~G.}\ \bibnamefont
  {Debenedetti}}\ and\ \bibinfo {author} {\bibfnamefont {H.~E.}\ \bibnamefont
  {Stanley}},\ }\bibfield  {title} {\enquote {\bibinfo {title} {Supercooled and
  {{Glassy Water}}},}\ }\href {https://doi.org/10.1063/1.1595053} {\bibfield
  {journal} {\bibinfo  {journal} {Phys. Today}\ }\textbf {\bibinfo {volume}
  {56}},\ \bibinfo {pages} {40--46} (\bibinfo {year} {2003})}\BibitemShut
  {NoStop}%
\bibitem [{\citenamefont {Tanaka}(2000)}]{tanaka2000Simple}%
  \BibitemOpen
  \bibfield  {author} {\bibinfo {author} {\bibfnamefont {H.}~\bibnamefont
  {Tanaka}},\ }\bibfield  {title} {\enquote {\bibinfo {title} {Simple physical
  model of liquid water},}\ }\href {https://doi.org/10.1063/1.480609}
  {\bibfield  {journal} {\bibinfo  {journal} {J. Chem. Phys.}\ }\textbf
  {\bibinfo {volume} {112}},\ \bibinfo {pages} {799--809} (\bibinfo {year}
  {2000})}\BibitemShut {NoStop}%
\bibitem [{\citenamefont {Holten}\ \emph {et~al.}(2012)\citenamefont {Holten},
  \citenamefont {Bertrand}, \citenamefont {Anisimov},\ and\ \citenamefont
  {Sengers}}]{holten2012Thermodynamics}%
  \BibitemOpen
  \bibfield  {author} {\bibinfo {author} {\bibfnamefont {V.}~\bibnamefont
  {Holten}}, \bibinfo {author} {\bibfnamefont {C.~E.}\ \bibnamefont
  {Bertrand}}, \bibinfo {author} {\bibfnamefont {M.~A.}\ \bibnamefont
  {Anisimov}},\ and\ \bibinfo {author} {\bibfnamefont {J.~V.}\ \bibnamefont
  {Sengers}},\ }\bibfield  {title} {\enquote {\bibinfo {title} {Thermodynamics
  of supercooled water},}\ }\href {https://doi.org/10.1063/1.3690497}
  {\bibfield  {journal} {\bibinfo  {journal} {J. Chem. Phys.}\ }\textbf
  {\bibinfo {volume} {136}},\ \bibinfo {pages} {094507} (\bibinfo {year}
  {2012})}\BibitemShut {NoStop}%
\bibitem [{\citenamefont {Poole}\ \emph {et~al.}(1992)\citenamefont {Poole},
  \citenamefont {Sciortino}, \citenamefont {Essmann},\ and\ \citenamefont
  {Stanley}}]{poole1992Phase}%
  \BibitemOpen
  \bibfield  {author} {\bibinfo {author} {\bibfnamefont {P.~H.}\ \bibnamefont
  {Poole}}, \bibinfo {author} {\bibfnamefont {F.}~\bibnamefont {Sciortino}},
  \bibinfo {author} {\bibfnamefont {U.}~\bibnamefont {Essmann}},\ and\ \bibinfo
  {author} {\bibfnamefont {H.~E.}\ \bibnamefont {Stanley}},\ }\bibfield
  {title} {\enquote {\bibinfo {title} {Phase behaviour of metastable water},}\
  }\href {https://doi.org/10.1038/360324a0} {\bibfield  {journal} {\bibinfo
  {journal} {Nature}\ }\textbf {\bibinfo {volume} {360}},\ \bibinfo {pages}
  {324--328} (\bibinfo {year} {1992})}\BibitemShut {NoStop}%
\bibitem [{\citenamefont {Handle}, \citenamefont {Loerting},\ and\
  \citenamefont {Sciortino}(2017)}]{handle2017Supercooled}%
  \BibitemOpen
  \bibfield  {author} {\bibinfo {author} {\bibfnamefont {P.~H.}\ \bibnamefont
  {Handle}}, \bibinfo {author} {\bibfnamefont {T.}~\bibnamefont {Loerting}},\
  and\ \bibinfo {author} {\bibfnamefont {F.}~\bibnamefont {Sciortino}},\
  }\bibfield  {title} {\enquote {\bibinfo {title} {Supercooled and glassy
  water: {{Metastable}} liquid(s), amorphous solid(s), and a no-man's land},}\
  }\href {https://doi.org/10.1073/pnas.1700103114} {\bibfield  {journal}
  {\bibinfo  {journal} {Proc. Natl. Acad. Sci. U.S.A.}\ }\textbf {\bibinfo
  {volume} {114}},\ \bibinfo {pages} {13336--13344} (\bibinfo {year}
  {2017})}\BibitemShut {NoStop}%
\bibitem [{\citenamefont {Palmer}\ \emph {et~al.}(2018)\citenamefont {Palmer},
  \citenamefont {Poole}, \citenamefont {Sciortino},\ and\ \citenamefont
  {Debenedetti}}]{palmer2018Advances}%
  \BibitemOpen
  \bibfield  {author} {\bibinfo {author} {\bibfnamefont {J.~C.}\ \bibnamefont
  {Palmer}}, \bibinfo {author} {\bibfnamefont {P.~H.}\ \bibnamefont {Poole}},
  \bibinfo {author} {\bibfnamefont {F.}~\bibnamefont {Sciortino}},\ and\
  \bibinfo {author} {\bibfnamefont {P.~G.}\ \bibnamefont {Debenedetti}},\
  }\bibfield  {title} {\enquote {\bibinfo {title} {Advances in {{Computational
  Studies}} of the {{Liquid}}--{{Liquid Transition}} in {{Water}} and
  {{Water-Like Models}}},}\ }\href
  {https://doi.org/10.1021/acs.chemrev.8b00228} {\bibfield  {journal} {\bibinfo
   {journal} {Chem. Rev.}\ }\textbf {\bibinfo {volume} {118}},\ \bibinfo
  {pages} {9129--9151} (\bibinfo {year} {2018})}\BibitemShut {NoStop}%
\bibitem [{\citenamefont {Debenedetti}, \citenamefont {Sciortino},\ and\
  \citenamefont {Zerze}(2020)}]{debenedetti2020Second}%
  \BibitemOpen
  \bibfield  {author} {\bibinfo {author} {\bibfnamefont {P.~G.}\ \bibnamefont
  {Debenedetti}}, \bibinfo {author} {\bibfnamefont {F.}~\bibnamefont
  {Sciortino}},\ and\ \bibinfo {author} {\bibfnamefont {G.~H.}\ \bibnamefont
  {Zerze}},\ }\bibfield  {title} {\enquote {\bibinfo {title} {Second critical
  point in two realistic models of water},}\ }\href
  {https://doi.org/10.1126/science.abb9796} {\bibfield  {journal} {\bibinfo
  {journal} {Science}\ }\textbf {\bibinfo {volume} {369}},\ \bibinfo {pages}
  {289--292} (\bibinfo {year} {2020})}\BibitemShut {NoStop}%
\bibitem [{\citenamefont {Mishima}, \citenamefont {Calvert},\ and\
  \citenamefont {Whalley}(1985)}]{mishima1985Apparently}%
  \BibitemOpen
  \bibfield  {author} {\bibinfo {author} {\bibfnamefont {O.}~\bibnamefont
  {Mishima}}, \bibinfo {author} {\bibfnamefont {L.~D.}\ \bibnamefont
  {Calvert}},\ and\ \bibinfo {author} {\bibfnamefont {E.}~\bibnamefont
  {Whalley}},\ }\bibfield  {title} {\enquote {\bibinfo {title} {An apparently
  first-order transition between two amorphous phases of ice induced by
  pressure},}\ }\href {https://doi.org/10.1038/314076a0} {\bibfield  {journal}
  {\bibinfo  {journal} {Nature}\ }\textbf {\bibinfo {volume} {314}},\ \bibinfo
  {pages} {76--78} (\bibinfo {year} {1985})}\BibitemShut {NoStop}%
\bibitem [{\citenamefont {Mishima}\ and\ \citenamefont
  {Stanley}(1998)}]{mishima1998Relationship}%
  \BibitemOpen
  \bibfield  {author} {\bibinfo {author} {\bibfnamefont {O.}~\bibnamefont
  {Mishima}}\ and\ \bibinfo {author} {\bibfnamefont {H.~E.}\ \bibnamefont
  {Stanley}},\ }\bibfield  {title} {\enquote {\bibinfo {title} {The
  relationship between liquid, supercooled and glassy water},}\ }\href
  {https://doi.org/10.1038/24540} {\bibfield  {journal} {\bibinfo  {journal}
  {Nature}\ }\textbf {\bibinfo {volume} {396}},\ \bibinfo {pages} {329--335}
  (\bibinfo {year} {1998})}\BibitemShut {NoStop}%
\bibitem [{\citenamefont {Kim}\ \emph {et~al.}(2017)\citenamefont {Kim},
  \citenamefont {Sp{\"a}h}, \citenamefont {Pathak}, \citenamefont {Perakis},
  \citenamefont {Mariedahl}, \citenamefont {{Amann-Winkel}}, \citenamefont
  {Sellberg}, \citenamefont {Lee}, \citenamefont {Kim}, \citenamefont {Park},
  \citenamefont {Nam}, \citenamefont {Katayama},\ and\ \citenamefont
  {Nilsson}}]{kim2017Maxima}%
  \BibitemOpen
  \bibfield  {author} {\bibinfo {author} {\bibfnamefont {K.~H.}\ \bibnamefont
  {Kim}}, \bibinfo {author} {\bibfnamefont {A.}~\bibnamefont {Sp{\"a}h}},
  \bibinfo {author} {\bibfnamefont {H.}~\bibnamefont {Pathak}}, \bibinfo
  {author} {\bibfnamefont {F.}~\bibnamefont {Perakis}}, \bibinfo {author}
  {\bibfnamefont {D.}~\bibnamefont {Mariedahl}}, \bibinfo {author}
  {\bibfnamefont {K.}~\bibnamefont {{Amann-Winkel}}}, \bibinfo {author}
  {\bibfnamefont {J.~A.}\ \bibnamefont {Sellberg}}, \bibinfo {author}
  {\bibfnamefont {J.~H.}\ \bibnamefont {Lee}}, \bibinfo {author} {\bibfnamefont
  {S.}~\bibnamefont {Kim}}, \bibinfo {author} {\bibfnamefont {J.}~\bibnamefont
  {Park}}, \bibinfo {author} {\bibfnamefont {K.~H.}\ \bibnamefont {Nam}},
  \bibinfo {author} {\bibfnamefont {T.}~\bibnamefont {Katayama}},\ and\
  \bibinfo {author} {\bibfnamefont {A.}~\bibnamefont {Nilsson}},\ }\bibfield
  {title} {\enquote {\bibinfo {title} {Maxima in the thermodynamic response and
  correlation functions of deeply supercooled water},}\ }\href
  {https://doi.org/10.1126/science.aap8269} {\bibfield  {journal} {\bibinfo
  {journal} {Science}\ }\textbf {\bibinfo {volume} {358}},\ \bibinfo {pages}
  {1589--1593} (\bibinfo {year} {2017})}\BibitemShut {NoStop}%
\bibitem [{\citenamefont {Kim}\ \emph {et~al.}(2020)\citenamefont {Kim},
  \citenamefont {{Amann-Winkel}}, \citenamefont {Giovambattista}, \citenamefont
  {Sp{\"a}h}, \citenamefont {Perakis}, \citenamefont {Pathak}, \citenamefont
  {Parada}, \citenamefont {Yang}, \citenamefont {Mariedahl}, \citenamefont
  {Eklund}, \citenamefont {Lane}, \citenamefont {You}, \citenamefont {Jeong},
  \citenamefont {Weston}, \citenamefont {Lee}, \citenamefont {Eom},
  \citenamefont {Kim}, \citenamefont {Park}, \citenamefont {Chun},
  \citenamefont {Poole},\ and\ \citenamefont {Nilsson}}]{kim2020Experimental}%
  \BibitemOpen
  \bibfield  {author} {\bibinfo {author} {\bibfnamefont {K.~H.}\ \bibnamefont
  {Kim}}, \bibinfo {author} {\bibfnamefont {K.}~\bibnamefont {{Amann-Winkel}}},
  \bibinfo {author} {\bibfnamefont {N.}~\bibnamefont {Giovambattista}},
  \bibinfo {author} {\bibfnamefont {A.}~\bibnamefont {Sp{\"a}h}}, \bibinfo
  {author} {\bibfnamefont {F.}~\bibnamefont {Perakis}}, \bibinfo {author}
  {\bibfnamefont {H.}~\bibnamefont {Pathak}}, \bibinfo {author} {\bibfnamefont
  {M.~L.}\ \bibnamefont {Parada}}, \bibinfo {author} {\bibfnamefont
  {C.}~\bibnamefont {Yang}}, \bibinfo {author} {\bibfnamefont {D.}~\bibnamefont
  {Mariedahl}}, \bibinfo {author} {\bibfnamefont {T.}~\bibnamefont {Eklund}},
  \bibinfo {author} {\bibfnamefont {{\relax Thomas}.~J.}\ \bibnamefont {Lane}},
  \bibinfo {author} {\bibfnamefont {S.}~\bibnamefont {You}}, \bibinfo {author}
  {\bibfnamefont {S.}~\bibnamefont {Jeong}}, \bibinfo {author} {\bibfnamefont
  {M.}~\bibnamefont {Weston}}, \bibinfo {author} {\bibfnamefont {J.~H.}\
  \bibnamefont {Lee}}, \bibinfo {author} {\bibfnamefont {I.}~\bibnamefont
  {Eom}}, \bibinfo {author} {\bibfnamefont {M.}~\bibnamefont {Kim}}, \bibinfo
  {author} {\bibfnamefont {J.}~\bibnamefont {Park}}, \bibinfo {author}
  {\bibfnamefont {S.~H.}\ \bibnamefont {Chun}}, \bibinfo {author}
  {\bibfnamefont {P.~H.}\ \bibnamefont {Poole}},\ and\ \bibinfo {author}
  {\bibfnamefont {A.}~\bibnamefont {Nilsson}},\ }\bibfield  {title} {\enquote
  {\bibinfo {title} {Experimental observation of the liquid-liquid transition
  in bulk supercooled water under pressure},}\ }\href
  {https://doi.org/10.1126/science.abb9385} {\bibfield  {journal} {\bibinfo
  {journal} {Science}\ }\textbf {\bibinfo {volume} {370}},\ \bibinfo {pages}
  {978--982} (\bibinfo {year} {2020})}\BibitemShut {NoStop}%
\bibitem [{\citenamefont {Sciortino}\ \emph {et~al.}(1997)\citenamefont
  {Sciortino}, \citenamefont {Poole}, \citenamefont {Essmann},\ and\
  \citenamefont {Stanley}}]{sciortino1997Line}%
  \BibitemOpen
  \bibfield  {author} {\bibinfo {author} {\bibfnamefont {F.}~\bibnamefont
  {Sciortino}}, \bibinfo {author} {\bibfnamefont {P.~H.}\ \bibnamefont
  {Poole}}, \bibinfo {author} {\bibfnamefont {U.}~\bibnamefont {Essmann}},\
  and\ \bibinfo {author} {\bibfnamefont {H.~E.}\ \bibnamefont {Stanley}},\
  }\bibfield  {title} {\enquote {\bibinfo {title} {Line of compressibility
  maxima in the phase diagram of supercooled water},}\ }\href
  {https://doi.org/10.1103/PhysRevE.55.727} {\bibfield  {journal} {\bibinfo
  {journal} {Phys. Rev. E}\ }\textbf {\bibinfo {volume} {55}},\ \bibinfo
  {pages} {727--737} (\bibinfo {year} {1997})}\BibitemShut {NoStop}%
\bibitem [{\citenamefont {Starr}, \citenamefont {Sciortino},\ and\
  \citenamefont {Stanley}(1999)}]{starr1999Dynamics}%
  \BibitemOpen
  \bibfield  {author} {\bibinfo {author} {\bibfnamefont {F.~W.}\ \bibnamefont
  {Starr}}, \bibinfo {author} {\bibfnamefont {F.}~\bibnamefont {Sciortino}},\
  and\ \bibinfo {author} {\bibfnamefont {H.~E.}\ \bibnamefont {Stanley}},\
  }\bibfield  {title} {\enquote {\bibinfo {title} {Dynamics of simulated water
  under pressure},}\ }\href {https://doi.org/10.1103/PhysRevE.60.6757}
  {\bibfield  {journal} {\bibinfo  {journal} {Phys. Rev. E}\ }\textbf {\bibinfo
  {volume} {60}},\ \bibinfo {pages} {6757--6768} (\bibinfo {year}
  {1999})}\BibitemShut {NoStop}%
\bibitem [{\citenamefont {Xu}\ \emph {et~al.}(2006)\citenamefont {Xu},
  \citenamefont {Ehrenberg}, \citenamefont {Buldyrev},\ and\ \citenamefont
  {Stanley}}]{xu2006Relationship}%
  \BibitemOpen
  \bibfield  {author} {\bibinfo {author} {\bibfnamefont {L.}~\bibnamefont
  {Xu}}, \bibinfo {author} {\bibfnamefont {I.}~\bibnamefont {Ehrenberg}},
  \bibinfo {author} {\bibfnamefont {S.~V.}\ \bibnamefont {Buldyrev}},\ and\
  \bibinfo {author} {\bibfnamefont {H.~E.}\ \bibnamefont {Stanley}},\
  }\bibfield  {title} {\enquote {\bibinfo {title} {Relationship between the
  liquid--liquid phase transition and dynamic behaviour in the {{Jagla}}
  model},}\ }\href {https://doi.org/10.1088/0953-8984/18/36/S01} {\bibfield
  {journal} {\bibinfo  {journal} {J. Phys.: Condens. Matter}\ }\textbf
  {\bibinfo {volume} {18}},\ \bibinfo {pages} {S2239--S2246} (\bibinfo {year}
  {2006})}\BibitemShut {NoStop}%
\bibitem [{\citenamefont {Cuthbertson}\ and\ \citenamefont
  {Poole}(2011)}]{cuthbertson2011Mixturelike}%
  \BibitemOpen
  \bibfield  {author} {\bibinfo {author} {\bibfnamefont {M.~J.}\ \bibnamefont
  {Cuthbertson}}\ and\ \bibinfo {author} {\bibfnamefont {P.~H.}\ \bibnamefont
  {Poole}},\ }\bibfield  {title} {\enquote {\bibinfo {title} {Mixturelike
  {{Behavior Near}} a {{Liquid-Liquid Phase Transition}} in {{Simulations}} of
  {{Supercooled Water}}},}\ }\href
  {https://doi.org/10.1103/PhysRevLett.106.115706} {\bibfield  {journal}
  {\bibinfo  {journal} {Phys. Rev. Lett.}\ }\textbf {\bibinfo {volume} {106}},\
  \bibinfo {pages} {115706} (\bibinfo {year} {2011})}\BibitemShut {NoStop}%
\bibitem [{\citenamefont {Holten}\ and\ \citenamefont
  {Anisimov}(2012)}]{holten2012Entropydriven}%
  \BibitemOpen
  \bibfield  {author} {\bibinfo {author} {\bibfnamefont {V.}~\bibnamefont
  {Holten}}\ and\ \bibinfo {author} {\bibfnamefont {M.~A.}\ \bibnamefont
  {Anisimov}},\ }\bibfield  {title} {\enquote {\bibinfo {title} {Entropy-driven
  liquid--liquid separation in supercooled water},}\ }\href
  {https://doi.org/10.1038/srep00713} {\bibfield  {journal} {\bibinfo
  {journal} {Sci. Rep.}\ }\textbf {\bibinfo {volume} {2}},\ \bibinfo {pages}
  {713} (\bibinfo {year} {2012})}\BibitemShut {NoStop}%
\bibitem [{\citenamefont {Overduin}\ and\ \citenamefont
  {Patey}(2013)}]{overduin2013Analysis}%
  \BibitemOpen
  \bibfield  {author} {\bibinfo {author} {\bibfnamefont {S.~D.}\ \bibnamefont
  {Overduin}}\ and\ \bibinfo {author} {\bibfnamefont {G.~N.}\ \bibnamefont
  {Patey}},\ }\bibfield  {title} {\enquote {\bibinfo {title} {An analysis of
  fluctuations in supercooled {{TIP4P}}/2005 water},}\ }\href
  {https://doi.org/10.1063/1.4803868} {\bibfield  {journal} {\bibinfo
  {journal} {J. Chem. Phys.}\ }\textbf {\bibinfo {volume} {138}},\ \bibinfo
  {pages} {184502} (\bibinfo {year} {2013})}\BibitemShut {NoStop}%
\bibitem [{\citenamefont {Yagasaki}, \citenamefont {Matsumoto},\ and\
  \citenamefont {Tanaka}(2014)}]{yagasaki2014Spontaneous}%
  \BibitemOpen
  \bibfield  {author} {\bibinfo {author} {\bibfnamefont {T.}~\bibnamefont
  {Yagasaki}}, \bibinfo {author} {\bibfnamefont {M.}~\bibnamefont
  {Matsumoto}},\ and\ \bibinfo {author} {\bibfnamefont {H.}~\bibnamefont
  {Tanaka}},\ }\bibfield  {title} {\enquote {\bibinfo {title} {Spontaneous
  liquid-liquid phase separation of water},}\ }\href
  {https://doi.org/10.1103/PhysRevE.89.020301} {\bibfield  {journal} {\bibinfo
  {journal} {Phys. Rev. E}\ }\textbf {\bibinfo {volume} {89}},\ \bibinfo
  {pages} {020301} (\bibinfo {year} {2014})}\BibitemShut {NoStop}%
\bibitem [{\citenamefont {Palmer}\ \emph {et~al.}(2014)\citenamefont {Palmer},
  \citenamefont {Martelli}, \citenamefont {Liu}, \citenamefont {Car},
  \citenamefont {Panagiotopoulos},\ and\ \citenamefont
  {Debenedetti}}]{palmer2014Metastable}%
  \BibitemOpen
  \bibfield  {author} {\bibinfo {author} {\bibfnamefont {J.~C.}\ \bibnamefont
  {Palmer}}, \bibinfo {author} {\bibfnamefont {F.}~\bibnamefont {Martelli}},
  \bibinfo {author} {\bibfnamefont {Y.}~\bibnamefont {Liu}}, \bibinfo {author}
  {\bibfnamefont {R.}~\bibnamefont {Car}}, \bibinfo {author} {\bibfnamefont
  {A.~Z.}\ \bibnamefont {Panagiotopoulos}},\ and\ \bibinfo {author}
  {\bibfnamefont {P.~G.}\ \bibnamefont {Debenedetti}},\ }\bibfield  {title}
  {\enquote {\bibinfo {title} {Metastable liquid--liquid transition in a
  molecular model of water},}\ }\href {https://doi.org/10.1038/nature13405}
  {\bibfield  {journal} {\bibinfo  {journal} {Nature}\ }\textbf {\bibinfo
  {volume} {510}},\ \bibinfo {pages} {385--388} (\bibinfo {year}
  {2014})}\BibitemShut {NoStop}%
\bibitem [{\citenamefont {Singh}\ \emph {et~al.}(2016)\citenamefont {Singh},
  \citenamefont {Biddle}, \citenamefont {Debenedetti},\ and\ \citenamefont
  {Anisimov}}]{singh2016Twostate}%
  \BibitemOpen
  \bibfield  {author} {\bibinfo {author} {\bibfnamefont {R.~S.}\ \bibnamefont
  {Singh}}, \bibinfo {author} {\bibfnamefont {J.~W.}\ \bibnamefont {Biddle}},
  \bibinfo {author} {\bibfnamefont {P.~G.}\ \bibnamefont {Debenedetti}},\ and\
  \bibinfo {author} {\bibfnamefont {M.~A.}\ \bibnamefont {Anisimov}},\
  }\bibfield  {title} {\enquote {\bibinfo {title} {Two-state thermodynamics and
  the possibility of a liquid-liquid phase transition in supercooled
  {{TIP4P}}/2005 water},}\ }\href {https://doi.org/10.1063/1.4944986}
  {\bibfield  {journal} {\bibinfo  {journal} {J. Chem. Phys.}\ }\textbf
  {\bibinfo {volume} {144}},\ \bibinfo {pages} {144504} (\bibinfo {year}
  {2016})}\BibitemShut {NoStop}%
\bibitem [{\citenamefont {Biddle}\ \emph {et~al.}(2017)\citenamefont {Biddle},
  \citenamefont {Singh}, \citenamefont {Sparano}, \citenamefont {Ricci},
  \citenamefont {Gonz{\'a}lez}, \citenamefont {Valeriani}, \citenamefont
  {Abascal}, \citenamefont {Debenedetti}, \citenamefont {Anisimov},\ and\
  \citenamefont {Caupin}}]{biddle2017Twostructure}%
  \BibitemOpen
  \bibfield  {author} {\bibinfo {author} {\bibfnamefont {J.~W.}\ \bibnamefont
  {Biddle}}, \bibinfo {author} {\bibfnamefont {R.~S.}\ \bibnamefont {Singh}},
  \bibinfo {author} {\bibfnamefont {E.~M.}\ \bibnamefont {Sparano}}, \bibinfo
  {author} {\bibfnamefont {F.}~\bibnamefont {Ricci}}, \bibinfo {author}
  {\bibfnamefont {M.~A.}\ \bibnamefont {Gonz{\'a}lez}}, \bibinfo {author}
  {\bibfnamefont {C.}~\bibnamefont {Valeriani}}, \bibinfo {author}
  {\bibfnamefont {J.~L.~F.}\ \bibnamefont {Abascal}}, \bibinfo {author}
  {\bibfnamefont {P.~G.}\ \bibnamefont {Debenedetti}}, \bibinfo {author}
  {\bibfnamefont {M.~A.}\ \bibnamefont {Anisimov}},\ and\ \bibinfo {author}
  {\bibfnamefont {F.}~\bibnamefont {Caupin}},\ }\bibfield  {title} {\enquote
  {\bibinfo {title} {Two-structure thermodynamics for the {{TIP4P}}/2005 model
  of water covering supercooled and deeply stretched regions},}\ }\href
  {https://doi.org/10.1063/1.4973546} {\bibfield  {journal} {\bibinfo
  {journal} {J. Chem. Phys.}\ }\textbf {\bibinfo {volume} {146}},\ \bibinfo
  {pages} {034502} (\bibinfo {year} {2017})}\BibitemShut {NoStop}%
\bibitem [{\citenamefont {Kawasaki}\ and\ \citenamefont
  {Kim}(2017)}]{kawasaki2017Identifying}%
  \BibitemOpen
  \bibfield  {author} {\bibinfo {author} {\bibfnamefont {T.}~\bibnamefont
  {Kawasaki}}\ and\ \bibinfo {author} {\bibfnamefont {K.}~\bibnamefont {Kim}},\
  }\bibfield  {title} {\enquote {\bibinfo {title} {Identifying time scales for
  violation/preservation of {{Stokes-Einstein}} relation in supercooled
  water},}\ }\href {https://doi.org/10.1126/sciadv.1700399} {\bibfield
  {journal} {\bibinfo  {journal} {Sci. Adv.}\ }\textbf {\bibinfo {volume}
  {3}},\ \bibinfo {pages} {e1700399} (\bibinfo {year} {2017})}\BibitemShut
  {NoStop}%
\bibitem [{\citenamefont {Guillaud}\ \emph {et~al.}(2017)\citenamefont
  {Guillaud}, \citenamefont {Merabia}, \citenamefont {{de Ligny}},\ and\
  \citenamefont {Joly}}]{guillaud2017Decoupling}%
  \BibitemOpen
  \bibfield  {author} {\bibinfo {author} {\bibfnamefont {E.}~\bibnamefont
  {Guillaud}}, \bibinfo {author} {\bibfnamefont {S.}~\bibnamefont {Merabia}},
  \bibinfo {author} {\bibfnamefont {D.}~\bibnamefont {{de Ligny}}},\ and\
  \bibinfo {author} {\bibfnamefont {L.}~\bibnamefont {Joly}},\ }\bibfield
  {title} {\enquote {\bibinfo {title} {Decoupling of viscosity and relaxation
  processes in supercooled water: A molecular dynamics study with the
  {{TIP4P}}/2005f model},}\ }\href {https://doi.org/10.1039/C6CP07863J}
  {\bibfield  {journal} {\bibinfo  {journal} {Phys. Chem. Chem. Phys.}\
  }\textbf {\bibinfo {volume} {19}},\ \bibinfo {pages} {2124--2130} (\bibinfo
  {year} {2017})}\BibitemShut {NoStop}%
\bibitem [{\citenamefont {Galamba}(2017)}]{galamba2017Hydrogenbond}%
  \BibitemOpen
  \bibfield  {author} {\bibinfo {author} {\bibfnamefont {N.}~\bibnamefont
  {Galamba}},\ }\bibfield  {title} {\enquote {\bibinfo {title} {On the
  hydrogen-bond network and the non-{{Arrhenius}} transport properties of
  water},}\ }\href {https://doi.org/10.1088/0953-8984/29/1/015101} {\bibfield
  {journal} {\bibinfo  {journal} {J. Phys.: Condens. Matter}\ }\textbf
  {\bibinfo {volume} {29}},\ \bibinfo {pages} {015101} (\bibinfo {year}
  {2017})}\BibitemShut {NoStop}%
\bibitem [{\citenamefont {Saito}, \citenamefont {Bagchi},\ and\ \citenamefont
  {Ohmine}(2018)}]{saito2018Crucial}%
  \BibitemOpen
  \bibfield  {author} {\bibinfo {author} {\bibfnamefont {S.}~\bibnamefont
  {Saito}}, \bibinfo {author} {\bibfnamefont {B.}~\bibnamefont {Bagchi}},\ and\
  \bibinfo {author} {\bibfnamefont {I.}~\bibnamefont {Ohmine}},\ }\bibfield
  {title} {\enquote {\bibinfo {title} {Crucial role of fragmented and isolated
  defects in persistent relaxation of deeply supercooled water},}\ }\href
  {https://doi.org/10.1063/1.5044458} {\bibfield  {journal} {\bibinfo
  {journal} {J. Chem. Phys.}\ }\textbf {\bibinfo {volume} {149}},\ \bibinfo
  {pages} {124504} (\bibinfo {year} {2018})}\BibitemShut {NoStop}%
\bibitem [{\citenamefont {Saito}\ and\ \citenamefont
  {Bagchi}(2019)}]{saito2019Thermodynamic}%
  \BibitemOpen
  \bibfield  {author} {\bibinfo {author} {\bibfnamefont {S.}~\bibnamefont
  {Saito}}\ and\ \bibinfo {author} {\bibfnamefont {B.}~\bibnamefont {Bagchi}},\
  }\bibfield  {title} {\enquote {\bibinfo {title} {Thermodynamic picture of
  vitrification of water through complex specific heat and entropy: {{A}}
  journey through ``no man's land''},}\ }\href
  {https://doi.org/10.1063/1.5079594} {\bibfield  {journal} {\bibinfo
  {journal} {J. Chem. Phys.}\ }\textbf {\bibinfo {volume} {150}},\ \bibinfo
  {pages} {054502} (\bibinfo {year} {2019})}\BibitemShut {NoStop}%
\bibitem [{\citenamefont {Martelli}(2019)}]{martelli2019Unravelling}%
  \BibitemOpen
  \bibfield  {author} {\bibinfo {author} {\bibfnamefont {F.}~\bibnamefont
  {Martelli}},\ }\bibfield  {title} {\enquote {\bibinfo {title} {Unravelling
  the contribution of local structures to the anomalies of water: {{The}}
  synergistic action of several factors},}\ }\href
  {https://doi.org/10.1063/1.5087471} {\bibfield  {journal} {\bibinfo
  {journal} {J. Chem. Phys.}\ }\textbf {\bibinfo {volume} {150}},\ \bibinfo
  {pages} {094506} (\bibinfo {year} {2019})}\BibitemShut {NoStop}%
\bibitem [{\citenamefont {Neophytou}, \citenamefont {Chakrabarti},\ and\
  \citenamefont {Sciortino}(2022)}]{neophytou2022Topological}%
  \BibitemOpen
  \bibfield  {author} {\bibinfo {author} {\bibfnamefont {A.}~\bibnamefont
  {Neophytou}}, \bibinfo {author} {\bibfnamefont {D.}~\bibnamefont
  {Chakrabarti}},\ and\ \bibinfo {author} {\bibfnamefont {F.}~\bibnamefont
  {Sciortino}},\ }\bibfield  {title} {\enquote {\bibinfo {title} {Topological
  nature of the liquid--liquid phase transition in tetrahedral liquids},}\
  }\href {https://doi.org/10.1038/s41567-022-01698-6} {\bibfield  {journal}
  {\bibinfo  {journal} {Nat. Phys.}\ ,\ \bibinfo {pages} {1--6}} (\bibinfo
  {year} {2022})}\BibitemShut {NoStop}%
\bibitem [{\citenamefont {Sciortino}, \citenamefont {Gartner},\ and\
  \citenamefont {Debenedetti}(2024)}]{sciortino2024Freeenergy}%
  \BibitemOpen
  \bibfield  {author} {\bibinfo {author} {\bibfnamefont {F.}~\bibnamefont
  {Sciortino}}, \bibinfo {author} {\bibfnamefont {T.~E.}\ \bibnamefont
  {Gartner}},\ and\ \bibinfo {author} {\bibfnamefont {P.~G.}\ \bibnamefont
  {Debenedetti}},\ }\bibfield  {title} {\enquote {\bibinfo {title} {Free-energy
  landscape and spinodals for the liquid--liquid transition of the
  {{TIP4P}}/2005 and {{TIP4P}}/{{Ice}} models of water},}\ }\href
  {https://doi.org/10.1063/5.0196964} {\bibfield  {journal} {\bibinfo
  {journal} {J. Chem. Phys.}\ }\textbf {\bibinfo {volume} {160}},\ \bibinfo
  {pages} {104501} (\bibinfo {year} {2024})}\BibitemShut {NoStop}%
\bibitem [{\citenamefont {Malosso}\ \emph {et~al.}(2025)\citenamefont
  {Malosso}, \citenamefont {Donkor}, \citenamefont {Baroni},\ and\
  \citenamefont {Hassanali}}]{malosso2025Dynamical}%
  \BibitemOpen
  \bibfield  {author} {\bibinfo {author} {\bibfnamefont {C.}~\bibnamefont
  {Malosso}}, \bibinfo {author} {\bibfnamefont {E.~D.}\ \bibnamefont {Donkor}},
  \bibinfo {author} {\bibfnamefont {S.}~\bibnamefont {Baroni}},\ and\ \bibinfo
  {author} {\bibfnamefont {A.}~\bibnamefont {Hassanali}},\ }\bibfield  {title}
  {\enquote {\bibinfo {title} {Dynamical heterogeneity in supercooled water and
  its spectroscopic fingerprints},}\ }\href {https://doi.org/10.1063/5.0288343}
  {\bibfield  {journal} {\bibinfo  {journal} {J. Chem. Phys.}\ }\textbf
  {\bibinfo {volume} {163}},\ \bibinfo {pages} {144508} (\bibinfo {year}
  {2025})}\BibitemShut {NoStop}%
\bibitem [{\citenamefont {{Dubou{\'e}-Dijon}}\ and\ \citenamefont
  {Laage}(2015)}]{duboue-dijon2015Characterization}%
  \BibitemOpen
  \bibfield  {author} {\bibinfo {author} {\bibfnamefont {E.}~\bibnamefont
  {{Dubou{\'e}-Dijon}}}\ and\ \bibinfo {author} {\bibfnamefont
  {D.}~\bibnamefont {Laage}},\ }\bibfield  {title} {\enquote {\bibinfo {title}
  {Characterization of the {{Local Structure}} in {{Liquid Water}} by {{Various
  Order Parameters}}},}\ }\href {https://doi.org/10.1021/acs.jpcb.5b02936}
  {\bibfield  {journal} {\bibinfo  {journal} {J. Phys. Chem. B}\ }\textbf
  {\bibinfo {volume} {119}},\ \bibinfo {pages} {8406--8418} (\bibinfo {year}
  {2015})}\BibitemShut {NoStop}%
\bibitem [{\citenamefont {Shi}\ and\ \citenamefont
  {Tanaka}(2018)}]{shi2018Microscopic}%
  \BibitemOpen
  \bibfield  {author} {\bibinfo {author} {\bibfnamefont {R.}~\bibnamefont
  {Shi}}\ and\ \bibinfo {author} {\bibfnamefont {H.}~\bibnamefont {Tanaka}},\
  }\bibfield  {title} {\enquote {\bibinfo {title} {Microscopic structural
  descriptor of liquid water},}\ }\href {https://doi.org/10.1063/1.5024565}
  {\bibfield  {journal} {\bibinfo  {journal} {J. Chem. Phys.}\ }\textbf
  {\bibinfo {volume} {148}},\ \bibinfo {pages} {124503} (\bibinfo {year}
  {2018})}\BibitemShut {NoStop}%
\bibitem [{\citenamefont {Tanaka}\ \emph {et~al.}(2019)\citenamefont {Tanaka},
  \citenamefont {Tong}, \citenamefont {Shi},\ and\ \citenamefont
  {Russo}}]{tanaka2019Revealing}%
  \BibitemOpen
  \bibfield  {author} {\bibinfo {author} {\bibfnamefont {H.}~\bibnamefont
  {Tanaka}}, \bibinfo {author} {\bibfnamefont {H.}~\bibnamefont {Tong}},
  \bibinfo {author} {\bibfnamefont {R.}~\bibnamefont {Shi}},\ and\ \bibinfo
  {author} {\bibfnamefont {J.}~\bibnamefont {Russo}},\ }\bibfield  {title}
  {\enquote {\bibinfo {title} {Revealing key structural features hidden in
  liquids and glasses},}\ }\href {https://doi.org/10.1038/s42254-019-0053-3}
  {\bibfield  {journal} {\bibinfo  {journal} {Nat. Rev. Phys.}\ }\textbf
  {\bibinfo {volume} {1}},\ \bibinfo {pages} {333--348} (\bibinfo {year}
  {2019})}\BibitemShut {NoStop}%
\bibitem [{\citenamefont {Verde}\ \emph {et~al.}(2019)\citenamefont {Verde},
  \citenamefont {{Montes de Oca}}, \citenamefont {Accordino}, \citenamefont
  {Alarc{\'o}n},\ and\ \citenamefont {Appignanesi}}]{verde2019Comparing}%
  \BibitemOpen
  \bibfield  {author} {\bibinfo {author} {\bibfnamefont {A.~R.}\ \bibnamefont
  {Verde}}, \bibinfo {author} {\bibfnamefont {J.~M.}\ \bibnamefont {{Montes de
  Oca}}}, \bibinfo {author} {\bibfnamefont {S.~R.}\ \bibnamefont {Accordino}},
  \bibinfo {author} {\bibfnamefont {L.~M.}\ \bibnamefont {Alarc{\'o}n}},\ and\
  \bibinfo {author} {\bibfnamefont {G.~A.}\ \bibnamefont {Appignanesi}},\
  }\bibfield  {title} {\enquote {\bibinfo {title} {Comparing the performance of
  two structural indicators for different water models while seeking for
  connections between structure and dynamics in the glassy regime},}\ }\href
  {https://doi.org/10.1063/1.5108796} {\bibfield  {journal} {\bibinfo
  {journal} {J. Chem. Phys.}\ }\textbf {\bibinfo {volume} {150}},\ \bibinfo
  {pages} {244504} (\bibinfo {year} {2019})}\BibitemShut {NoStop}%
\bibitem [{\citenamefont {Verde}, \citenamefont {Alarc{\'o}n},\ and\
  \citenamefont {Appignanesi}(2022)}]{verde2022Journey}%
  \BibitemOpen
  \bibfield  {author} {\bibinfo {author} {\bibfnamefont {A.~R.}\ \bibnamefont
  {Verde}}, \bibinfo {author} {\bibfnamefont {L.~M.}\ \bibnamefont
  {Alarc{\'o}n}},\ and\ \bibinfo {author} {\bibfnamefont {G.~A.}\ \bibnamefont
  {Appignanesi}},\ }\bibfield  {title} {\enquote {\bibinfo {title} {A journey
  into the local structural order of liquid water: From the insights earned by
  geometrically-inspired descriptors to the development of a brand new
  energy-based indicator},}\ }\href
  {https://doi.org/10.1140/epjp/s13360-022-03318-x} {\bibfield  {journal}
  {\bibinfo  {journal} {Eur. Phys. J. Plus}\ }\textbf {\bibinfo {volume}
  {137}},\ \bibinfo {pages} {1112} (\bibinfo {year} {2022})}\BibitemShut
  {NoStop}%
\bibitem [{\citenamefont
  {Matubayasi}(1994)}]{matubayasi1994MatchingMismatching}%
  \BibitemOpen
  \bibfield  {author} {\bibinfo {author} {\bibfnamefont {N.}~\bibnamefont
  {Matubayasi}},\ }\bibfield  {title} {\enquote {\bibinfo {title}
  {Matching-{{Mismatching}} of {{Water Geometry}} and {{Hydrophobic
  Hydration}}},}\ }\href {https://doi.org/10.1021/ja00083a033} {\bibfield
  {journal} {\bibinfo  {journal} {J. Am. Chem. Soc.}\ }\textbf {\bibinfo
  {volume} {116}},\ \bibinfo {pages} {1450--1456} (\bibinfo {year}
  {1994})}\BibitemShut {NoStop}%
\bibitem [{\citenamefont {Chau}\ and\ \citenamefont
  {Hardwick}(1998)}]{chau1998New}%
  \BibitemOpen
  \bibfield  {author} {\bibinfo {author} {\bibfnamefont {P.-L.}\ \bibnamefont
  {Chau}}\ and\ \bibinfo {author} {\bibfnamefont {A.~J.}\ \bibnamefont
  {Hardwick}},\ }\bibfield  {title} {\enquote {\bibinfo {title} {A new order
  parameter for tetrahedral configurations},}\ }\href
  {https://doi.org/10.1080/002689798169195} {\bibfield  {journal} {\bibinfo
  {journal} {Mol. Phys.}\ }\textbf {\bibinfo {volume} {93}},\ \bibinfo {pages}
  {511--518} (\bibinfo {year} {1998})}\BibitemShut {NoStop}%
\bibitem [{\citenamefont {Errington}\ and\ \citenamefont
  {Debenedetti}(2001)}]{errington2001Relationship}%
  \BibitemOpen
  \bibfield  {author} {\bibinfo {author} {\bibfnamefont {J.~R.}\ \bibnamefont
  {Errington}}\ and\ \bibinfo {author} {\bibfnamefont {P.~G.}\ \bibnamefont
  {Debenedetti}},\ }\bibfield  {title} {\enquote {\bibinfo {title}
  {Relationship between structural order and the anomalies of liquid water},}\
  }\href {https://doi.org/10.1038/35053024} {\bibfield  {journal} {\bibinfo
  {journal} {Nature}\ }\textbf {\bibinfo {volume} {409}},\ \bibinfo {pages}
  {318--321} (\bibinfo {year} {2001})}\BibitemShut {NoStop}%
\bibitem [{\citenamefont {Russo}\ and\ \citenamefont
  {Tanaka}(2014)}]{russo2014Understanding}%
  \BibitemOpen
  \bibfield  {author} {\bibinfo {author} {\bibfnamefont {J.}~\bibnamefont
  {Russo}}\ and\ \bibinfo {author} {\bibfnamefont {H.}~\bibnamefont {Tanaka}},\
  }\bibfield  {title} {\enquote {\bibinfo {title} {Understanding water's
  anomalies with locally favoured structures},}\ }\href
  {https://doi.org/10.1038/ncomms4556} {\bibfield  {journal} {\bibinfo
  {journal} {Nat. Commun.}\ }\textbf {\bibinfo {volume} {5}},\ \bibinfo {pages}
  {3556} (\bibinfo {year} {2014})}\BibitemShut {NoStop}%
\bibitem [{\citenamefont {Shiratani}\ and\ \citenamefont
  {Sasai}(1996)}]{shiratani1996Growth}%
  \BibitemOpen
  \bibfield  {author} {\bibinfo {author} {\bibfnamefont {E.}~\bibnamefont
  {Shiratani}}\ and\ \bibinfo {author} {\bibfnamefont {M.}~\bibnamefont
  {Sasai}},\ }\bibfield  {title} {\enquote {\bibinfo {title} {Growth and
  collapse of structural patterns in the hydrogen bond network in liquid
  water},}\ }\href {https://doi.org/10.1063/1.471475} {\bibfield  {journal}
  {\bibinfo  {journal} {J. Chem. Phys.}\ }\textbf {\bibinfo {volume} {104}},\
  \bibinfo {pages} {7671--7680} (\bibinfo {year} {1996})}\BibitemShut {NoStop}%
\bibitem [{\citenamefont {Shiratani}\ and\ \citenamefont
  {Sasai}(1998)}]{shiratani1998Molecular}%
  \BibitemOpen
  \bibfield  {author} {\bibinfo {author} {\bibfnamefont {E.}~\bibnamefont
  {Shiratani}}\ and\ \bibinfo {author} {\bibfnamefont {M.}~\bibnamefont
  {Sasai}},\ }\bibfield  {title} {\enquote {\bibinfo {title} {Molecular scale
  precursor of the liquid--liquid phase transition of water},}\ }\href
  {https://doi.org/10.1063/1.475723} {\bibfield  {journal} {\bibinfo  {journal}
  {J. Chem. Phys.}\ }\textbf {\bibinfo {volume} {108}},\ \bibinfo {pages}
  {3264--3276} (\bibinfo {year} {1998})}\BibitemShut {NoStop}%
\bibitem [{\citenamefont {Faccio}\ \emph {et~al.}(2022)\citenamefont {Faccio},
  \citenamefont {Benzi}, \citenamefont {{Zanetti-Polzi}},\ and\ \citenamefont
  {Daidone}}]{faccio2022Low}%
  \BibitemOpen
  \bibfield  {author} {\bibinfo {author} {\bibfnamefont {C.}~\bibnamefont
  {Faccio}}, \bibinfo {author} {\bibfnamefont {M.}~\bibnamefont {Benzi}},
  \bibinfo {author} {\bibfnamefont {L.}~\bibnamefont {{Zanetti-Polzi}}},\ and\
  \bibinfo {author} {\bibfnamefont {I.}~\bibnamefont {Daidone}},\ }\bibfield
  {title} {\enquote {\bibinfo {title} {Low- and high-density forms of liquid
  water revealed by a new medium-range order descriptor},}\ }\href
  {https://doi.org/10.1016/j.molliq.2022.118922} {\bibfield  {journal}
  {\bibinfo  {journal} {J. Mol. Liq.}\ }\textbf {\bibinfo {volume} {355}},\
  \bibinfo {pages} {118922} (\bibinfo {year} {2022})}\BibitemShut {NoStop}%
\bibitem [{\citenamefont {Foffi}\ and\ \citenamefont
  {Sciortino}(2023)}]{foffi2023Correlated}%
  \BibitemOpen
  \bibfield  {author} {\bibinfo {author} {\bibfnamefont {R.}~\bibnamefont
  {Foffi}}\ and\ \bibinfo {author} {\bibfnamefont {F.}~\bibnamefont
  {Sciortino}},\ }\bibfield  {title} {\enquote {\bibinfo {title} {Correlated
  {{Fluctuations}} of {{Structural Indicators Close}} to the
  {{Liquid}}--{{Liquid Transition}} in {{Supercooled Water}}},}\ }\href
  {https://doi.org/10.1021/acs.jpcb.2c07169} {\bibfield  {journal} {\bibinfo
  {journal} {J. Phys. Chem. B}\ }\textbf {\bibinfo {volume} {127}},\ \bibinfo
  {pages} {378--386} (\bibinfo {year} {2023})}\BibitemShut {NoStop}%
\bibitem [{\citenamefont {Foffi}\ and\ \citenamefont
  {Sciortino}(2024)}]{foffi2024Identification}%
  \BibitemOpen
  \bibfield  {author} {\bibinfo {author} {\bibfnamefont {R.}~\bibnamefont
  {Foffi}}\ and\ \bibinfo {author} {\bibfnamefont {F.}~\bibnamefont
  {Sciortino}},\ }\bibfield  {title} {\enquote {\bibinfo {title}
  {Identification of local structures in water from supercooled to ambient
  conditions},}\ }\href {https://doi.org/10.1063/5.0188764} {\bibfield
  {journal} {\bibinfo  {journal} {J. Chem. Phys.}\ }\textbf {\bibinfo {volume}
  {160}},\ \bibinfo {pages} {094504} (\bibinfo {year} {2024})}\BibitemShut
  {NoStop}%
\bibitem [{\citenamefont {Xu}\ \emph {et~al.}(2005)\citenamefont {Xu},
  \citenamefont {Kumar}, \citenamefont {Buldyrev}, \citenamefont {Chen},
  \citenamefont {Poole}, \citenamefont {Sciortino},\ and\ \citenamefont
  {Stanley}}]{xu2005Relation}%
  \BibitemOpen
  \bibfield  {author} {\bibinfo {author} {\bibfnamefont {L.}~\bibnamefont
  {Xu}}, \bibinfo {author} {\bibfnamefont {P.}~\bibnamefont {Kumar}}, \bibinfo
  {author} {\bibfnamefont {S.~V.}\ \bibnamefont {Buldyrev}}, \bibinfo {author}
  {\bibfnamefont {S.-H.}\ \bibnamefont {Chen}}, \bibinfo {author}
  {\bibfnamefont {P.~H.}\ \bibnamefont {Poole}}, \bibinfo {author}
  {\bibfnamefont {F.}~\bibnamefont {Sciortino}},\ and\ \bibinfo {author}
  {\bibfnamefont {H.~E.}\ \bibnamefont {Stanley}},\ }\bibfield  {title}
  {\enquote {\bibinfo {title} {Relation between the {{Widom}} line and the
  dynamic crossover in systems with a liquid-liquid phase transition},}\ }\href
  {https://doi.org/10.1073/pnas.0507870102} {\bibfield  {journal} {\bibinfo
  {journal} {Proc. Natl. Acad. Sci. U.S.A.}\ }\textbf {\bibinfo {volume}
  {102}},\ \bibinfo {pages} {16558--16562} (\bibinfo {year}
  {2005})}\BibitemShut {NoStop}%
\bibitem [{\citenamefont {Stanley}\ \emph {et~al.}(2008)\citenamefont
  {Stanley}, \citenamefont {Kumar}, \citenamefont {Franzese}, \citenamefont
  {Xu}, \citenamefont {Yan}, \citenamefont {Mazza}, \citenamefont {Buldyrev},
  \citenamefont {Chen},\ and\ \citenamefont {Mallamace}}]{stanley2008Liquid}%
  \BibitemOpen
  \bibfield  {author} {\bibinfo {author} {\bibfnamefont {H.~E.}\ \bibnamefont
  {Stanley}}, \bibinfo {author} {\bibfnamefont {P.}~\bibnamefont {Kumar}},
  \bibinfo {author} {\bibfnamefont {G.}~\bibnamefont {Franzese}}, \bibinfo
  {author} {\bibfnamefont {L.}~\bibnamefont {Xu}}, \bibinfo {author}
  {\bibfnamefont {Z.}~\bibnamefont {Yan}}, \bibinfo {author} {\bibfnamefont
  {M.~G.}\ \bibnamefont {Mazza}}, \bibinfo {author} {\bibfnamefont {S.~V.}\
  \bibnamefont {Buldyrev}}, \bibinfo {author} {\bibfnamefont {S.-H.}\
  \bibnamefont {Chen}},\ and\ \bibinfo {author} {\bibfnamefont
  {F.}~\bibnamefont {Mallamace}},\ }\bibfield  {title} {\enquote {\bibinfo
  {title} {Liquid polyamorphism: {{Possible}} relation to the anomalous
  behaviour of water},}\ }\href {https://doi.org/10.1140/epjst/e2008-00746-3}
  {\bibfield  {journal} {\bibinfo  {journal} {Eur. Phys. J. Spec. Top.}\
  }\textbf {\bibinfo {volume} {161}},\ \bibinfo {pages} {1--17} (\bibinfo
  {year} {2008})}\BibitemShut {NoStop}%
\bibitem [{\citenamefont {Geiger}\ and\ \citenamefont
  {Dellago}(2013)}]{geiger2013Neural}%
  \BibitemOpen
  \bibfield  {author} {\bibinfo {author} {\bibfnamefont {P.}~\bibnamefont
  {Geiger}}\ and\ \bibinfo {author} {\bibfnamefont {C.}~\bibnamefont
  {Dellago}},\ }\bibfield  {title} {\enquote {\bibinfo {title} {Neural networks
  for local structure detection in polymorphic systems},}\ }\href
  {https://doi.org/10.1063/1.4825111} {\bibfield  {journal} {\bibinfo
  {journal} {J. Chem. Phys.}\ }\textbf {\bibinfo {volume} {139}},\ \bibinfo
  {pages} {164105} (\bibinfo {year} {2013})}\BibitemShut {NoStop}%
\bibitem [{\citenamefont {Boattini}, \citenamefont {Dijkstra},\ and\
  \citenamefont {Filion}(2019)}]{boattini2019Unsupervised}%
  \BibitemOpen
  \bibfield  {author} {\bibinfo {author} {\bibfnamefont {E.}~\bibnamefont
  {Boattini}}, \bibinfo {author} {\bibfnamefont {M.}~\bibnamefont {Dijkstra}},\
  and\ \bibinfo {author} {\bibfnamefont {L.}~\bibnamefont {Filion}},\
  }\bibfield  {title} {\enquote {\bibinfo {title} {Unsupervised learning for
  local structure detection in colloidal systems},}\ }\href
  {https://doi.org/10.1063/1.5118867} {\bibfield  {journal} {\bibinfo
  {journal} {J. Chem. Phys.}\ }\textbf {\bibinfo {volume} {151}},\ \bibinfo
  {pages} {154901} (\bibinfo {year} {2019})}\BibitemShut {NoStop}%
\bibitem [{\citenamefont {Boattini}\ \emph {et~al.}(2020)\citenamefont
  {Boattini}, \citenamefont {{Mar{\'i}n-Aguilar}}, \citenamefont {Mitra},
  \citenamefont {Foffi}, \citenamefont {Smallenburg},\ and\ \citenamefont
  {Filion}}]{boattini2020Autonomously}%
  \BibitemOpen
  \bibfield  {author} {\bibinfo {author} {\bibfnamefont {E.}~\bibnamefont
  {Boattini}}, \bibinfo {author} {\bibfnamefont {S.}~\bibnamefont
  {{Mar{\'i}n-Aguilar}}}, \bibinfo {author} {\bibfnamefont {S.}~\bibnamefont
  {Mitra}}, \bibinfo {author} {\bibfnamefont {G.}~\bibnamefont {Foffi}},
  \bibinfo {author} {\bibfnamefont {F.}~\bibnamefont {Smallenburg}},\ and\
  \bibinfo {author} {\bibfnamefont {L.}~\bibnamefont {Filion}},\ }\bibfield
  {title} {\enquote {\bibinfo {title} {Autonomously revealing hidden local
  structures in supercooled liquids},}\ }\href
  {https://doi.org/10.1038/s41467-020-19286-8} {\bibfield  {journal} {\bibinfo
  {journal} {Nat. Commun.}\ }\textbf {\bibinfo {volume} {11}},\ \bibinfo
  {pages} {5479} (\bibinfo {year} {2020})}\BibitemShut {NoStop}%
\bibitem [{\citenamefont {Martelli}\ \emph {et~al.}(2020)\citenamefont
  {Martelli}, \citenamefont {Leoni}, \citenamefont {Sciortino},\ and\
  \citenamefont {Russo}}]{martelli2020Connection}%
  \BibitemOpen
  \bibfield  {author} {\bibinfo {author} {\bibfnamefont {F.}~\bibnamefont
  {Martelli}}, \bibinfo {author} {\bibfnamefont {F.}~\bibnamefont {Leoni}},
  \bibinfo {author} {\bibfnamefont {F.}~\bibnamefont {Sciortino}},\ and\
  \bibinfo {author} {\bibfnamefont {J.}~\bibnamefont {Russo}},\ }\bibfield
  {title} {\enquote {\bibinfo {title} {Connection between liquid and
  non-crystalline solid phases in water},}\ }\href
  {https://doi.org/10.1063/5.0018923} {\bibfield  {journal} {\bibinfo
  {journal} {J. Chem. Phys.}\ }\textbf {\bibinfo {volume} {153}},\ \bibinfo
  {pages} {104503} (\bibinfo {year} {2020})}\BibitemShut {NoStop}%
\bibitem [{\citenamefont {Doi}, \citenamefont {Takahashi},\ and\ \citenamefont
  {Aoyagi}(2021)}]{doi2021Searching}%
  \BibitemOpen
  \bibfield  {author} {\bibinfo {author} {\bibfnamefont {H.}~\bibnamefont
  {Doi}}, \bibinfo {author} {\bibfnamefont {K.~Z.}\ \bibnamefont {Takahashi}},\
  and\ \bibinfo {author} {\bibfnamefont {T.}~\bibnamefont {Aoyagi}},\
  }\bibfield  {title} {\enquote {\bibinfo {title} {Searching local order
  parameters to classify water structures of ice {{Ih}}, {{Ic}}, and liquid},}\
  }\href {https://doi.org/10.1063/5.0049258} {\bibfield  {journal} {\bibinfo
  {journal} {J. Chem. Phys.}\ }\textbf {\bibinfo {volume} {154}},\ \bibinfo
  {pages} {164505} (\bibinfo {year} {2021})}\BibitemShut {NoStop}%
\bibitem [{\citenamefont {Yoshikawa}\ \emph {et~al.}(2025)\citenamefont
  {Yoshikawa}, \citenamefont {Yano}, \citenamefont {Goto}, \citenamefont
  {Kim},\ and\ \citenamefont {Matubayasi}}]{yoshikawa2025Graph}%
  \BibitemOpen
  \bibfield  {author} {\bibinfo {author} {\bibfnamefont {K.}~\bibnamefont
  {Yoshikawa}}, \bibinfo {author} {\bibfnamefont {K.}~\bibnamefont {Yano}},
  \bibinfo {author} {\bibfnamefont {S.}~\bibnamefont {Goto}}, \bibinfo {author}
  {\bibfnamefont {K.}~\bibnamefont {Kim}},\ and\ \bibinfo {author}
  {\bibfnamefont {N.}~\bibnamefont {Matubayasi}},\ }\bibfield  {title}
  {\enquote {\bibinfo {title} {Graph neural network-based structural
  classification of glass-forming liquids and its interpretation via
  self-attention mechanism},}\ }\href {https://doi.org/10.1063/5.0277279}
  {\bibfield  {journal} {\bibinfo  {journal} {J. Chem. Phys.}\ }\textbf
  {\bibinfo {volume} {163}},\ \bibinfo {pages} {024508} (\bibinfo {year}
  {2025})}\BibitemShut {NoStop}%
\bibitem [{\citenamefont {Adadi}\ and\ \citenamefont
  {Berrada}(2018)}]{adadi2018Peeking}%
  \BibitemOpen
  \bibfield  {author} {\bibinfo {author} {\bibfnamefont {A.}~\bibnamefont
  {Adadi}}\ and\ \bibinfo {author} {\bibfnamefont {M.}~\bibnamefont
  {Berrada}},\ }\bibfield  {title} {\enquote {\bibinfo {title} {Peeking
  {{Inside}} the {{Black-Box}}: {{A Survey}} on {{Explainable Artificial
  Intelligence}} ({{XAI}})},}\ }\href
  {https://doi.org/10.1109/ACCESS.2018.2870052} {\bibfield  {journal} {\bibinfo
   {journal} {IEEE Access}\ }\textbf {\bibinfo {volume} {6}},\ \bibinfo {pages}
  {52138--52160} (\bibinfo {year} {2018})}\BibitemShut {NoStop}%
\bibitem [{\citenamefont {Molnar}(2020)}]{molnar2020Interpretable}%
  \BibitemOpen
  \bibfield  {author} {\bibinfo {author} {\bibfnamefont {C.}~\bibnamefont
  {Molnar}},\ }\href@noop {} {\emph {\bibinfo {title} {{Interpretable Machine
  Learning}}}}\ (\bibinfo  {publisher} {Lulu.com},\ \bibinfo {address}
  {Morisville, North Carolina},\ \bibinfo {year} {2020})\BibitemShut {NoStop}%
\bibitem [{\citenamefont {Holzinger}\ \emph {et~al.}(2022)\citenamefont
  {Holzinger}, \citenamefont {Saranti}, \citenamefont {Molnar}, \citenamefont
  {Biecek},\ and\ \citenamefont {Samek}}]{holzinger2022Explainable}%
  \BibitemOpen
  \bibfield  {author} {\bibinfo {author} {\bibfnamefont {A.}~\bibnamefont
  {Holzinger}}, \bibinfo {author} {\bibfnamefont {A.}~\bibnamefont {Saranti}},
  \bibinfo {author} {\bibfnamefont {C.}~\bibnamefont {Molnar}}, \bibinfo
  {author} {\bibfnamefont {P.}~\bibnamefont {Biecek}},\ and\ \bibinfo {author}
  {\bibfnamefont {W.}~\bibnamefont {Samek}},\ }\bibfield  {title} {\enquote
  {\bibinfo {title} {Explainable {{AI Methods}} - {{A Brief Overview}}},}\ }in\
  \href {https://doi.org/10.1007/978-3-031-04083-2_2} {\emph {\bibinfo
  {booktitle} {{{xxAI}} - {{Beyond Explainable AI}}}}},\ Vol.\ \bibinfo
  {volume} {13200},\ \bibinfo {editor} {edited by\ \bibinfo {editor}
  {\bibfnamefont {A.}~\bibnamefont {Holzinger}}, \bibinfo {editor}
  {\bibfnamefont {R.}~\bibnamefont {Goebel}}, \bibinfo {editor} {\bibfnamefont
  {R.}~\bibnamefont {Fong}}, \bibinfo {editor} {\bibfnamefont {T.}~\bibnamefont
  {Moon}}, \bibinfo {editor} {\bibfnamefont {K.-R.}\ \bibnamefont
  {M{\"u}ller}},\ and\ \bibinfo {editor} {\bibfnamefont {W.}~\bibnamefont
  {Samek}}}\ (\bibinfo  {publisher} {Springer International Publishing},\
  \bibinfo {address} {Cham},\ \bibinfo {year} {2022})\ pp.\ \bibinfo {pages}
  {13--38}\BibitemShut {NoStop}%
\bibitem [{\citenamefont {Ribeiro}, \citenamefont {Singh},\ and\ \citenamefont
  {Guestrin}(2016)}]{ribeiro2016Why}%
  \BibitemOpen
  \bibfield  {author} {\bibinfo {author} {\bibfnamefont {M.~T.}\ \bibnamefont
  {Ribeiro}}, \bibinfo {author} {\bibfnamefont {S.}~\bibnamefont {Singh}},\
  and\ \bibinfo {author} {\bibfnamefont {C.}~\bibnamefont {Guestrin}},\
  }\bibfield  {title} {\enquote {\bibinfo {title} {"{{Why Should I Trust
  You}}?": {{Explaining}} the {{Predictions}} of {{Any Classifier}}},}\
  }\href@noop {} {\  (\bibinfo {year} {2016})},\ \Eprint
  {https://arxiv.org/abs/arXiv:1602.04938} {arXiv:arXiv:1602.04938 [cs, stat]}
  \BibitemShut {NoStop}%
\bibitem [{\citenamefont {Ishiai}, \citenamefont {Endo},\ and\ \citenamefont
  {Yasuoka}(2023)}]{ishiai2023Graph}%
  \BibitemOpen
  \bibfield  {author} {\bibinfo {author} {\bibfnamefont {S.}~\bibnamefont
  {Ishiai}}, \bibinfo {author} {\bibfnamefont {K.}~\bibnamefont {Endo}},\ and\
  \bibinfo {author} {\bibfnamefont {K.}~\bibnamefont {Yasuoka}},\ }\bibfield
  {title} {\enquote {\bibinfo {title} {Graph neural networks classify molecular
  geometry and design novel order parameters of crystal and liquid},}\ }\href
  {https://doi.org/10.1063/5.0156203} {\bibfield  {journal} {\bibinfo
  {journal} {J. Chem. Phys.}\ }\textbf {\bibinfo {volume} {159}},\ \bibinfo
  {pages} {064103} (\bibinfo {year} {2023})}\BibitemShut {NoStop}%
\bibitem [{\citenamefont {Ishiai}\ \emph {et~al.}(2024)\citenamefont {Ishiai},
  \citenamefont {Yasuda}, \citenamefont {Endo},\ and\ \citenamefont
  {Yasuoka}}]{ishiai2024GraphNeuralNetworkBased}%
  \BibitemOpen
  \bibfield  {author} {\bibinfo {author} {\bibfnamefont {S.}~\bibnamefont
  {Ishiai}}, \bibinfo {author} {\bibfnamefont {I.}~\bibnamefont {Yasuda}},
  \bibinfo {author} {\bibfnamefont {K.}~\bibnamefont {Endo}},\ and\ \bibinfo
  {author} {\bibfnamefont {K.}~\bibnamefont {Yasuoka}},\ }\bibfield  {title}
  {\enquote {\bibinfo {title} {Graph-{{Neural-Network-Based Unsupervised
  Learning}} of the {{Temporal Similarity}} of {{Structural Features Observed}}
  in {{Molecular Dynamics Simulations}}},}\ }\href
  {https://doi.org/10.1021/acs.jctc.3c00995} {\bibfield  {journal} {\bibinfo
  {journal} {J. Chem. Theory Comput.}\ }\textbf {\bibinfo {volume} {20}},\
  \bibinfo {pages} {819--831} (\bibinfo {year} {2024})}\BibitemShut {NoStop}%
\bibitem [{\citenamefont {Donkor}, \citenamefont {Laio},\ and\ \citenamefont
  {Hassanali}(2023)}]{donkor2023MachineLearning}%
  \BibitemOpen
  \bibfield  {author} {\bibinfo {author} {\bibfnamefont {E.~D.}\ \bibnamefont
  {Donkor}}, \bibinfo {author} {\bibfnamefont {A.}~\bibnamefont {Laio}},\ and\
  \bibinfo {author} {\bibfnamefont {A.}~\bibnamefont {Hassanali}},\ }\bibfield
  {title} {\enquote {\bibinfo {title} {Do {{Machine-Learning Atomic
  Descriptors}} and {{Order Parameters Tell}} the {{Same Story}}? {{The Case}}
  of {{Liquid Water}}},}\ }\href {https://doi.org/10.1021/acs.jctc.2c01205}
  {\bibfield  {journal} {\bibinfo  {journal} {J. Chem. Theory Comput.}\
  }\textbf {\bibinfo {volume} {19}},\ \bibinfo {pages} {4596--4605} (\bibinfo
  {year} {2023})}\BibitemShut {NoStop}%
\bibitem [{\citenamefont {Donkor}\ \emph {et~al.}(2024)\citenamefont {Donkor},
  \citenamefont {{Offei-Danso}}, \citenamefont {Rodriguez}, \citenamefont
  {Sciortino},\ and\ \citenamefont {Hassanali}}]{donkor2024Local}%
  \BibitemOpen
  \bibfield  {author} {\bibinfo {author} {\bibfnamefont {E.~D.}\ \bibnamefont
  {Donkor}}, \bibinfo {author} {\bibfnamefont {A.}~\bibnamefont
  {{Offei-Danso}}}, \bibinfo {author} {\bibfnamefont {A.}~\bibnamefont
  {Rodriguez}}, \bibinfo {author} {\bibfnamefont {F.}~\bibnamefont
  {Sciortino}},\ and\ \bibinfo {author} {\bibfnamefont {A.}~\bibnamefont
  {Hassanali}},\ }\bibfield  {title} {\enquote {\bibinfo {title} {Beyond
  {{Local Structures}} in {{Critical Supercooled Water}} through {{Unsupervised
  Learning}}},}\ }\href {https://doi.org/10.1021/acs.jpclett.4c00383}
  {\bibfield  {journal} {\bibinfo  {journal} {J. Phys. Chem. Lett.}\ }\textbf
  {\bibinfo {volume} {15}},\ \bibinfo {pages} {3996--4005} (\bibinfo {year}
  {2024})}\BibitemShut {NoStop}%
\bibitem [{\citenamefont {Abascal}\ and\ \citenamefont
  {Vega}(2005)}]{abascal2005General}%
  \BibitemOpen
  \bibfield  {author} {\bibinfo {author} {\bibfnamefont {J.~L.~F.}\
  \bibnamefont {Abascal}}\ and\ \bibinfo {author} {\bibfnamefont
  {C.}~\bibnamefont {Vega}},\ }\bibfield  {title} {\enquote {\bibinfo {title}
  {A general purpose model for the condensed phases of water:
  {{TIP4P}}/2005},}\ }\href {https://doi.org/10.1063/1.2121687} {\bibfield
  {journal} {\bibinfo  {journal} {J. Chem. Phys.}\ }\textbf {\bibinfo {volume}
  {123}},\ \bibinfo {pages} {234505} (\bibinfo {year} {2005})}\BibitemShut
  {NoStop}%
\bibitem [{\citenamefont {Abascal}\ and\ \citenamefont
  {Vega}(2010)}]{abascal2010Widom}%
  \BibitemOpen
  \bibfield  {author} {\bibinfo {author} {\bibfnamefont {J.~L.~F.}\
  \bibnamefont {Abascal}}\ and\ \bibinfo {author} {\bibfnamefont
  {C.}~\bibnamefont {Vega}},\ }\bibfield  {title} {\enquote {\bibinfo {title}
  {Widom line and the liquid--liquid critical point for the {{TIP4P}}/2005
  water model},}\ }\href {https://doi.org/10.1063/1.3506860} {\bibfield
  {journal} {\bibinfo  {journal} {J. Chem. Phys.}\ }\textbf {\bibinfo {volume}
  {133}},\ \bibinfo {pages} {234502} (\bibinfo {year} {2010})}\BibitemShut
  {NoStop}%
\bibitem [{\citenamefont {Abraham}\ \emph {et~al.}(2015)\citenamefont
  {Abraham}, \citenamefont {Murtola}, \citenamefont {Schulz}, \citenamefont
  {P{\'a}ll}, \citenamefont {Smith}, \citenamefont {Hess},\ and\ \citenamefont
  {Lindahl}}]{abraham2015GROMACS}%
  \BibitemOpen
  \bibfield  {author} {\bibinfo {author} {\bibfnamefont {M.~J.}\ \bibnamefont
  {Abraham}}, \bibinfo {author} {\bibfnamefont {T.}~\bibnamefont {Murtola}},
  \bibinfo {author} {\bibfnamefont {R.}~\bibnamefont {Schulz}}, \bibinfo
  {author} {\bibfnamefont {S.}~\bibnamefont {P{\'a}ll}}, \bibinfo {author}
  {\bibfnamefont {J.~C.}\ \bibnamefont {Smith}}, \bibinfo {author}
  {\bibfnamefont {B.}~\bibnamefont {Hess}},\ and\ \bibinfo {author}
  {\bibfnamefont {E.}~\bibnamefont {Lindahl}},\ }\bibfield  {title} {\enquote
  {\bibinfo {title} {{{GROMACS}}: {{High}} performance molecular simulations
  through multi-level parallelism from laptops to supercomputers},}\ }\href
  {https://doi.org/10.1016/j.softx.2015.06.001} {\bibfield  {journal} {\bibinfo
   {journal} {SoftwareX}\ }\textbf {\bibinfo {volume} {1--2}},\ \bibinfo
  {pages} {19--25} (\bibinfo {year} {2015})}\BibitemShut {NoStop}%
\bibitem [{\citenamefont {Luzar}\ and\ \citenamefont
  {Chandler}(1996{\natexlab{a}})}]{luzar1996Effect}%
  \BibitemOpen
  \bibfield  {author} {\bibinfo {author} {\bibfnamefont {A.}~\bibnamefont
  {Luzar}}\ and\ \bibinfo {author} {\bibfnamefont {D.}~\bibnamefont
  {Chandler}},\ }\bibfield  {title} {\enquote {\bibinfo {title} {Effect of
  {{Environment}} on {{Hydrogen Bond Dynamics}} in {{Liquid Water}}},}\ }\href
  {https://doi.org/10.1103/PhysRevLett.76.928} {\bibfield  {journal} {\bibinfo
  {journal} {Phys. Rev. Lett.}\ }\textbf {\bibinfo {volume} {76}},\ \bibinfo
  {pages} {928--931} (\bibinfo {year} {1996}{\natexlab{a}})}\BibitemShut
  {NoStop}%
\bibitem [{\citenamefont {Luzar}\ and\ \citenamefont
  {Chandler}(1996{\natexlab{b}})}]{luzar1996Hydrogenbond}%
  \BibitemOpen
  \bibfield  {author} {\bibinfo {author} {\bibfnamefont {A.}~\bibnamefont
  {Luzar}}\ and\ \bibinfo {author} {\bibfnamefont {D.}~\bibnamefont
  {Chandler}},\ }\bibfield  {title} {\enquote {\bibinfo {title} {Hydrogen-bond
  kinetics in liquid water},}\ }\href {https://doi.org/10.1038/379055a0}
  {\bibfield  {journal} {\bibinfo  {journal} {Nature}\ }\textbf {\bibinfo
  {volume} {379}},\ \bibinfo {pages} {55--57} (\bibinfo {year}
  {1996}{\natexlab{b}})}\BibitemShut {NoStop}%
\bibitem [{\citenamefont {Rapaport}(1983)}]{rapaport1983Hydrogen}%
  \BibitemOpen
  \bibfield  {author} {\bibinfo {author} {\bibfnamefont {D.}~\bibnamefont
  {Rapaport}},\ }\bibfield  {title} {\enquote {\bibinfo {title} {Hydrogen bonds
  in water: {{Network}} organization and lifetimes},}\ }\href
  {https://doi.org/10.1080/00268978300102931} {\bibfield  {journal} {\bibinfo
  {journal} {Mol. Phys.}\ }\textbf {\bibinfo {volume} {50}},\ \bibinfo {pages}
  {1151--1162} (\bibinfo {year} {1983})}\BibitemShut {NoStop}%
\bibitem [{\citenamefont {Kumar}, \citenamefont {Schmidt},\ and\ \citenamefont
  {Skinner}(2007)}]{kumar2007Hydrogen}%
  \BibitemOpen
  \bibfield  {author} {\bibinfo {author} {\bibfnamefont {R.}~\bibnamefont
  {Kumar}}, \bibinfo {author} {\bibfnamefont {J.~R.}\ \bibnamefont {Schmidt}},\
  and\ \bibinfo {author} {\bibfnamefont {J.~L.}\ \bibnamefont {Skinner}},\
  }\bibfield  {title} {\enquote {\bibinfo {title} {Hydrogen bonding definitions
  and dynamics in liquid water},}\ }\href {https://doi.org/10.1063/1.2742385}
  {\bibfield  {journal} {\bibinfo  {journal} {J. Chem. Phys.}\ }\textbf
  {\bibinfo {volume} {126}},\ \bibinfo {pages} {204107} (\bibinfo {year}
  {2007})}\BibitemShut {NoStop}%
\bibitem [{\citenamefont {Soper}\ and\ \citenamefont
  {Ricci}(2000)}]{soper2000Structures}%
  \BibitemOpen
  \bibfield  {author} {\bibinfo {author} {\bibfnamefont {A.~K.}\ \bibnamefont
  {Soper}}\ and\ \bibinfo {author} {\bibfnamefont {M.~A.}\ \bibnamefont
  {Ricci}},\ }\bibfield  {title} {\enquote {\bibinfo {title} {Structures of
  {{High-Density}} and {{Low-Density Water}}},}\ }\href
  {https://doi.org/10.1103/PhysRevLett.84.2881} {\bibfield  {journal} {\bibinfo
   {journal} {Phys. Rev. Lett.}\ }\textbf {\bibinfo {volume} {84}},\ \bibinfo
  {pages} {2881--2884} (\bibinfo {year} {2000})}\BibitemShut {NoStop}%
\bibitem [{\citenamefont {Yan}\ \emph {et~al.}(2007)\citenamefont {Yan},
  \citenamefont {Buldyrev}, \citenamefont {Kumar}, \citenamefont
  {Giovambattista}, \citenamefont {Debenedetti},\ and\ \citenamefont
  {Stanley}}]{yan2007Structure}%
  \BibitemOpen
  \bibfield  {author} {\bibinfo {author} {\bibfnamefont {Z.}~\bibnamefont
  {Yan}}, \bibinfo {author} {\bibfnamefont {S.~V.}\ \bibnamefont {Buldyrev}},
  \bibinfo {author} {\bibfnamefont {P.}~\bibnamefont {Kumar}}, \bibinfo
  {author} {\bibfnamefont {N.}~\bibnamefont {Giovambattista}}, \bibinfo
  {author} {\bibfnamefont {P.~G.}\ \bibnamefont {Debenedetti}},\ and\ \bibinfo
  {author} {\bibfnamefont {H.~E.}\ \bibnamefont {Stanley}},\ }\bibfield
  {title} {\enquote {\bibinfo {title} {Structure of the first- and
  second-neighbor shells of simulated water: {{Quantitative}} relation to
  translational and orientational order},}\ }\href
  {https://doi.org/10.1103/PhysRevE.76.051201} {\bibfield  {journal} {\bibinfo
  {journal} {Phys. Rev. E}\ }\textbf {\bibinfo {volume} {76}},\ \bibinfo
  {pages} {051201} (\bibinfo {year} {2007})}\BibitemShut {NoStop}%
\bibitem [{\citenamefont {Martelli}\ \emph {et~al.}(2018)\citenamefont
  {Martelli}, \citenamefont {Ko}, \citenamefont {O{\u g}uz},\ and\
  \citenamefont {Car}}]{martelli2018Localorder}%
  \BibitemOpen
  \bibfield  {author} {\bibinfo {author} {\bibfnamefont {F.}~\bibnamefont
  {Martelli}}, \bibinfo {author} {\bibfnamefont {H.-Y.}\ \bibnamefont {Ko}},
  \bibinfo {author} {\bibfnamefont {E.~C.}\ \bibnamefont {O{\u g}uz}},\ and\
  \bibinfo {author} {\bibfnamefont {R.}~\bibnamefont {Car}},\ }\bibfield
  {title} {\enquote {\bibinfo {title} {Local-order metric for condensed-phase
  environments},}\ }\href {https://doi.org/10.1103/PhysRevB.97.064105}
  {\bibfield  {journal} {\bibinfo  {journal} {Phys. Rev. B}\ }\textbf {\bibinfo
  {volume} {97}},\ \bibinfo {pages} {064105} (\bibinfo {year}
  {2018})}\BibitemShut {NoStop}%
\bibitem [{\citenamefont {Santis}\ \emph {et~al.}(2024)\citenamefont {Santis},
  \citenamefont {Herman}, \citenamefont {Heindel},\ and\ \citenamefont
  {Xantheas}}]{santis2024Descriptors}%
  \BibitemOpen
  \bibfield  {author} {\bibinfo {author} {\bibfnamefont {G.~D.}\ \bibnamefont
  {Santis}}, \bibinfo {author} {\bibfnamefont {K.~M.}\ \bibnamefont {Herman}},
  \bibinfo {author} {\bibfnamefont {J.~P.}\ \bibnamefont {Heindel}},\ and\
  \bibinfo {author} {\bibfnamefont {S.~S.}\ \bibnamefont {Xantheas}},\
  }\bibfield  {title} {\enquote {\bibinfo {title} {Descriptors of water
  aggregation},}\ }\href {https://doi.org/10.1063/5.0179815} {\bibfield
  {journal} {\bibinfo  {journal} {J. Chem. Phys.}\ }\textbf {\bibinfo {volume}
  {160}},\ \bibinfo {pages} {054306} (\bibinfo {year} {2024})}\BibitemShut
  {NoStop}%
\bibitem [{\citenamefont {Shi}, \citenamefont {Russo},\ and\ \citenamefont
  {Tanaka}(2018)}]{shi2018Common}%
  \BibitemOpen
  \bibfield  {author} {\bibinfo {author} {\bibfnamefont {R.}~\bibnamefont
  {Shi}}, \bibinfo {author} {\bibfnamefont {J.}~\bibnamefont {Russo}},\ and\
  \bibinfo {author} {\bibfnamefont {H.}~\bibnamefont {Tanaka}},\ }\bibfield
  {title} {\enquote {\bibinfo {title} {Common microscopic structural origin for
  water's thermodynamic and dynamic anomalies},}\ }\href
  {https://doi.org/10.1063/1.5055908} {\bibfield  {journal} {\bibinfo
  {journal} {J. Chem. Phys.}\ }\textbf {\bibinfo {volume} {149}},\ \bibinfo
  {pages} {224502} (\bibinfo {year} {2018})}\BibitemShut {NoStop}%
\bibitem [{\citenamefont {{Montes de Oca}}, \citenamefont {Sciortino},\ and\
  \citenamefont {Appignanesi}(2020)}]{montesdeoca2020Structural}%
  \BibitemOpen
  \bibfield  {author} {\bibinfo {author} {\bibfnamefont {J.~M.}\ \bibnamefont
  {{Montes de Oca}}}, \bibinfo {author} {\bibfnamefont {F.}~\bibnamefont
  {Sciortino}},\ and\ \bibinfo {author} {\bibfnamefont {G.~A.}\ \bibnamefont
  {Appignanesi}},\ }\bibfield  {title} {\enquote {\bibinfo {title} {A
  structural indicator for water built upon potential energy considerations},}\
  }\href {https://doi.org/10.1063/5.0010895} {\bibfield  {journal} {\bibinfo
  {journal} {J. Chem. Phys.}\ }\textbf {\bibinfo {volume} {152}},\ \bibinfo
  {pages} {244503} (\bibinfo {year} {2020})}\BibitemShut {NoStop}%
\bibitem [{\citenamefont {Loubet}\ \emph {et~al.}(2024)\citenamefont {Loubet},
  \citenamefont {Verde}, \citenamefont {Accordino}, \citenamefont
  {Alarc{\'o}n},\ and\ \citenamefont {Appignanesi}}]{loubet2024Role}%
  \BibitemOpen
  \bibfield  {author} {\bibinfo {author} {\bibfnamefont {N.~A.}\ \bibnamefont
  {Loubet}}, \bibinfo {author} {\bibfnamefont {A.~R.}\ \bibnamefont {Verde}},
  \bibinfo {author} {\bibfnamefont {S.~R.}\ \bibnamefont {Accordino}}, \bibinfo
  {author} {\bibfnamefont {L.~M.}\ \bibnamefont {Alarc{\'o}n}},\ and\ \bibinfo
  {author} {\bibfnamefont {G.~A.}\ \bibnamefont {Appignanesi}},\ }\bibfield
  {title} {\enquote {\bibinfo {title} {Role of hydrogen-bond coordination
  defects in the structural relaxation of supercooled water},}\ }\href
  {https://doi.org/10.1103/PhysRevE.110.054601} {\bibfield  {journal} {\bibinfo
   {journal} {Phys. Rev. E}\ }\textbf {\bibinfo {volume} {110}},\ \bibinfo
  {pages} {054601} (\bibinfo {year} {2024})}\BibitemShut {NoStop}%
\bibitem [{\citenamefont {Foffi}, \citenamefont {Russo},\ and\ \citenamefont
  {Sciortino}(2021)}]{foffi2021Structural}%
  \BibitemOpen
  \bibfield  {author} {\bibinfo {author} {\bibfnamefont {R.}~\bibnamefont
  {Foffi}}, \bibinfo {author} {\bibfnamefont {J.}~\bibnamefont {Russo}},\ and\
  \bibinfo {author} {\bibfnamefont {F.}~\bibnamefont {Sciortino}},\ }\bibfield
  {title} {\enquote {\bibinfo {title} {Structural and topological changes
  across the liquid--liquid transition in water},}\ }\href
  {https://doi.org/10.1063/5.0049299} {\bibfield  {journal} {\bibinfo
  {journal} {J. Chem. Phys.}\ }\textbf {\bibinfo {volume} {154}},\ \bibinfo
  {pages} {184506} (\bibinfo {year} {2021})}\BibitemShut {NoStop}%
\bibitem [{\citenamefont {Foffi}\ and\ \citenamefont
  {Sciortino}(2021)}]{foffi2021Structure}%
  \BibitemOpen
  \bibfield  {author} {\bibinfo {author} {\bibfnamefont {R.}~\bibnamefont
  {Foffi}}\ and\ \bibinfo {author} {\bibfnamefont {F.}~\bibnamefont
  {Sciortino}},\ }\bibfield  {title} {\enquote {\bibinfo {title} {Structure of
  {{High-Pressure Supercooled}} and {{Glassy Water}}},}\ }\href
  {https://doi.org/10.1103/PhysRevLett.127.175502} {\bibfield  {journal}
  {\bibinfo  {journal} {Phys. Rev. Lett.}\ }\textbf {\bibinfo {volume} {127}},\
  \bibinfo {pages} {175502} (\bibinfo {year} {2021})}\BibitemShut {NoStop}%
\bibitem [{\citenamefont {Steinhardt}, \citenamefont {Nelson},\ and\
  \citenamefont {Ronchetti}(1983)}]{steinhardt1983Bondorientational}%
  \BibitemOpen
  \bibfield  {author} {\bibinfo {author} {\bibfnamefont {P.~J.}\ \bibnamefont
  {Steinhardt}}, \bibinfo {author} {\bibfnamefont {D.~R.}\ \bibnamefont
  {Nelson}},\ and\ \bibinfo {author} {\bibfnamefont {M.}~\bibnamefont
  {Ronchetti}},\ }\bibfield  {title} {\enquote {\bibinfo {title}
  {Bond-orientational order in liquids and glasses},}\ }\href
  {https://doi.org/10.1103/PhysRevB.28.784} {\bibfield  {journal} {\bibinfo
  {journal} {Phys. Rev. B}\ }\textbf {\bibinfo {volume} {28}},\ \bibinfo
  {pages} {784--805} (\bibinfo {year} {1983})}\BibitemShut {NoStop}%
\bibitem [{\citenamefont {Leocmach}()}]{Pyboo}%
  \BibitemOpen
  \bibfield  {author} {\bibinfo {author} {\bibfnamefont {M.}~\bibnamefont
  {Leocmach}},\ }\href@noop {} {\enquote {\bibinfo {title} {Pyboo},}\ }\bibinfo
  {howpublished} {\url{https://github.com/MathieuLeocmach/pyboo}}\BibitemShut
  {NoStop}%
\bibitem [{\citenamefont {Montoro}\ and\ \citenamefont
  {Abascal}(1993)}]{montoro1993Voronoi}%
  \BibitemOpen
  \bibfield  {author} {\bibinfo {author} {\bibfnamefont {J.~C.~G.}\
  \bibnamefont {Montoro}}\ and\ \bibinfo {author} {\bibfnamefont {J.~L.~F.}\
  \bibnamefont {Abascal}},\ }\bibfield  {title} {\enquote {\bibinfo {title}
  {The {{Voronoi}} polyhedra as tools for structure determination in simple
  disordered systems},}\ }\href {https://doi.org/10.1021/j100118a044}
  {\bibfield  {journal} {\bibinfo  {journal} {J. Phys. Chem.}\ }\textbf
  {\bibinfo {volume} {97}},\ \bibinfo {pages} {4211--4215} (\bibinfo {year}
  {1993})}\BibitemShut {NoStop}%
\bibitem [{\citenamefont {Shih}, \citenamefont {Sheu},\ and\ \citenamefont
  {Mou}(1994)}]{shih1994Voronoi}%
  \BibitemOpen
  \bibfield  {author} {\bibinfo {author} {\bibfnamefont {J.-P.}\ \bibnamefont
  {Shih}}, \bibinfo {author} {\bibfnamefont {S.-Y.}\ \bibnamefont {Sheu}},\
  and\ \bibinfo {author} {\bibfnamefont {C.-Y.}\ \bibnamefont {Mou}},\
  }\bibfield  {title} {\enquote {\bibinfo {title} {A {{Voronoi}} polyhedra
  analysis of structures of liquid water},}\ }\href
  {https://doi.org/10.1063/1.466517} {\bibfield  {journal} {\bibinfo  {journal}
  {J. Chem. Phys.}\ }\textbf {\bibinfo {volume} {100}},\ \bibinfo {pages}
  {2202--2212} (\bibinfo {year} {1994})}\BibitemShut {NoStop}%
\bibitem [{\citenamefont {Jordan}()}]{Pyvoro}%
  \BibitemOpen
  \bibfield  {author} {\bibinfo {author} {\bibfnamefont {J.}~\bibnamefont
  {Jordan}},\ }\href@noop {} {\enquote {\bibinfo {title} {Pyvoro},}\ }\bibinfo
  {howpublished} {\url{https://github.com/joe-jordan/pyvoro}}\BibitemShut
  {NoStop}%
\bibitem [{\citenamefont {Abadi}\ \emph {et~al.}(2015)\citenamefont {Abadi},
  \citenamefont {Agarwal}, \citenamefont {Barham}, \citenamefont {Brevdo},
  \citenamefont {Chen}, \citenamefont {Citro}, \citenamefont {Corrado},
  \citenamefont {Davis}, \citenamefont {Dean}, \citenamefont {Devin},
  \citenamefont {Ghemawat}, \citenamefont {Goodfellow}, \citenamefont {Harp},
  \citenamefont {Irving}, \citenamefont {Isard}, \citenamefont {Jia},
  \citenamefont {Jozefowicz}, \citenamefont {Kaiser}, \citenamefont {Kudlur},
  \citenamefont {Levenberg}, \citenamefont {Man\'{e}}, \citenamefont {Monga},
  \citenamefont {Moore}, \citenamefont {Murray}, \citenamefont {Olah},
  \citenamefont {Schuster}, \citenamefont {Shlens}, \citenamefont {Steiner},
  \citenamefont {Sutskever}, \citenamefont {Talwar}, \citenamefont {Tucker},
  \citenamefont {Vanhoucke}, \citenamefont {Vasudevan}, \citenamefont
  {Vi\'{e}gas}, \citenamefont {Vinyals}, \citenamefont {Warden}, \citenamefont
  {Wattenberg}, \citenamefont {Wicke}, \citenamefont {Yu},\ and\ \citenamefont
  {Zheng}}]{tensorflow2015-whitepaper}%
  \BibitemOpen
  \bibfield  {author} {\bibinfo {author} {\bibfnamefont {M.}~\bibnamefont
  {Abadi}}, \bibinfo {author} {\bibfnamefont {A.}~\bibnamefont {Agarwal}},
  \bibinfo {author} {\bibfnamefont {P.}~\bibnamefont {Barham}}, \bibinfo
  {author} {\bibfnamefont {E.}~\bibnamefont {Brevdo}}, \bibinfo {author}
  {\bibfnamefont {Z.}~\bibnamefont {Chen}}, \bibinfo {author} {\bibfnamefont
  {C.}~\bibnamefont {Citro}}, \bibinfo {author} {\bibfnamefont {G.~S.}\
  \bibnamefont {Corrado}}, \bibinfo {author} {\bibfnamefont {A.}~\bibnamefont
  {Davis}}, \bibinfo {author} {\bibfnamefont {J.}~\bibnamefont {Dean}},
  \bibinfo {author} {\bibfnamefont {M.}~\bibnamefont {Devin}}, \bibinfo
  {author} {\bibfnamefont {S.}~\bibnamefont {Ghemawat}}, \bibinfo {author}
  {\bibfnamefont {I.}~\bibnamefont {Goodfellow}}, \bibinfo {author}
  {\bibfnamefont {A.}~\bibnamefont {Harp}}, \bibinfo {author} {\bibfnamefont
  {G.}~\bibnamefont {Irving}}, \bibinfo {author} {\bibfnamefont
  {M.}~\bibnamefont {Isard}}, \bibinfo {author} {\bibfnamefont
  {Y.}~\bibnamefont {Jia}}, \bibinfo {author} {\bibfnamefont {R.}~\bibnamefont
  {Jozefowicz}}, \bibinfo {author} {\bibfnamefont {L.}~\bibnamefont {Kaiser}},
  \bibinfo {author} {\bibfnamefont {M.}~\bibnamefont {Kudlur}}, \bibinfo
  {author} {\bibfnamefont {J.}~\bibnamefont {Levenberg}}, \bibinfo {author}
  {\bibfnamefont {D.}~\bibnamefont {Man\'{e}}}, \bibinfo {author}
  {\bibfnamefont {R.}~\bibnamefont {Monga}}, \bibinfo {author} {\bibfnamefont
  {S.}~\bibnamefont {Moore}}, \bibinfo {author} {\bibfnamefont
  {D.}~\bibnamefont {Murray}}, \bibinfo {author} {\bibfnamefont
  {C.}~\bibnamefont {Olah}}, \bibinfo {author} {\bibfnamefont {M.}~\bibnamefont
  {Schuster}}, \bibinfo {author} {\bibfnamefont {J.}~\bibnamefont {Shlens}},
  \bibinfo {author} {\bibfnamefont {B.}~\bibnamefont {Steiner}}, \bibinfo
  {author} {\bibfnamefont {I.}~\bibnamefont {Sutskever}}, \bibinfo {author}
  {\bibfnamefont {K.}~\bibnamefont {Talwar}}, \bibinfo {author} {\bibfnamefont
  {P.}~\bibnamefont {Tucker}}, \bibinfo {author} {\bibfnamefont
  {V.}~\bibnamefont {Vanhoucke}}, \bibinfo {author} {\bibfnamefont
  {V.}~\bibnamefont {Vasudevan}}, \bibinfo {author} {\bibfnamefont
  {F.}~\bibnamefont {Vi\'{e}gas}}, \bibinfo {author} {\bibfnamefont
  {O.}~\bibnamefont {Vinyals}}, \bibinfo {author} {\bibfnamefont
  {P.}~\bibnamefont {Warden}}, \bibinfo {author} {\bibfnamefont
  {M.}~\bibnamefont {Wattenberg}}, \bibinfo {author} {\bibfnamefont
  {M.}~\bibnamefont {Wicke}}, \bibinfo {author} {\bibfnamefont
  {Y.}~\bibnamefont {Yu}},\ and\ \bibinfo {author} {\bibfnamefont
  {X.}~\bibnamefont {Zheng}},\ }\href@noop {} {\enquote {\bibinfo {title}
  {{TensorFlow}: Large-scale machine learning on heterogeneous systems},}\
  }\bibinfo {howpublished} {\url{https://www.tensorflow.org/}} (\bibinfo {year}
  {2015})\BibitemShut {NoStop}%
\bibitem [{\citenamefont {Ribeiro}()}]{ribeiro2026Marcotcr}%
  \BibitemOpen
  \bibfield  {author} {\bibinfo {author} {\bibfnamefont {M.~T.~C.}\
  \bibnamefont {Ribeiro}},\ }\href@noop {} {\enquote {\bibinfo {title}
  {Lime},}\ }\bibinfo {howpublished}
  {\url{https://github.com/marcotcr/lime}}\BibitemShut {NoStop}%
\end{thebibliography}
\end{document}


\setcounter{equation}{0}
\setcounter{figure}{0}
\setcounter{table}{0}
\setcounter{page}{1}

\renewcommand{\theequation}{S.\arabic{equation}}
\renewcommand{\theHequation}{SI.\arabic{equation}}

\renewcommand{\figurename}{Supplementary Fig.}
\renewcommand{\thefigure}{\arabic{figure}}
\renewcommand{\theHfigure}{SI.\arabic{figure}}

\renewcommand{\tablename}{Supplementary Table}
\renewcommand{\thetable}{\arabic{table}}
\renewcommand{\theHtable}{SI.\arabic{table}}

\noindent{\bf\Large Supplementary Information}
\vspace{5mm}
\begin{center}
\textbf{\large  
Machine learning evaluation of structural descriptors for supercooled water
\\
}

\vspace{5mm}
 
{Kohei Yoshikawa, Kokoro Shikata, Kang Kim, and Nobuyuki Matubayasi}
\\

\vspace{5mm}

\noindent
\textit{
Division of Chemical Engineering, Graduate School of Engineering Science, The University of Osaka, Osaka 560-8531, Japan}

\end{center}

\begin{figure}[H]
\centering
\includegraphics[width=\linewidth]{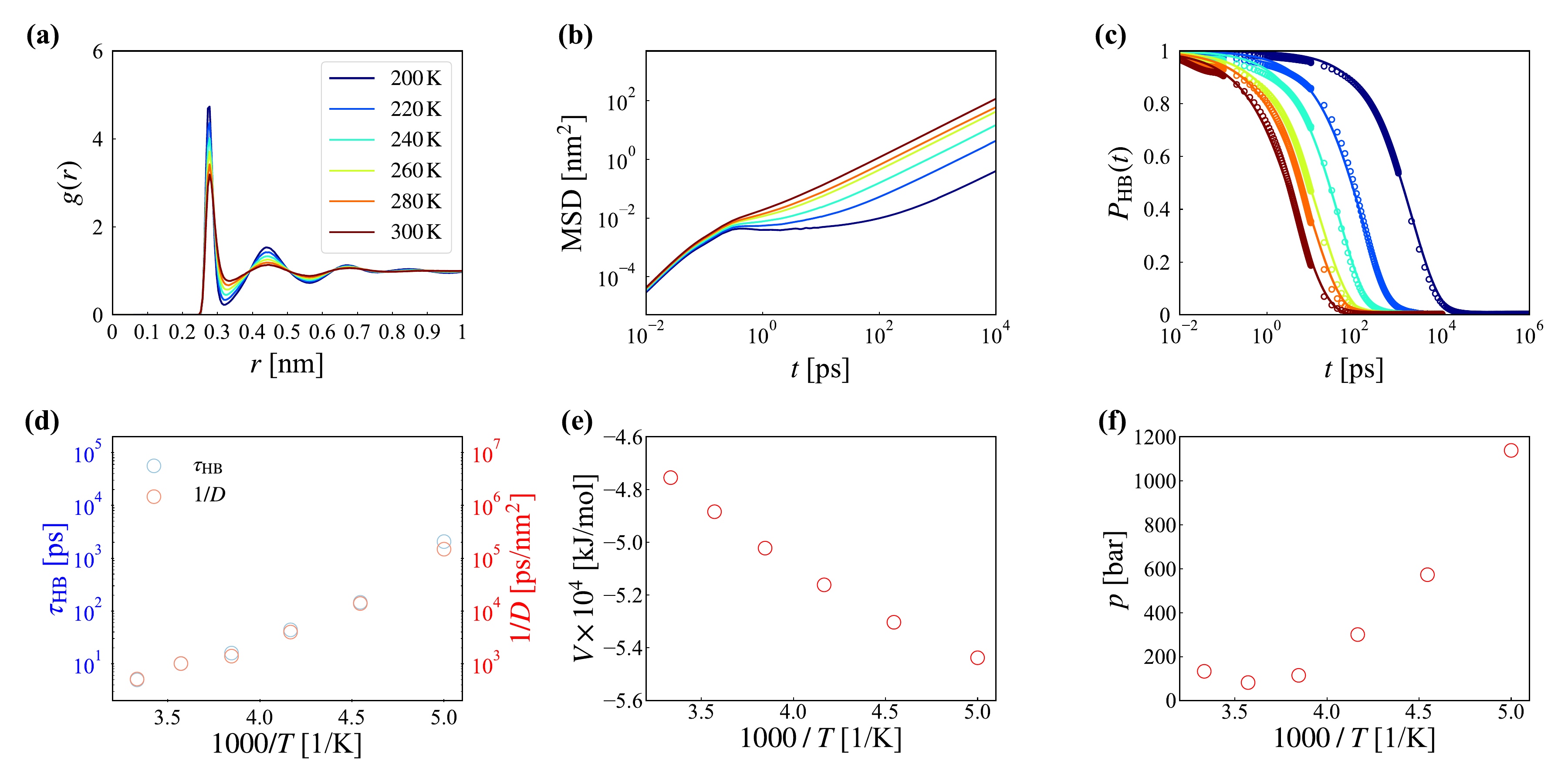}
\caption{Temperature dependence of the O-O radial distribution function $g(r)$
 (a), mean square displacement (MSD) of O atoms (b), hydroben-bond (H-Bond)
 time-correlation function $P_\mathrm{HB}(t)$ (c), H-bond lifetime
 $\tau_\mathrm{HB}$ (blue) and inverse of diffusion coefficient $1/D$
 (red) (d), potential energy $V$ (e), and pressure $p$ (f) under
 isochoric conditions at a mass density of 1 g/cm$^5$.
In panel (c), circles represent results from MD simulations, while solid
 curves represent fits to the Kohlrausch--Williams--Watts
 (KWW) function, $P_\mathrm{HB}(t)
=\exp[-(t/\tau_\mathrm{KWW})^{\beta_\mathrm{KWW}}]$.
}
\label{fig:property_NVT}
\end{figure}

\newpage
\begin{figure}[H]
\centering
\includegraphics[width=\linewidth]{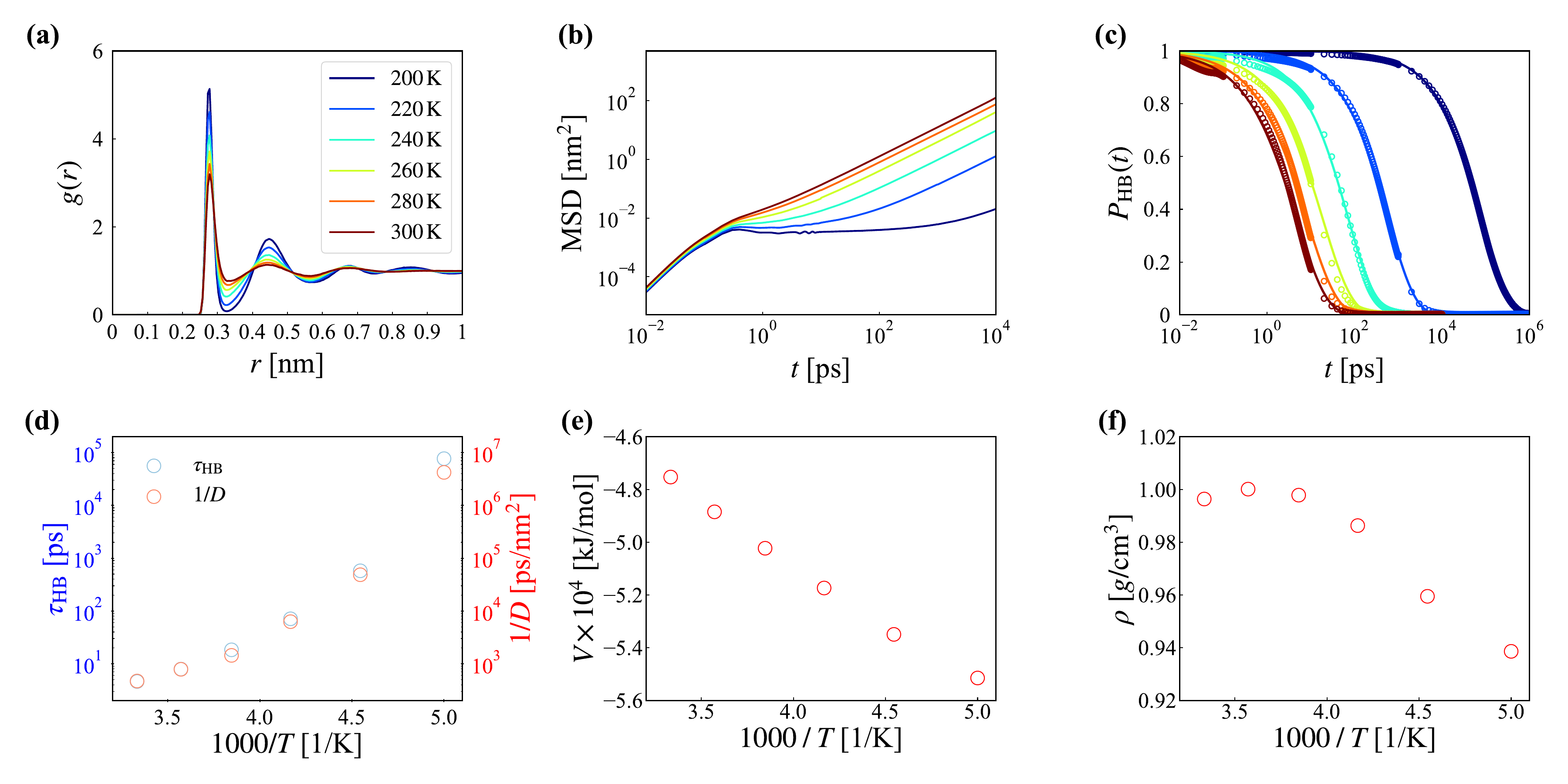}
\caption{Temperature dependence of the O-O radial distribution function $g(r)$
 (a), mean square displacement (MSD) of O atoms (b), hydroben-bond (H-bond)
 time-correlation function $P_\mathrm{HB}(t)$ (c), H-bond lifetime
 $\tau_\mathrm{HB}$ (blue) and inverse of diffusion coefficient $1/D$
 (red) (d), potential energy $V$ (e), and mass density $\rho$ (f) under
 isobaric conditions at a pressure of 1 bar.
In panel (c), circles represent results from MD simulations, while solid
 curves represent fits to the Kohlrausch--Williams--Watts
 (KWW) function, $P_\mathrm{HB}(t)
=\exp[-(t/\tau_\mathrm{KWW})^{\beta_\mathrm{KWW}}]$.
}
\label{fig:property_NPT}
\end{figure}

\newpage
\begin{figure}[H]
\centering
\includegraphics[width=\linewidth]{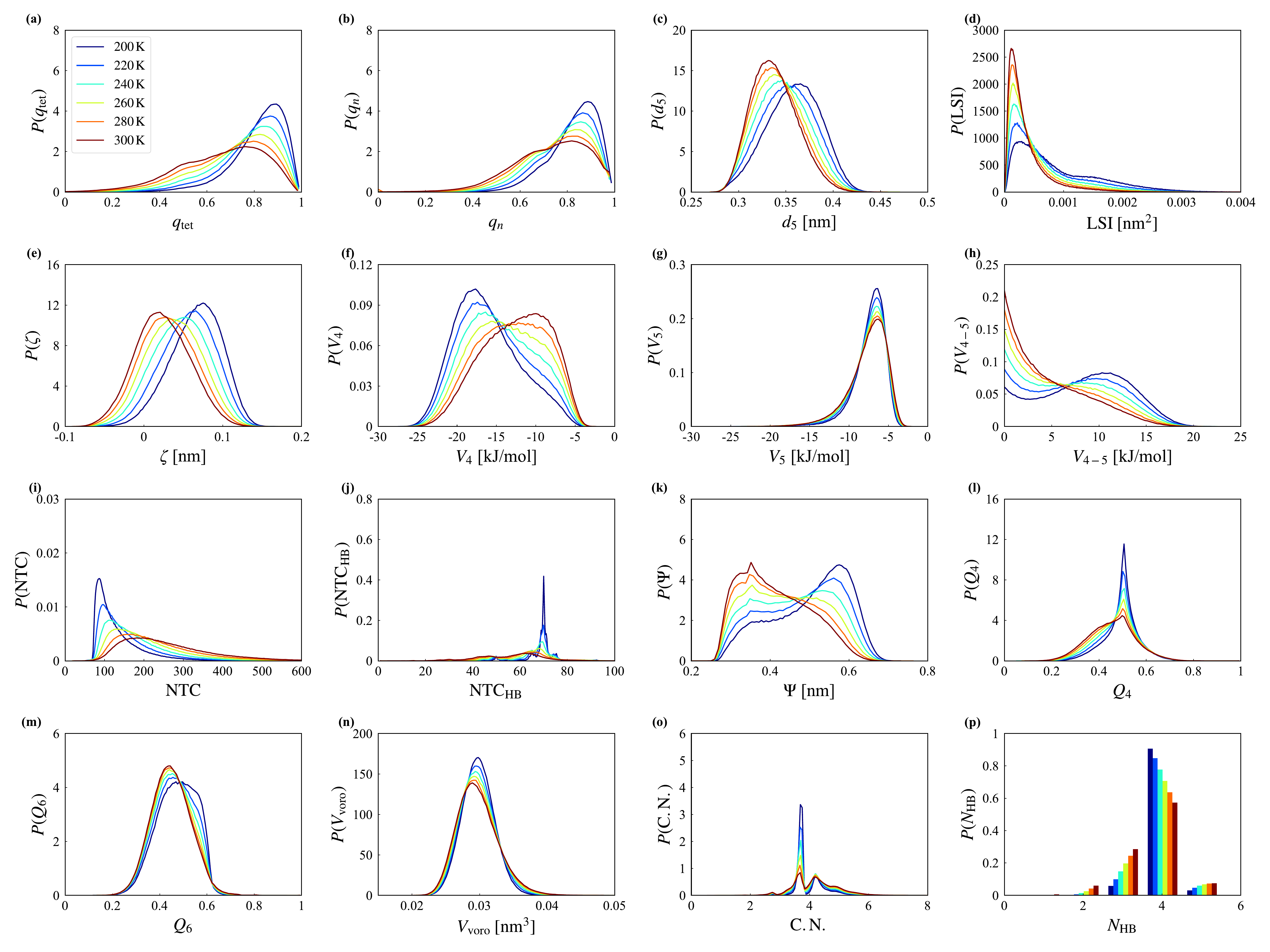}
\caption{Temperature-dependent distributions of structural descriptors,
 $q_\mathrm{tet}$ (a), $q_n$ (b), $d_5$ (c),  LSI (d), $\zeta$(e), 
 $V_4$ (f), $V_5$ (g), $V_{4-5}$ (h), NTC (i), NTC$_\mathrm{HB}$ (j), 
 $\Psi$ (k), $Q_4$ (l), $Q_6$ (m), $V_\mathrm{voro}$ (n), C.N. (o), 
 $N_\mathrm{HB}$ (p) under isochoric conditions at a mass density of 1 g/cm$^3$.
Panels (a)-(o) show the probability density, whereas panel (p) shows the
 probability.
}
\label{fig:descriptor_NVT}
\end{figure}

\newpage
\begin{figure}[H]
\centering
\includegraphics[width=\linewidth]{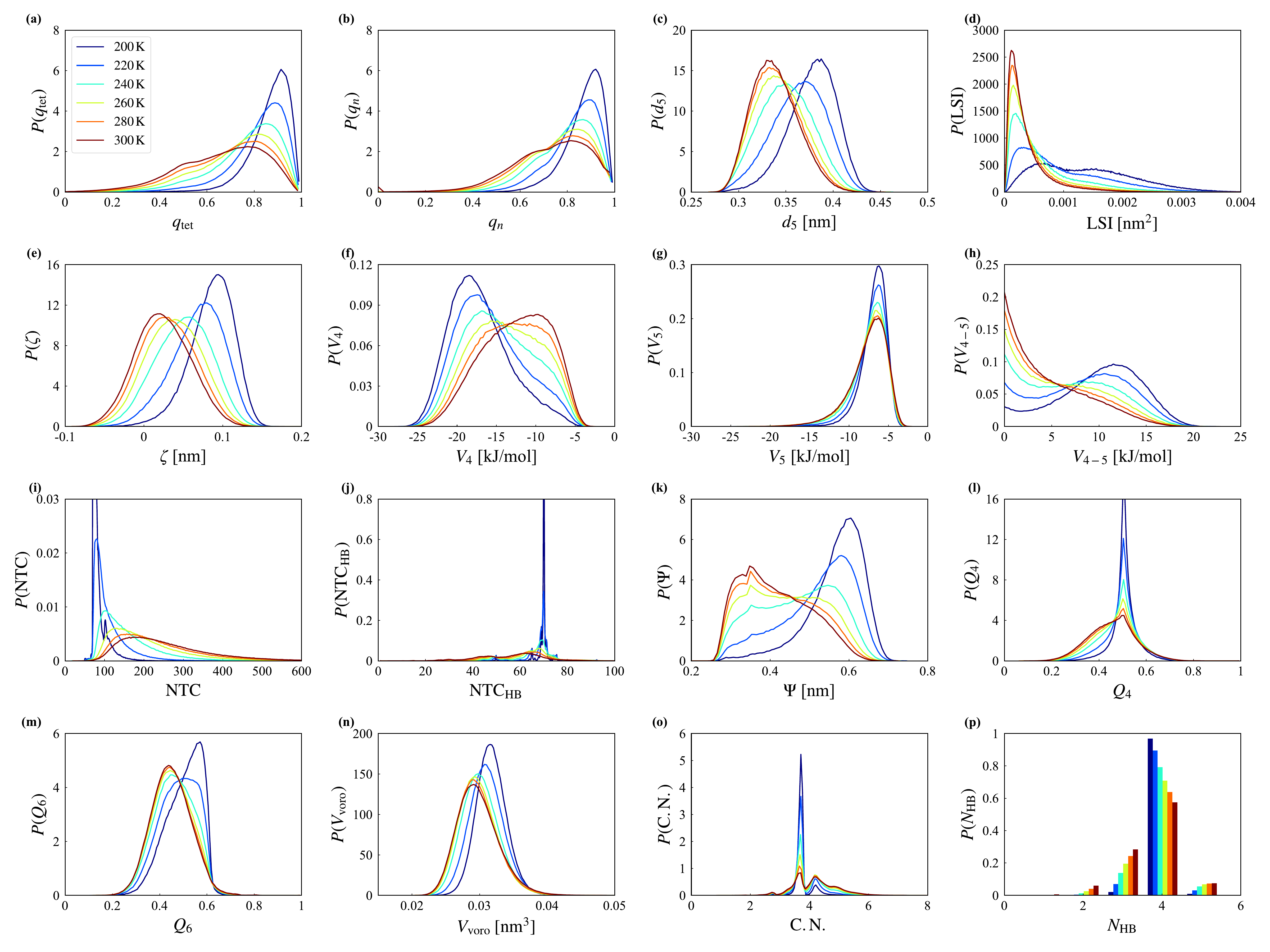}
\caption{Temperature-dependent distributions of structural descriptors,
 $q_\mathrm{tet}$ (a), $q_n$ (b), $d_5$ (c),  LSI (d), $\zeta$(e), 
 $V_4$ (f), $V_5$ (g), $V_{4-5}$ (h), NTC (i), NTC$_\mathrm{HB}$ (j), 
 $\Psi$ (k), $Q_4$ (l), $Q_6$ (m), $V_\mathrm{voro}$ (n), C.N. (o), 
 $N_\mathrm{HB}$ (p) under isobaric conditions at a pressure of 1 bar.
Panels (a)-(o) show the probability density, whereas panel (p) shows the
 probability.
}
\label{fig:descriptor_NPT}
\end{figure}

\newpage
\begin{figure}[H]
\centering
\includegraphics[width=0.5\linewidth]{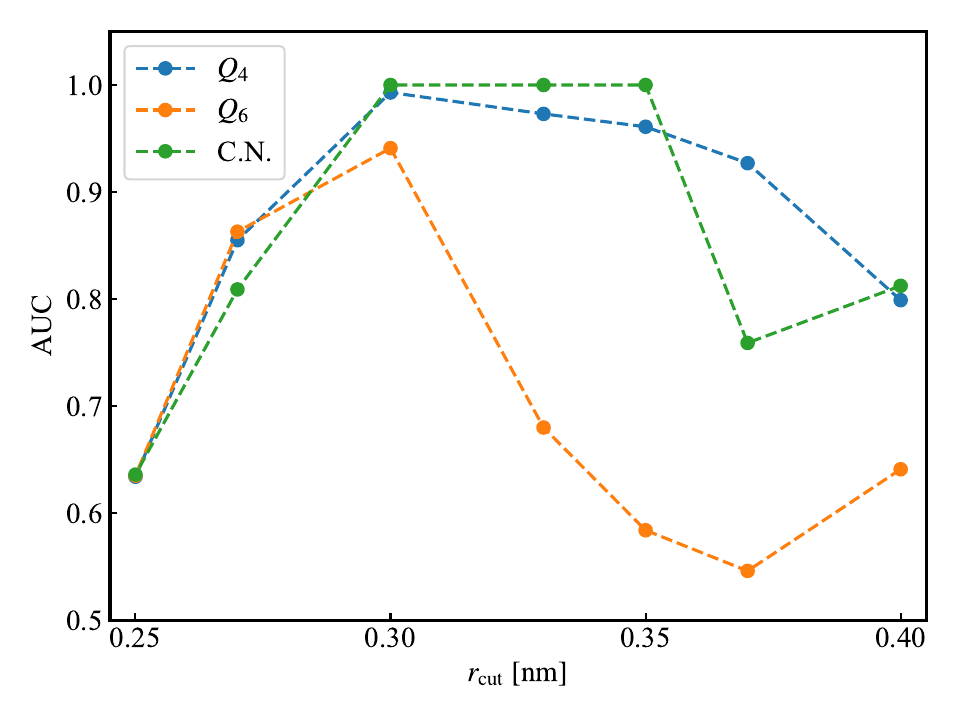}
\caption{Dependence of the AUC for $Q_4$, $Q_6$, and C.N. on the cutoff
 distance $r_\mathrm{cut}$ for the temperature combination of 200 K and
 300 K under isochoric conditions.
}
\label{fig:Q4_Q6_CN}
\end{figure}

\begin{figure}[H]
\centering
\includegraphics[width=\linewidth]{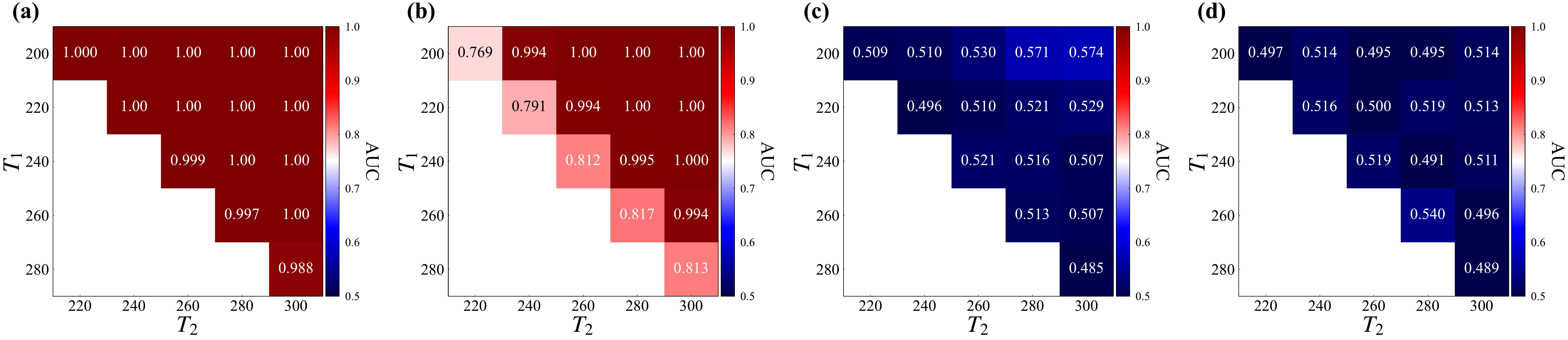}
\caption{Classification performance obtained using logistic regression for the
 structural descriptors (a) LSI, (b) $\zeta$, (c) NTC$_\mathrm{HB}$,
 and (d) $N_\mathrm{HB}$ under isochoric conditions.
Definitions and color scales are identical to those in
 Fig.~3 in the main text.
}
\label{fig:linear_NVT}
\end{figure}

\begin{figure}[H]
\centering
\includegraphics[width=\linewidth]{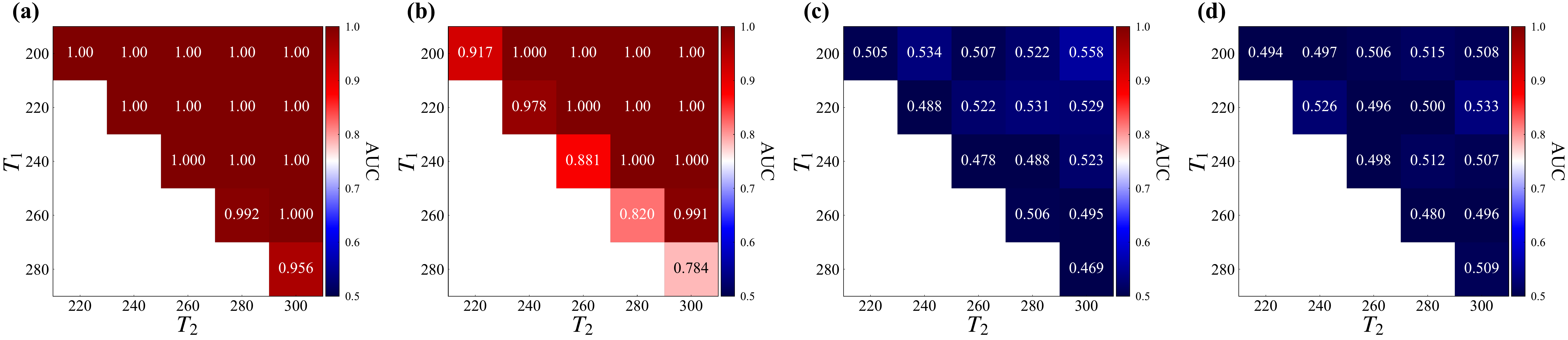}
\caption{Classification performance obtained using logistic regression for the
 structural descriptors (a) LSI, (b) $\zeta$, (c) NTC$_\mathrm{HB}$,
 and (d) $N_\mathrm{HB}$ under isobaric conditions.
Definitions and color scales are identical to those in
 Fig.~3 in the main text.}
\label{fig:linear_NPT}
\end{figure}

\newpage
\begin{figure}[H]
\centering
\includegraphics[width=\linewidth]{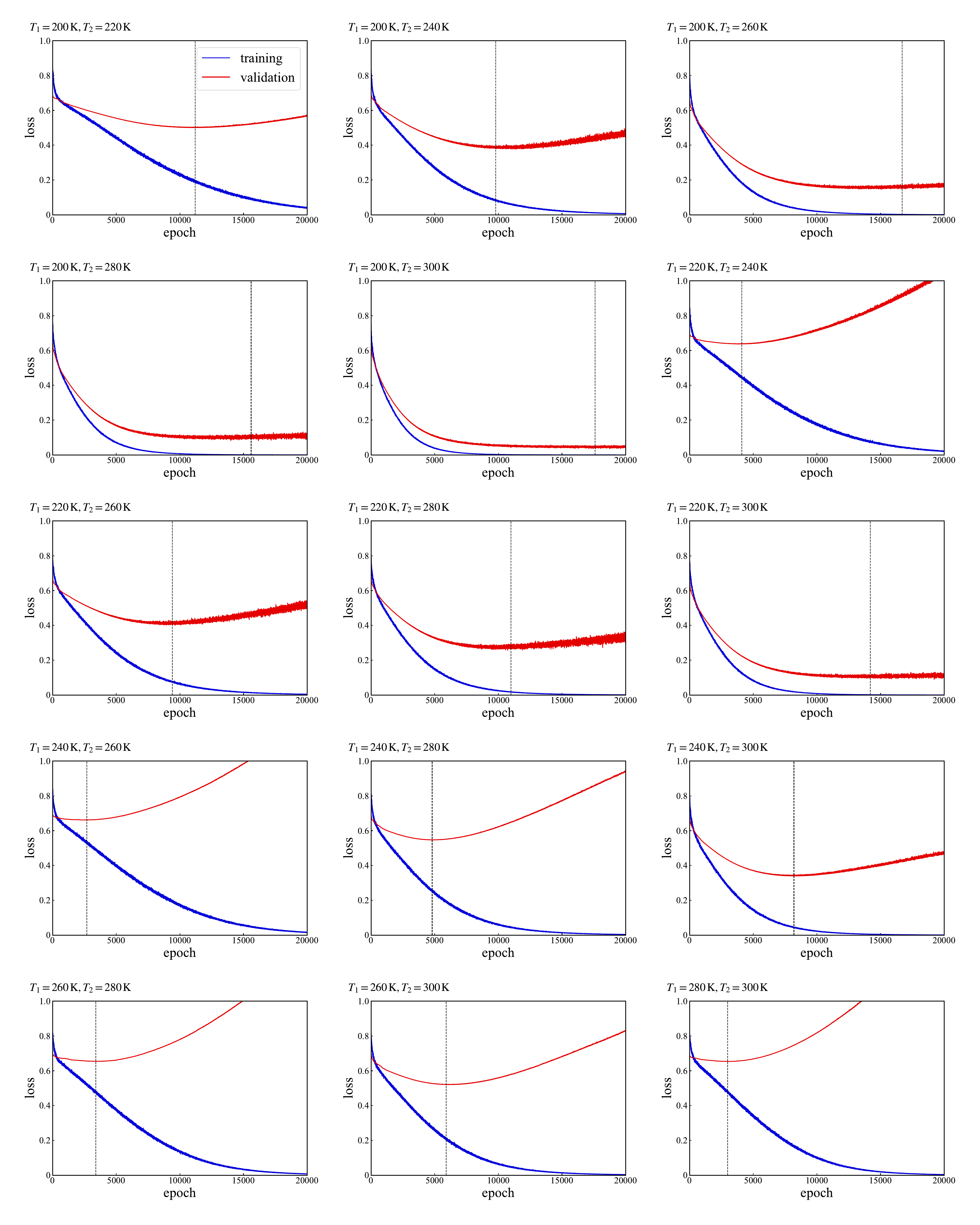}
\caption{Epoch dependence of the loss function for training data (blue) and
 validation data (red) when using $q_\mathrm{tet}$ across various temperature
 combinations under isochoric conditions.
Horizontal lines indicate the epoch at which the validation loss 
 reaches its minimum, and the AUC was calculated using
 the neural network at that point.
}
\label{fig:loss_NVT}
\end{figure}

\newpage
\begin{figure}[H]
\centering
\includegraphics[width=\linewidth]{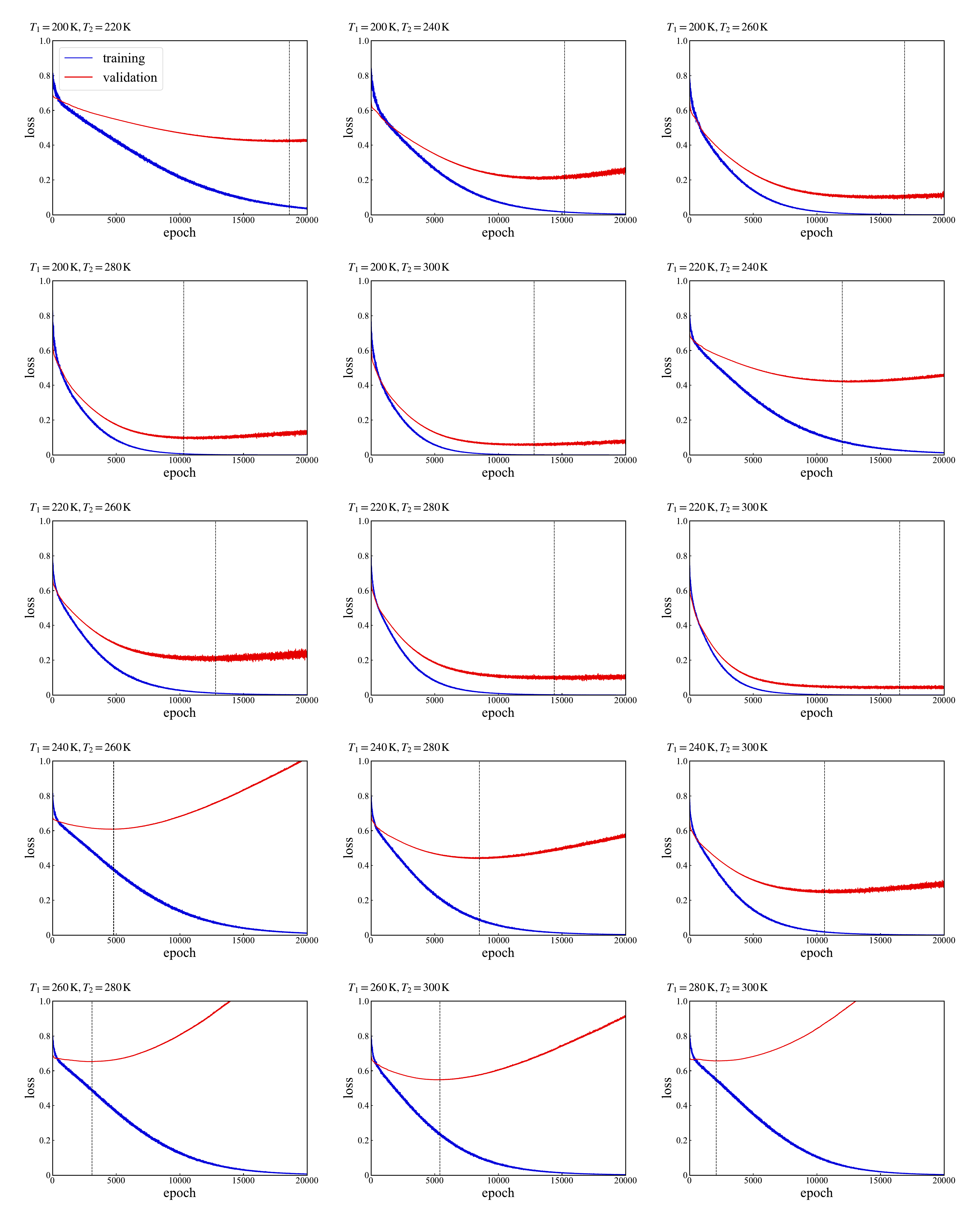}
\caption{Epoch dependence of the loss function for training data (blue) and
 validation data (red) when using $q_\mathrm{tet}$ across various temperature
 combinations under isobaric conditions.
Horizontal lines indicate the epoch at which the validation loss 
 reaches its minimum, and the AUC was calculated using
 the neural network at that point.
}
\label{fig:loss_NPT}
\end{figure}

\newpage
\begin{figure}[H]
\centering
\includegraphics[width=0.4\linewidth]{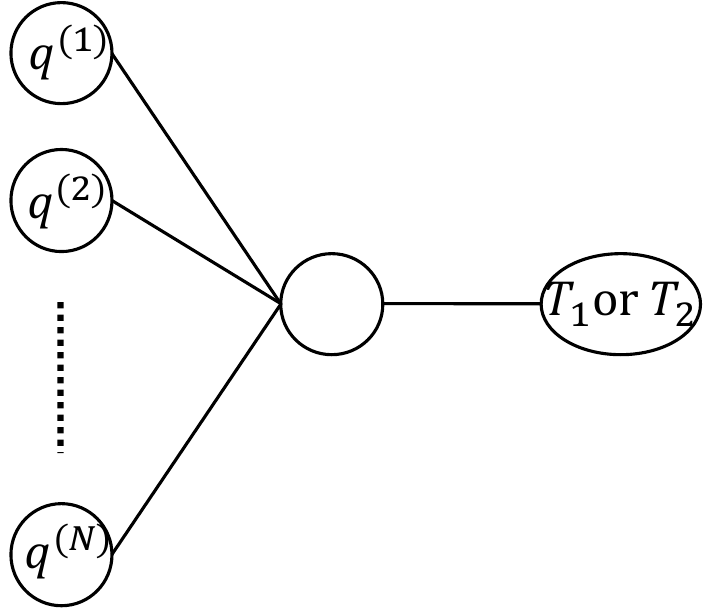}
\caption{Shematic diagram of the logistic regression model used for the binary classification. 
From the descriptor value $q^{(i)}$ assigned to molecule $i$, the output
 is computed as $\sum_{i=1}^M w_i q^{(i)}+ w_0$, where $w_i$ are the
 corresponding coefficients.
The output is then passed through a sigmoid activation function
 to 
predict the
 temperature, $T_1$ or $T_2$.
Note that the node representing the bias term $w_0$ is omitted from the diagram.
}
\label{fig:LR}
\end{figure}